\documentclass[11pt]{article}
\usepackage{hyperref,url}
\usepackage{latexsym,graphicx,mathrsfs,amsfonts,subfig}
\usepackage{slashed, color}
\usepackage{multirow}
\usepackage{tikz}
\usepackage{empheq}
\usepackage{amsmath}
\allowdisplaybreaks[2]
\usepackage[multiple]{footmisc}
\usepackage{textcomp}
\usepackage{amssymb}
\usetikzlibrary{positioning}
\usepackage{amscd}
\usepackage{amsthm}
\usepackage{array} 

\renewcommand{\thefootnote}{\fnsymbol{footnote}}
\numberwithin{equation}{section}
\usepackage{dsfont}
\usepackage[toc,page]{appendix}

\newcommand*\widefbox[1]{\fbox{\hspace{2em}#1\hspace{2em}}}

\DeclareFontFamily{U}{MnSymbolC}{}
\DeclareSymbolFont{MnSyC}{U}{MnSymbolC}{m}{n}
\DeclareFontShape{U}{MnSymbolC}{m}{n}{
	<-6>  MnSymbolC5
	<6-7>  MnSymbolC6
	<7-8>  MnSymbolC7
	<8-9>  MnSymbolC8
	<9-10> MnSymbolC9
	<10-12> MnSymbolC10
	<12->   MnSymbolC12}{}
\DeclareMathSymbol{\intprod}{\mathbin}{MnSyC}{'270}
\newcommand{\ov}{\overline}
\newcommand{\C}{\mathbb{C}}

\newcommand{\del}{\partial}

\newcommand{\M}{\mathbb{M}}
\newcommand{\til}{\widetilde}

\let\nc\newcommand
\let\renc\renewcommand
\nc{\wbar}{\overline}
\let\td\tilde
\let\wtd\widetilde
\let\wht\widehat
\let\mcl\mathcal

\nc{\ab}{{\bar{a}}} \nc{\at}{\tilde{a}} \nc{\ah}{\hat{a}}
\nc{\bb}{{\bar{b}}} 
\nc{\bh}{\hat{b}}
\nc{\cb}{{\bar{c}}} \nc{\ct}{\tilde{c}} 
\nc{\db}{{\bar{d}}} \nc{\dt}{\tilde{d}} \renc{\dh}{\hat{d}}
\nc{\eb}{{\bar{e}}} \nc{\et}{\tilde{e}} \nc{\eh}{\hat{e}}
\nc{\fb}{{\bar{f}}} \nc{\ft}{\tilde{f}} \nc{\fh}{\hat{f}}
\nc{\ib}{{\bar{\imath}}} \nc{\ih}{\hat{\imath}} 
\nc{\jb}{{\bar{\jmath}}} \nc{\jt}{\tilde{\jmath}} \nc{\jh}{\hat{\jmath}}
\nc{\kb}{{\bar{k}}} \nc{\kt}{\tilde{k}} \nc{\kh}{\hat{k}}
\nc{\lb}{{\bar{l}}} \nc{\lt}{\tilde{l}} \nc{\lh}{\hat{l}}
\nc{\mb}{{\bar{m}}} \nc{\mt}{\tilde{m}} \nc{\mh}{\hat{m}}
\nc{\nb}{{\bar{n}}} \nc{\nt}{\tilde{n}} \nc{\nh}{\hat{n}}
\nc{\ob}{{\bar{o}}} \nc{\ot}{\tilde{o}} \nc{\oh}{\hat{o}}
\nc{\pb}{{\bar{p}}} \nc{\pt}{\tilde{p}} \nc{\ph}{\hat{p}}
\nc{\qb}{{\bar{q}}} \nc{\qt}{\tilde{q}} \nc{\qh}{\hat{q}}
\nc{\rb}{{\bar{r}}} \nc{\rt}{\tilde{r}} \nc{\rh}{\hat{r}}
\renc{\sb}{{\bar{s}}} \nc{\st}{\tilde{s}} \nc{\sh}{\hat{s}}
\nc{\tb}{{\bar{t}}} \renc{\th}{\hat{t}} 
\nc{\ub}{{\bar{u}}} \nc{\ut}{\tilde{u}} \nc{\uh}{\hat{u}}
\nc{\vb}{{\bar{v}}} \nc{\vt}{\tilde{v}} \nc{\vh}{\hat{v}}
\nc{\wt}{\tilde{w}} \nc{\wh}{\hat{w}}
\nc{\xb}{{\bar{x}}} \nc{\xt}{\tilde{x}} \nc{\xh}{\hat{x}}
\nc{\yb}{{\bar{y}}} \nc{\yt}{\tilde{y}} \nc{\yh}{\hat{y}}
\nc{\zb}{{\bar{z}}} \nc{\zt}{\tilde{z}} 

\nc{\Ab}{\wbar{A}} \nc{\At}{\wtd{A}} \nc{\Ah}{\wht{A}}
\nc{\Bb}{\wbar{B}} \nc{\Bt}{\wtd{B}} \nc{\Bh}{\wht{B}}
\nc{\Cb}{\wbar{C}} \nc{\Ct}{\wtd{C}} \nc{\Ch}{\wht{C}}
\nc{\Db}{\wbar{D}} \nc{\Dt}{\wtd{D}} \nc{\Dh}{\wht{D}}
\nc{\Eb}{\wbar{E}} \nc{\Et}{\wtd{E}} \nc{\Eh}{\wht{E}}
\nc{\Fb}{\wbar{F}} \nc{\Ft}{\wtd{F}} \nc{\Fh}{\wht{F}}
\nc{\Gb}{\wbar{G}} \nc{\Gt}{\wtd{G}} \nc{\Gh}{\wht{G}}
\nc{\Hb}{\wbar{H}} \nc{\Ht}{\wtd{H}} \nc{\Hh}{\wht{H}}
\nc{\Ib}{\wbar{I}} \nc{\It}{\wtd{I}} \nc{\Ih}{\wht{I}}
\nc{\Jb}{\wbar{J}} \nc{\Jt}{\wtd{J}} \nc{\Jh}{\wht{J}}
\nc{\Kb}{\wbar{K}} \nc{\Kt}{\wtd{K}} \nc{\Kh}{\wht{K}}
\nc{\Lb}{\wbar{L}} \nc{\Lt}{\wtd{L}} \nc{\Lh}{\wht{L}}
\nc{\Mb}{\wbar{M}} \nc{\Mt}{\wtd{M}} \nc{\Mh}{\wht{M}}
\nc{\Nb}{\wbar{N}} \nc{\Nt}{\wtd{N}} \nc{\Nh}{\wht{N}}
\nc{\Ob}{\wbar{O}} \nc{\Ot}{\wtd{O}} \nc{\Oh}{\wht{O}}
\nc{\Pb}{\wbar{P}} \nc{\Pt}{\wtd{P}} \nc{\Ph}{\wht{P}}
\nc{\Qb}{\wbar{Q}} \nc{\Qt}{\wtd{Q}} \nc{\Qh}{\wht{Q}}
\nc{\Rb}{\wbar{R}} \nc{\Rt}{\wtd{R}} \nc{\Rh}{\wht{R}}
\nc{\Sb}{\wbar{S}} \nc{\St}{\wtd{S}} \nc{\Sh}{\wht{S}}
\nc{\Tb}{\wbar{T}} \nc{\Tt}{\wtd{T}} \nc{\Th}{\wht{T}}
\nc{\Ub}{\wbar{U}} \nc{\Ut}{\wtd{U}} \nc{\Uh}{\wht{U}}
\nc{\Vb}{\wbar{V}} \nc{\Vt}{\wtd{V}} \nc{\Vh}{\wht{V}}
\nc{\Wb}{\wbar{W}} \nc{\Wt}{\wtd{W}} \nc{\Wh}{\wht{W}}
\nc{\Xb}{\wbar{X}} \nc{\Xt}{\wtd{X}} \nc{\Xh}{\wht{X}}
\nc{\Yb}{\wbar{Y}} \nc{\Yt}{\wtd{Y}} \nc{\Yh}{\wht{Y}}
\nc{\Zb}{\wbar{Z}} \nc{\Zt}{\wtd{Z}} \nc{\Zh}{\wht{Z}}

\nc{\CA}{\mcl{A}} \nc{\CAb}{\wbar{\CA}} \nc{\CAt}{\wtd{\CA}} \nc{\CAh}{\wht{\CA}}
\nc{\CB}{\mcl{B}} \nc{\CBb}{\wbar{\CB}} \nc{\CBt}{\wtd{\CB}} \nc{\CBh}{\wht{\CB}}
\nc{\CC}{\mcl{C}} \nc{\CCb}{\wbar{\CC}} \nc{\CCt}{\wtd{\CC}} \nc{\CCh}{\wht{\CC}}
\nc{\cDt}{\wtd{\cC}} \nc{\cDh}{\wht{\cD}}
\nc{\CE}{\mcl{E}} \nc{\CEb}{\wbar{\CE}} \nc{\CEt}{\wtd{\CE}} \nc{\CEh}{\wht{\CE}}
\nc{\CF}{\mcl{F}} \nc{\CFb}{\wbar{\CF}} \nc{\CFt}{\wtd{\CF}} \nc{\CFh}{\wht{\CF}}
\nc{\CG}{\mcl{G}} \nc{\CGb}{\wbar{\CG}} \nc{\CGt}{\wtd{\CG}} \nc{\CGh}{\wht{\CG}}
\nc{\CH}{\mcl{H}} \nc{\CHb}{\wbar{\CH}} \nc{\CHt}{\wtd{\CH}} \nc{\CHh}{\wht{\CH}}
\nc{\CI}{\mcl{I}} \nc{\CIb}{\wbar{\CI}} \nc{\CIt}{\wtd{\CI}} \nc{\CIh}{\wht{\CI}}
\nc{\CJ}{\mcl{J}} \nc{\CJb}{\wbar{\CJ}} \nc{\CJt}{\wtd{\CJ}} \nc{\CJh}{\wht{\CJ}}
\nc{\CK}{\mcl{K}} \nc{\CKb}{\wbar{\CK}} \nc{\CKt}{\wtd{\CK}} \nc{\CKh}{\wht{\CK}}
\nc{\CL}{\mcl{L}} \nc{\CLb}{\wbar{\CL}} \nc{\CLt}{\wtd{\CL}} \nc{\CLh}{\wht{\CL}}
\nc{\CM}{\mcl{M}} \nc{\CMb}{\wbar{\CM}} \nc{\CMt}{\wtd{\CM}} \nc{\CMh}{\wht{\CM}}
\nc{\CN}{\mcl{N}} \nc{\CNb}{\wbar{\CN}} \nc{\CNt}{\wtd{\CN}} \nc{\CNh}{\wht{\CN}}
\nc{\CO}{\mcl{O}} \nc{\COb}{\wbar{\CO}} \nc{\COt}{\wtd{\CO}} \nc{\COh}{\wht{\CO}}
\nc{\CQ}{\mcl{Q}} \nc{\CQb}{\wbar{\CQ}} \nc{\CQt}{\wtd{\CQ}} \nc{\CQh}{\wht{\CQ}}
\nc{\CR}{\mcl{R}} \nc{\CRb}{\wbar{\CR}} \nc{\CRt}{\wtd{\CR}} \nc{\CRh}{\wht{\CR}}
\nc{\CS}{\mcl{S}} \nc{\CSb}{\wbar{\CS}} \nc{\CSt}{\wtd{\CS}} \nc{\CSh}{\wht{\CS}}
\nc{\CT}{\mcl{T}} \nc{\CTb}{\wbar{\CT}} \nc{\CTt}{\wtd{\CT}} \nc{\CTh}{\wht{\CT}}
\nc{\CU}{\mcl{U}} \nc{\CUb}{\wbar{\CU}} \nc{\CUt}{\wtd{\CU}} \nc{\CUh}{\wht{\CU}}
\nc{\CV}{\mcl{V}} \nc{\CVb}{\wbar{\CV}} \nc{\CVt}{\wtd{\CV}} \nc{\CVh}{\wht{\CV}}
\nc{\CW}{\mcl{W}} \nc{\CWb}{\wbar{\CW}} \nc{\CWt}{\wtd{\CW}} \nc{\CWh}{\wht{\CW}}
\nc{\CX}{\mcl{X}} \nc{\CXb}{\wbar{\CX}} \nc{\CXt}{\wtd{\CX}} \nc{\CXh}{\wht{\CX}}
\nc{\CY}{\mcl{Y}} \nc{\CYb}{\wbar{\CY}} \nc{\CYt}{\wtd{\CY}} \nc{\CYh}{\wht{\CY}}
\nc{\CZ}{\mcl{Z}} \nc{\CZb}{\wbar{\CZ}} \nc{\CZt}{\wtd{\CZ}} \nc{\CZh}{\wht{\CZ}}

\let\eps\epsilon
\let\ups\upsilon
\let\veps\varepsilon
\let\vtht\vartheta
\let\vsgm\varsigma
\let\vphi\varphi
\let\vrho\varrho

\nc{\alphab}{\bar{\alpha}} \nc{\alphat}{\td{\alpha}} \nc{\alphah}{\hat{\alpha}}
\nc{\betab}{\bar{\beta}}   \nc{\betat}{\td{\beta}}   \nc{\betah}{\hat{\beta}} 
\nc{\gammab}{\bar{\gamma}} \nc{\gammat}{\td{\gamma}} \nc{\gammah}{\hat{\gamma}} 
\nc{\deltab}{\bar{\delta}} \nc{\deltat}{\td{\delta}} \nc{\deltah}{\hat{\delta}} 
\nc{\epsilonb}{\bar{\eps}} \nc{\epsilont}{\td{\eps}} \nc{\epsilonh}{\hat{\eps}} 
\nc{\vepsb}{\bar{\veps}}   \nc{\vepst}{\td{\veps}}   \nc{\vepsh}{\hat{\veps}} 
\nc{\zetab}{\bar{\zeta}}   \nc{\zetat}{\td{\zeta}}   \nc{\zetah}{\hat{\zeta}} 
\nc{\etab}{\bar{\eta}}     
\nc{\etah}{\hat{\eta}} 
\nc{\thetab}{\bar{\theta}} \nc{\thetat}{\td{\theta}} \nc{\thetah}{\hat{\theta}} 
\nc{\vthetab}{\bar{\vtht}} \nc{\vthetat}{\td{\vtht}} \nc{\vthetah}{\hat{\vtht}} 
\nc{\lambdat}{\td{\lambda}} \nc{\lambdah}{\hat{\lambda}} 
\nc{\iotab}{\bar{\iota}}   \nc{\iotat}{\td{\iota}}   \nc{\iotah}{\hat{\iota}} 
\nc{\kappab}{\bar{\kappa}} \nc{\kappat}{\td{\kappa}} \nc{\kappah}{\hat{\kappa}} 
\nc{\lmdb}{\bar{\lmd}}     \nc{\lmdt}{\td{\lmd}}     \nc{\lmdh}{\hat{\lmd}} 
\nc{\mub}{\bar{\mu}}       \nc{\mut}{\td{\mu}}       \nc{\muh}{\hat{\mu}} 
\nc{\nub}{\bar{\nu}}       \nc{\nut}{\td{\nu}}       \nc{\nuh}{\hat{\nu}} 
\nc{\xib}{\bar{\xi}}       \nc{\xit}{\td{\xi}}       \nc{\xih}{\hat{\xi}} 
\nc{\pib}{\bar{\pi}}       \nc{\pit}{\td{\pi}}       \nc{\pih}{\hat{\pi}} 
\nc{\vpib}{\bar{\vpi}}     \nc{\vpit}{\td{\vpi}}     \nc{\vpih}{\hat{\vpi}} 
\nc{\rhob}{\bar{\rho}}     \nc{\rhot}{\td{\rho}}     \nc{\rhoh}{\hat{\rho}} 
\nc{\vrhob}{\bar{\vrho}}   \nc{\vrhot}{\td{\vrho}}   \nc{\vrhoh}{\hat{\vrho}} 
\nc{\sigmab}{\bar{\sigma}} \nc{\sigmat}{\td{\sigma}} \nc{\sigmah}{\hat{\sigma}} 
\nc{\vsigmab}{\bar{\vsgm}} \nc{\vsigmat}{\td{\vsgm}} \nc{\vsigmah}{\hat{\vsgm}} 
\nc{\taub}{\bar{\tau}}     \nc{\taut}{\td{\tau}}     \nc{\tauh}{\hat{\tau}} 
\nc{\upsb}{\bar{\ups}} \nc{\upst}{\td{\ups}} \nc{\upsh}{\hat{\ups}} 
\nc{\phib}{\bar{\phi}}     \nc{\phit}{\td{\phi}}     \nc{\phih}{\hat{\phi}} 
\nc{\varphib}{\bar{\vphi}}   \nc{\varphit}{\td{\vphi}}   \nc{\varphih}{\hat{\vphi}} 
\nc{\chib}{\bar{\chi}}     
\nc{\chih}{\hat{\chi}} 
\nc{\psib}{\bar{\psi}}     
\nc{\psih}{\hat{\psi}} 
\nc{\omegab}{\bar{\omega}} \nc{\omegat}{\td{\omega}} \nc{\omegah}{\hat{\omega}} 

\nc{\Gammab}{\wbar{\Gamma}}     \nc{\Gammat}{\wtd{\Gamma}}     \nc{\Gammah}{\wht{\Gamma}}
\nc{\Deltab}{\wbar{\Delta}}     \nc{\Deltat}{\wtd{\Delta}}     \nc{\Deltah}{\wht{\Delta}}
\nc{\Thetab}{\wbar{\Theta}}     \nc{\Thetat}{\wtd{\Theta}}     \nc{\Thetah}{\wht{\Theta}}
\nc{\Lambdab}{\wbar{\Lambda}}   \nc{\Lambdat}{\wtd{\Lambda}}   \nc{\Lambdah}{\wht{\Lambda}}
\nc{\Xib}{\wbar{\Xi}}           \nc{\Xit}{\wtd{\Xi}}           \nc{\Xih}{\wht{\Xi}}
\nc{\Pib}{\wbar{\Pi}}           \nc{\Pit}{\wtd{\Pi}}           \nc{\Pih}{\wht{\Pi}}
\nc{\Sigmab}{\wbar{\Sigma}}     \nc{\Sigmat}{\wtd{\Sigma}}     \nc{\Sigmah}{\wht{\Sigma}}
\nc{\Upsilonb}{\wbar{\Upsilon}} \nc{\Upsilont}{\wtd{\Upsilon}} \nc{\Upsilonh}{\wht{\Upsilon}}
\nc{\Phib}{\wbar{\Phi}}         \nc{\Phit}{\wtd{\Phi}}         \nc{\Phih}{\wht{\Phi}}
\nc{\Psib}{\wbar{\Psi}}         \nc{\Psit}{\wtd{\Psi}}         \nc{\Psih}{\wht{\Psi}}
\nc{\Omegab}{\wbar{\Omega}}     \nc{\Omegat}{\wtd{\Omega}}     \nc{\Omegah}{\wht{\Omega}}
\nc{\Varepsilon}{\mathcal{E}}

\newcommand{\cD}{{\cal D}}

\nc{\balpha}{\bar{\alpha}}
\nc{\bbeta}{\bar{\beta}}
\nc{\bgamma}{\bar{\gamma}}
\nc{\bm}{\bar{m}}
\nc{\bn}{\bar{n}}
\nc{\bp}{\bar{p}}
\nc{\al}{\alpha}
\nc{\bt}{\beta}
\nc{\gm}{\gamma}
\nc{\zh}{\wht{z}}
\nc{\zhb}{\ov{\wht{z}}}
\nc{\mbh}{\wht{\ov{m}}}
\nc{\bc}{|_{x^2=0}}

\nc{\tal}{\til{\al}}
\nc{\tbt}{\til{\bt}}
\nc{\tgm}{\til{\gm}}

\nc{\wb}{\ov{w}}
\nc{\teta}{\til{\eta}}
\nc{\tpsi}{\til{\psi}}

\def\IL{\relax{\rm I\kern-.18em L}}
\def\IH{\relax{\rm I\kern-.18em H}}
\def\IB{\relax{\rm I\kern-.18em B}}
\def\ID{\relax{\rm I\kern-.18em D}}
\def\IE{\relax{\rm I\kern-.18em E}}
\def\IF{\relax{\rm I\kern-.18em F}}

\def\IG{\relax\hbox{$\inbar\kern-.3em{\rm G}$}}
\def\IGa{\relax\hbox{${\rm I}\kern-.18em\Gamma$}}
\def\IH{\relax{\rm I\kern-.18em H}}
\def\II{\relax{\rm I\kern-.18em I}}
\def\IK{\relax{\rm I\kern-.18em K}}
\def\IP{\relax{\rm I\kern-.18em P}}
\def\IQ{\relax\hbox{$\inbar\kern-.3em{\rm Q}$}}

\def\hat{\widehat}
\def\CM {{\cal M}}
\def\CN {{\cal N}}
\def\CR {{\cal R}}

\def\CF {{\cal F}}
\def\CJ {{\cal J}}

\def\CL {{\cal L}}
\def\CV {{\cal V}}
\def\CO {{\cal O}}
\def\CZ {{\cal Z}}
\def\CE {{\cal E}}
\def\CG {{\cal G}}
\def\CH {{\cal H}}
\def\CC {{\cal C}}
\def\CB {{\cal B}}
\def\CS {{\cal S}}
\def\CA{{\cal A}}
\def\CK{{\cal K}}
\def\CQ{{\cal Q}}

\def\p{\partial}
\def\pb{{\bar \p}}

\def\vt#1#2#3{ {\vartheta[{#1 \atop  #2}](#3\vert \tau)} }

\def\jb{{\bar j}}

\def\inbar{\,\vrule height1.5ex width.4pt depth0pt}

\nc{\hTheta}{\hat{\Theta}}
\nc{\vp}{\varphi}
\nc{\tg}{\widetilde{g}}

\let\OLDthebibliography\thebibliography
\renewcommand\thebibliography[1]{
	\OLDthebibliography{#1}
	\setlength{\parskip}{5pt}
	\setlength{\itemsep}{0pt plus 0.3ex}
}
\usepackage{titlesec}
\titleformat*{\section}{\bfseries\large}
\usepackage{tikz}                                       %
\usepackage{tikz-cd}                                    %
\usetikzlibrary{cd}                                     %
\usetikzlibrary{calc}                                   %
\usepackage{lipsum}
\usepackage[sorting=none]{biblatex}                     %
\usepackage{hyperref}                                   %
\bibliography{references}                               %

\begin{document}
\addtolength{\baselineskip}{1.5mm}

\thispagestyle{empty}

\vbox{}
\vspace{3.0cm}

\begin{center}
	\centerline{\LARGE{Vafa-Witten Theory: Invariants, Floer Homologies, Higgs Bundles,}}
	\bigskip
	\centerline{\LARGE{a Geometric Langlands Correspondence, and Categorification}} 

\vspace{2.5cm}

	{Zhi-Cong~Ong\footnote{E-mail: e0338243@u.nus.edu} and Meng-Chwan~Tan\footnote{E-mail: mctan@nus.edu.sg}}
	\\[2mm]
	{\it Department of Physics\\
		National University of Singapore \\
		2 Science Drive 3, Singapore 117551} \\[1mm] 
\end{center}

\vspace{2.0cm}

\centerline{\bf Abstract}\smallskip \noindent

We revisit Vafa-Witten theory in the more general setting whereby the underlying moduli space is not that of instantons, but of the full Vafa-Witten equations. We physically derive (i) a novel Vafa-Witten four-manifold invariant associated with this moduli space, (ii) their relation to Gromov-Witten invariants, (iii) a novel Vafa-Witten Floer homology assigned to three-manifold boundaries, (iv) a novel Vafa-Witten Atiyah-Floer correspondence, (v) a proof and generalization of a conjecture by Abouzaid-Manolescu in~\cite{abouzaid2020sheaf} about the hypercohomology of a perverse sheaf of vanishing cycles, (vi) a Langlands duality of these invariants, Floer homologies and hypercohomology, and (vii) a quantum geometric Langlands correspondence with purely imaginary parameter that specializes to the classical correspondence in the zero-coupling limit, where Higgs bundles feature in (ii), (iv), (vi) and (vii). We also explain how these invariants and homologies will be categorified in the process, and discuss their higher categorification. We thereby relate differential and enumerative geometry, topology and geometric representation theory in mathematics, via a maximally-supersymmetric topological quantum field theory with electric-magnetic duality in physics.

\newpage

\renewcommand{\thefootnote}{\arabic{footnote}}
\setcounter{footnote}{0}

\tableofcontents
\section{Introduction, Summary and Acknowledgements}

\bigskip\noindent\textit{Introduction}
\vspace*{0.5em}\\
For an $\mathcal{N}=4$ SYM theory on a Euclidean four-manifold $M_4$ with gauge group $G$, where $G$ is a real, simple, compact Lie group, one can perform topological twisting in three different ways \cite{yamron1988topological}, allowing one to end up with three different twisted theories. The multiplet of an $\mathcal{N}=4$  theory contains a single gauge gauge boson $A_\mu$ ($\mu=1,2,3,4$) with spin 1, gauge fermions $\lambda_{\alpha}^{i}$ and $ \lambda_{\dot{\alpha}}^{i}$ ($\dot{\alpha},\,\alpha=1,2$) with spin $\frac{1}{2}$, and six adjoint-valued bosonic scalars $\phi_{ij}=-\phi_{ji}$ ($i=1,2,3,4$) with spin 0 in the six-dimensional representation of its $SU(4)_{\cal R}$ $R$-symmetry. Here, $\mu$ represents spacetime indices; $\alpha$, $\dot{\alpha}$ represents spinor indices of $SU(2)_L \otimes SU(2)_R$ of spacetime; and $i$, $j$ represent the internal indices of $SU(4)_{\cal R}$.\footnote{One can obtain $\mathcal{N}=4$ SYM in 4d from a dimensional reduction of $\mathcal{N}=1$ supersymmetry in 10d, by compactifying along six dimensions. This explains the adjoint-valued bosonic scalar fields being in the six-dimensional representation of the internal $SU(4)_{\cal R}$ $R$-symmetry.}

The idea of twisting in order to shift the spin of the supercharges such that they behave as scalars whence they are insensitive to the geometry of $M_4$, was pioneered by Witten in~\cite{witten1988topological}. To explain twisting, first notice that the sixteen fermions $\lambda_{\alpha}^{i}$, $ \lambda_{\dot{\alpha}}^{i}$ and thus, the sixteen supercharges $\mathcal{Q}_{\alpha}^{i}$, $\mathcal{Q}_{\dot{\alpha}}^{i}$, transform under $SU(4)_{\cal R}$. Then, twisting just involves making a choice of homomorphism  $SO(4) \to SU(4)_{\cal R}$ of the spacetime symmetry group to the $R$-symmetry group, whence the aforementioned shift in the spin of the supercharges can be effected. This will modify the spins of not just the  $\lambda^i$'s and ${\cal Q}^i$'s (where they are necessarily shifted in the same way), but also that of the $\phi_i$'s (as they also transform under $SU(4)_{\cal R}$). Amongst the sixteen supercharges with shifted spins, one can always find scalar supercharges $\mathcal{Q}$ such that $\mathcal{Q}^2=0$. Because $\mathcal{Q}$ is insensitive to the geometry of $M_4$, it is a topological supercharge whereby the generated supersymmetry remains unbroken under smooth metric deformations of $M_4$. This `shifted' theory with topological supercharge $\cal Q$, is also known as a (cohomological) Topological Quantum Field Theory (TQFT).


A feature of such a TQFT is that the action can be expressed as
\begin{equation} \label{Stop}
    S = \{\mathcal{Q}, \mathcal{V} \} + \text{topological term},
\end{equation}
where $\mathcal{V}$ is called a gauge fermion. This allows us to rescale $\cal V$ whilst leaving the path integral invariant (since the expectation value of any operator of the form $\{Q, \dots\}$ is zero), whence we can compute the path integral exactly using a convenient rescaling of $\cal V$ for which its contributions localize to a finite-dimensional moduli space.

That such a TQFT is independent of the metric can be seen from the fact that its energy-momentum tensor $ {\delta S / \delta g_{\mu\nu}} = T_{\mu\nu}$ is $\mathcal{Q}$-exact, i.e., it can be  written as $T_{\mu\nu}=\{\mathcal{Q}, G_{\mu\nu}\}$ for a certain fermionic symmetric tensor $G_{\mu\nu}$, whence a variation of the metric would leave the path integral invariant (according to our explanation in the last paragraph). That being said, it is only in this sense that the word `topological' holds, since TQFT's are \textit{not} independent of all non-topological information. We will see that there are dependencies on symplectic structures when dimensional reduction via a deformation of the metric is performed later.



Another feature of such a TQFT, is that the nilpotency of $\mathcal{Q}$ means that one can define its spectrum to be the $\cal Q$-cohomology of $\cal Q$-closed operators which are not $\cal Q$-exact that therefore have nonvanishing expectation values. These $\cal Q$-supersymmetric operators correspond to certain BPS states of the original $\mathcal{N}=4$ theory. Moreover, their correlation functions are topological invariants of $M_4$, whence they have useful mathematical applications. 

Last but not least, note that anything that is $\mathcal{Q}$-exact is cohomologous to zero. That the action can be expressed as \eqref{Stop} means that it is actually zero in  $\cal Q$-cohomology. This just reflects the fact that there are no field dynamics of the theory (since supersymmetry will allow us to integrate out non-zero modes up to a factor of $\pm1$ in the path integral). In other words, the crux of any TQFT is in the structure of its zero modes. This is also consistent with the observation that the Hamiltonian of TQFT's, $H \sim T_{00} = \{{\cal Q}, G_{00} \}$, is also zero in  $\cal Q$-cohomology, i.e., only ground states are relevant in the spectrum of a TQFT.

In this paper, we will concern ourselves with the twist leading to the theory studied in \cite{vafa1994strong}, also known as Vafa-Witten (VW) theory. Unlike in \cite{vafa1994strong}, we will consider the more general setting whereby the underlying moduli space is not that of instantons, but of the full Vafa-Witten equations. We will explore and elucidate the mathematical implications of this theory by exploiting its invariance under metric deformations of the underlying $M_4$, and its electric-magnetic S-duality. 

Let us now give a brief plan and summary of the paper.


\bigskip\noindent\textit{A Brief Plan and Summary of the Paper}
\vspace*{0.5em}

In $\S$\ref{vwgeneral}, we discuss general aspects of the VW twist leading up to the action with complexified gauge coupling parameter $\tau$, where
the theory will localize on a virtually zero-dimensional moduli space of configurations satisfying the VW equations. 
We then give a physical, path integral derivation of a novel $\tau$-dependent Vafa-Witten invariant of $M_4$, as the partition function of VW theory, in (\ref{vwk}):
\begin{equation} \label{vwk0}
  \boxed {  \mathcal{Z}_{\text{VW},M_4}(\tau, G) = \sum_{k}  a_{k} q^{m_k} }
\end{equation}
Here, $q = e^{2\pi i \tau}$, $k$ denotes the $k^{\text{th}}$ sector of the moduli space $\mathcal{M}_{\text{VW}}$ of the VW equations in (\ref{4dbps}), the number $a_k$ is given in \eqref{ak} as
\begin{equation} \label{ak-0}
\boxed{a_k =  \int_{\mathcal{M}^k_{\text{VW}}} \Omega^0 \wedge e({T_{\mathcal{M}^k_{\text{VW}}}}), \quad \text{where \, $\Omega^0(\mathcal{M}^k_{\text{VW}})=(1 + B^4)^{dim_{\mathbb{C}}\mathcal{M}^k_{\text{VW}}}$}
}
\end{equation}
$B$ is a coordinate on $\mathcal{M}^k_{\text{VW}}(A,B)$, $e$ is the signed Euler class of the tangent bundle ${T_{\mathcal{M}^k_{\text{VW}}}}$,  and $m_k$ is the corresponding VW number given in \eqref{m_k} as
\begin{equation} \label{m_k-0}
  \boxed{  m_k =  \frac{ 1}{8\pi^2}\int_{M_4} \text{Tr}\, \bigg(F_{(k)}\wedge F_{(k)}  +  dB_{(k)}\wedge \star DB_{(k)} + B_{(k)}\wedge d(\star DB_{(k)})\bigg) }
\end{equation} 
where $A_{(k)}$ is a one-form $G$-connection with two-form curvature $F_{(k)}$, and $B_{(k)}$ is a self-dual two-form.  

When $B = 0$, $a_k$ will become the Euler characteristic $\chi(\mathcal{M}^k_{\text{inst}})$, while $m_k$ will become the instanton number. Then, $\mathcal{Z}_{\text{VW},M_4}$ will just become the usual partition function for instantons first derived in~\cite{vafa1994strong}, as expected. 
 
 In $\S$\ref{sigmareduction}, we compactify VW theory on $M_4=\Sigma\times C$ along $C$, where both $\Sigma$ and $C$ are closed Riemann surfaces, and $C$ has a genus $g \geq 2$. This allows us to arrive at an $A$-model in complex structure $I$ on $\Sigma$ with $\mathcal{N}=(4,4)$ supersymmetry and target space $\mathcal{M}^G_H(C)$, the moduli space of Hitchin's equations on $C$. In complex structure $I$, $\mathcal{M}^G_H(C)$ can be identified with $\mathcal{M}^G_{\text{Higgs}}(C)$, the moduli space of stable Higgs $G$-bundles on $C$. 
 We then show that the partition function of the $A$-model in \eqref{ZAclosed} gives a $\tau$-dependent Gromov-Witten (GW) invariant in (\ref{GW invariant}): 
 \begin{equation}\label{GW invariant0}
    \boxed{\mathcal{Z}_{\text{GW},\Sigma}(\tau, \mathcal{M}^G_{\text{Higgs}}(C)) = \sum_l {\tilde a}_l q^{{\tilde m}_l} }  
\end{equation}
where $l$ denotes the $l^{\text{th}}$ sector of the moduli space $\mathcal{M}_{\text{maps}}$ of holomorphic maps described in (\ref{Mmaps final}) for \emph{genus one}, the rational number ${\tilde a}_l$ is given in \eqref{al} as
 \begin{equation} \label{al-0}
 \boxed{{\tilde a}_l = \int_{{\cal M}^l_{\text {maps}}} e(\mathcal{V}) }
 \end{equation}
where $e$ is the signed Euler class of the vector bundle $\mathcal{V}$ with fiber $H^0(\Sigma, K \otimes \Phi^*T^*{{\cal M}^l_{\text {maps}}})$ and canonical bundle $K$ on $\Sigma$,
and ${\tilde m}_l$ is the corresponding worldsheet instanton number given in \eqref{q_l} as
\begin{equation} \label{q_l-0}
    \boxed{{\tilde m}_l = \frac{1}{2\pi}\int_{\Sigma}\,\Phi^{*}_l(\omega_I)}
\end{equation}
Here, $\omega_I$ is the symplectic two-form of $\mathcal{M}^G_{\text{Higgs}}(C))$.  In turn, the topological invariance of VW theory will mean that we have a 4d-2d correspondence of partition functions in (\ref{4d2dpartition}), whence we have a correspondence between the VW and GW invariants in (\ref{VW=GW}):
 \begin{equation}\label{VW=GW0}
    \boxed{\mathcal{Z}_{\text{VW},M_4}(\tau, G) = \mathcal{Z}_{\text{GW},\Sigma}(\tau, \mathcal{M}^G_{\text{Higgs}}(C)) }  
\end{equation}
In other words, we have a correspondence between the VW invariant of $M_4 = \Sigma \times C$ and the GW invariant of $\mathcal{M}^G_{\text{Higgs}}(C))$. 
In fact, \eqref{VW=GW} means that we have, in \eqref{ak=al}, 
\begin{equation}
\label{ak=al-0}
 \boxed{   a_k = \tilde{a}_l}
\end{equation}
Thus, one can also determine the $a_k$'s, the VW invariants of $T^2 \times C$, via the signed Euler class of a bundle $\mathcal{V}$ over $\mathcal{M}^l_\text{maps}$.


In $\S$\ref{vwsqm}, we consider boundary VW theory on $M_4=M_3 \times \mathbb{R}^+$, with $M_3$ a closed three-manifold, where in temporal gauge, one can now interpret $A$ and $B$ as one-forms on $M_3$. Then, we will recast the 4d theory as 1d supersymmetric quantum mechanics (SQM) on $\frak A$, the space of all complexified connections ${\cal A} = A + i B$ of a $G_{\mathbb{C}}$-bundle on $M_3$,\footnote{
$G_{\mathbb{C}}$ denotes a complex (algebraic reductive) group, that is a complexification of $G$.
}
with potential being the complex Chern-Simons functional, which action is \eqref{vwsqmcomplex-final}.
This will in turn allow us to compute the partition function as \eqref{4d3dpartitionfinal}:
\begin{equation}\label{4d3dpartitionfinal-0}
    \boxed{\mathcal{Z}_{\text{VW},M_4}(\tau, G)  = \sum_k {\cal F}^{G,\tau}_{\text{VW}}(\Psi_{M_3}^k)
    =\sum_k \text{HF}_{d_k}^{\text{VW}}(M_3, G, \tau) = \mathcal{Z}^{\text{Floer}}_{\text{VW},M_3}(\tau,G)}
\end{equation}
Here, $\text{HF}_*^{\text{VW}}(M_3, G, \tau)$ is a \emph{novel} Vafa-Witten Floer homology assigned to $M_3$ defined by the Morse functional in \eqref{HFvw functional}:
\begin{equation} \label{HFvw functional-0}
 \boxed{  CS(\mathcal{A})= -\frac{1}{4\pi^2}\int_{M_3}\text{Tr}\bigg(\mathcal{A}\wedge d\mathcal{A} + \frac{2}{3}\mathcal{A}\wedge \mathcal{A}\wedge \mathcal{A}\bigg) }
\end{equation}
with Floer differential described by the gradient flow equation \eqref{HFvw grad flow}:
\begin{equation}\label{HFvw grad flow-0}
\boxed  {  {d{\mathcal{A}}^i \over dt} =- s g_{\frak A}^{ij}\frac{\partial CS({\cal A})}{\partial {\cal A}^{j}}}
\end{equation}
where the $\tau$-dependence is due to a factor of $q^{s_k}$ that is present in the $k^{\text{th}}$ term of the above summation, and the number $s_k$ is given in \eqref{sk} as 
\begin{equation} \label{sk-0}
    \boxed{s_k = \frac{1}{8\pi^2}\int_{M_3} \text{Tr}\bigg(A_{(k)}\wedge dA_{(k)} + \frac{2}{3}A_{(k)}\wedge A_{(k)}\wedge A_{(k)} + B_{(k)}\wedge \star D B_{(k)}\bigg)}
\end{equation}
Here, $(A_{(k)},B_{(k)})$ are the $k^{\text{th}}$ time-invariant solution to the VW equations on $M_3 \times \mathbb{R}^+$ restricted to $M_3$.



In $\S$\ref{vwafconj}, we continue with an $M_4=M_3 \times \mathbb{R}^+$ and perform a Heegaard split of $M_3$ along the Riemann surface $C$. Topological invariance of VW theory then allows us to compactify $C$ and equate the resulting theory with the original uncompactified theory. Via the calculations in $\S$\ref{sigmareduction}, we find that the resulting theory is an open $A$-model with boundaries given by Lagrangian $(A,B,A)$-branes $L_0$ and $L_1$ in $\mathcal{M}^G_{\text{Higgs}}(C)$, where they represent solutions to the relevant equations on the left and right Heegaard split pieces of $M_3$, respectively. Then, via the expression \eqref{4d3dpartitionfinal-0} for the original theory, we will be able to obtain a \emph{novel} Vafa-Witten Atiyah-Floer correspondence in~\eqref{atconj} as
\begin{equation}\label{atconj0}
    \boxed{\text{HF}_{*}^{\text{VW}}(M_3, G, \tau) \cong \text{HF}_{*}^{\text{Lagr}}\big(\mathcal{M}^G_{\text{Higgs}}(C), L_0, L_1, \tau\big)} 
\end{equation}
where $\text{HF}^{\text{Lagr}}_*$ is the Lagrangian Floer homology of $L_0$ and $L_1$ in $\mathcal{M}^G_{\text{Higgs}}(C)$.

Also, a hypercohomology $\text{HP}^*(M_3)$ of a perverse sheaf of vanishing cycles in the moduli space of irreducible flat $SL(2, \mathbb{C})$-connections on $M_3$ was constructed by Abouzaid-Manolescu in \cite{abouzaid2020sheaf}, where it was conjectured to be isomorphic to instanton Floer homology assigned to $M_3$ for the complex gauge group $SL(2, \mathbb{C})$. We proceed further in this section to physically prove this conjecture. To this end, we first physically realize the result of \cite[Remark 6.15]{brav2012symmetries} in \eqref{HP - HF^lag} as 
\begin{equation} \label{HP - HF^lag summary}
  \boxed{ \text{HP}^*(M_3) \cong   \text{HF}_{*}^{\text{Lagr}}\big(X_{\text {irr}}(C), L_0, L_1, \tau\big) }
 \end{equation} 
where $X_{\text {irr}}(C)$ is the moduli space of irreducible flat $SL(2, \mathbb{C})$-connections on $C$. Next, from \eqref{HFvw functional-0} and \eqref{HFvw grad flow-0}, and the fact that the $(A,B,A)$-branes $L_0$ and $L_1$ can also be interpreted as Lagrangian branes in $\mathcal{M}^{G}_H(C)$ in complex structure $K$ i.e., $\mathcal{M}^{G_{\mathbb{C}}}_{\text{flat}}(C)$, the moduli space of irreducible flat $G_\mathbb{C}$-connections on $C$, we find that \eqref{atconj0} can also be expressed as an Atiyah-Floer correspondence for $G_{\mathbb{C}}$-instantons, whence for $G_\mathbb{C} = SL(2, \mathbb{C})$, the RHS of \eqref{HP - HF^lag summary} can be identified with the LHS of \eqref{atconj0}, such that we will have in \eqref{AB-Mano conj}
\begin{equation} \label{AB-Mano conj summary}
 \boxed{   \text{HP}^*(M_3) \cong \text{HF}_{*}^{\text{inst}}(M_3, SL(2,{\mathbb{C}}), \tau) }
\end{equation}
for some value of $\tau$. This is exactly the aforementioned conjecture by Abouzaid-Manolescu in \cite{abouzaid2020sheaf}. 

Clearly, since the underlying VW theory is defined for general $G$, the above results for $SL(2, \mathbb{C})$ can be generalized to $G_{\mathbb{C}}$. In particular, we have, in \eqref{HP - HF^lag general},
\begin{equation} \label{HP - HF^lag general summary}
  \boxed{   \text{HP}^*(M_3, G_\mathbb{C}) \cong \text{HF}_{*}^{\text{Lagr}}\big({\cal M}^{G_\mathbb{C}}_{\text{flat}}(C), L_0, L_1, \tau\big)}
 \end{equation} 
which again physically realizes the result of \cite[Remark 6.15]{brav2012symmetries}, and, in \eqref{AB-Mano conj generalized},
\begin{equation} \label{AB-Mano conj general summary}
 \boxed{   \text{HP}^*(M_3, G_{\mathbb{C}}) \cong \text{HF}_{*}^{\text{inst}}(M_3, G_{\mathbb{C}}, \tau) }
\end{equation}
which is a $G_\mathbb{C}$ generalization of the Abouzaid-Manolescu conjecture. Our physically derived generalization is also consistent with their arguments in \cite[sect.~9.1]{abouzaid2020sheaf} which show that a generalization to $SL(N, {C})$ is mathematically possible.

In $\S$\ref{sec:langlands duality}, we will show that $S$-duality of VW theory will result in a Langlands duality of the invariants, Floer homologies and hypercohomology stated hitherto. Specifically, we have, in \eqref{vwdual},
\begin{equation}\label{vwdual - 0}
  \boxed{  \mathcal{Z}_{\text{VW},M_4}(\tau, G) \longleftrightarrow \mathcal{Z}_{\text{VW},M_4}\Big(-\frac{1}{n_{\mathfrak{g}}\tau},\, ^LG \Big)}
\end{equation}
a Langlands duality of VW invariants of $M_4$. In \eqref{ZGW=ZGW},
\begin{equation}\label{ZGW=ZGW - 0}
   \boxed{ \mathcal{Z}_{\text{GW},\Sigma}\big(\tau, \mathcal{M}^{G}_{\text{Higgs}}(C)\big) 	\longleftrightarrow \mathcal{Z}_{\text{GW},\Sigma}\Big(-\frac{1}{n_{\mathfrak{g}}\tau}, \,\mathcal{M}^{^LG}_{\text{Higgs}}(C)\Big)} \end{equation}
a Langlands duality of GW invariants 
that can be interpreted as a mirror symmetry of Higgs bundles. In \eqref{HFVWG to HFVWLG},
\begin{equation}\label{HFVWG to HFVWLG - 0}
  \boxed { \text{HF}^{\text{VW}}_*(M_3, G, \tau) \longleftrightarrow \text{HF}^{\text{VW}}_*\bigg(M_3, {^LG}, -\frac{1}{n_{\mathfrak{g}}\tau}\bigg)}
\end{equation}
a Langlands duality of VW Floer homologies assigned to $M_3$. In \eqref{HFL to HFL},
\begin{equation} \label{HFL to HFL - 0}
\boxed{\text{HF}_{*}^{\text{Lagr}}\big(\mathcal{M}^G_{\text{Higgs}}(C), L_0, L_1, \tau\big) \longleftrightarrow \text{HF}_{*}^{\text{Lagr}}\bigg(\mathcal{M}^{^LG}_{\text{Higgs}}(C), L_0, L_1, -\frac{1}{n_{\mathfrak{g}}\tau}\bigg)}
\end{equation} 
a Langlands duality of Lagrangian Floer homologies of Higgs bundles. And lastly, in \eqref{Abou-Mano Langlands},
\begin{equation} \label{Abou-Mano Langlands-0}
 \boxed{   \text{HP}^*(M_3, G_\mathbb{C}, \tau) \longleftrightarrow \text{HP}^*(M_3, ^LG_{\mathbb{C}}, - 1/n_\frak g\tau) }
\end{equation}
a Langlands duality of the Abouzaid-Manolescu hypercohomology of a perverse sheaf of vanishing cycles in the moduli space of irreducible flat complex connections on $M_3$. 

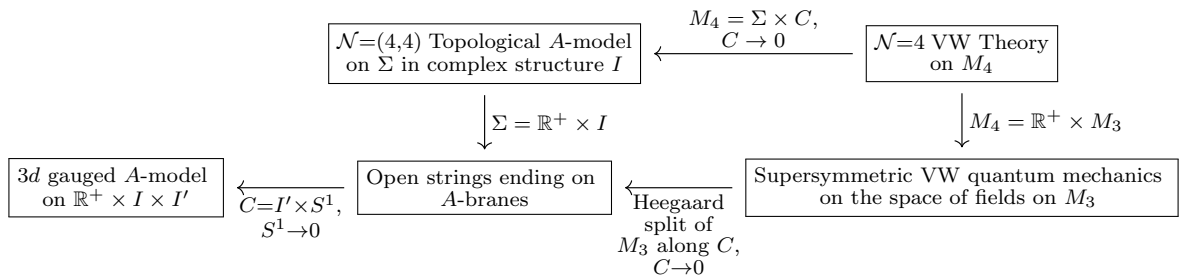
\begin{figure}
\begin{center}
  \begin{tikzcd}[column sep=30pt,row sep=20pt]
    &\boxed{\substack{\mathcal{N}=(4,4)\; \text{Topological $A$-model}\\ \text{on $\Sigma$ in complex structure $I$}  } }
    \arrow[d, "\text{$\Sigma=\mathbb{R}^{+}\times I $}"] 
    &  \boxed{\substack{\mathcal{N}=4 \;\text{VW Theory}\\ \text{on }M_{4}}}  
    \arrow[d, "\text{$M_4=\mathbb{R}^{+}\times M_3 $}"] 
    \arrow[l, " \substack{\text{$M_4=\Sigma \times C$,}\\ \text{$C\to 0$}}"']\\
    \boxed{\substack{3d\text{ gauged $A$-model }\\\text{on $\mathbb{R}^+ \times I \times I'$}}}
    &\boxed{\substack{\text{Open strings ending on }\\ \text{ $A$-branes }\\}}
    \arrow[l,  "\substack{C = I' \times S^1,\\S^1 \to 0}"]
    &  \boxed{\substack{\text{Supersymmetric VW quantum mechanics }\\{\text{on the space of fields on $M_3$ }}}}
    \arrow[l,  " \substack{\text{Heegaard}\\\text{ split of }\\ \text{$M_3$ along $C$,}\\C\to 0}"]
    \end{tikzcd}
\caption{The physical approach taken in this paper.} 
\label{fig:ideaflow}
\end{center}
\end{figure}

In $\S$\ref{sec:geometric langlands}, we will show that $S$-duality of VW theory will also result in a geometric Langlands correspondence. Specifically, we have, in \eqref{dualcata},
\begin{equation}\label{dualcata - 0}
    \boxed{\text{Cat}_{\text{$A$-branes}}\big(\tau, \mathcal{M}^{G}_{\text{Higgs}}(C)\big) \longleftrightarrow
    \text{Cat}_{\text{$A$-branes}}\Big(-\frac{1}{n_{\mathfrak{g}}\tau}, \,\mathcal{M}^{^LG}_{\text{Higgs}}(C)\Big) }
\end{equation}
a homological mirror symmetry of a $\tau$-dependent (derived) category of $A$-branes on the space of Higgs bundles, where if $\text{Re}(\tau) = 0$, we have, in \eqref{dualdmod},
\begin{equation}\label{dualdmod - 0}
    \boxed{{\cal D}^{\textbf{c}}_{-h^\vee}\text{-mod}\big(q, {\text{Bun}_{G_{\mathbb{C}}}}\big) \longleftrightarrow {\cal D}^{\textbf{c}}_{-^Lh^\vee}\text{-mod}\Big(-\frac{1}{n_{\mathfrak{g}}q}, \,{\text{Bun}_{{^LG}_{\mathbb{C}}}}\Big)}
\end{equation}
a quantum geometric Langlands correspondence for complex group $G_{\mathbb{C}}$ with complex curve $C$ and purely imaginary parameter $q$. Furthermore, in the zero-coupling, `classical' limit of VW theory in $G$ where  $\text{Im} (\tau) \to \infty$ whence $q \to \infty$, we have, in \eqref{classical GL},
\begin{equation}\label{classical GL - 0}
    \boxed{\text{Cat}_{\text{coh}}\big (\mathcal{M}_{\text{flat}}^{G_{\mathbb{C}}}(C) \big ) \longleftrightarrow {\cal D}^{\textbf{c}}_{-^Lh^\vee}\text{-mod}\Big(0, \,{\text{Bun}_{{^LG}_{\mathbb{C}}}}\Big)}
    \end{equation}
    a classical geometric Langlands correspondence for $G_{\mathbb{C}}$ with complex curve $C$.\\

\begin{figure}[hbt!]
    \centering
    \begin{tikzcd}[row sep=25, column sep=-15]
\boxed{\substack{\text{Lagrangian Floer}\\\text{homology of}\\\mathcal{M}^G_{\text{Higgs}}(C)\text{, with }  \tau} } 
\arrow[rrr, leftrightarrow, dashed, "\substack{\text{Lagrangian Floer}\\\text{Langlands duality}}"]
\arrow[dr, leftrightarrow,  " \substack{\text{VW Atiyah-Floer}\\\text{correspondence}}", " \substack{\textbf{d}:\\\text{ Heegaard split}\\ \text{of $M_3$ along $C$,}\\C\to 0}"']
&&&\boxed{\substack{\text{Lagrangian Floer}\\\text{homology of}\\\mathcal{M}^G_{\text{Higgs}}(C)\text{, with }  -1/n_{\mathfrak{g}}\tau } } 
\arrow[dr, leftrightarrow,  " \substack{\text{VW Atiyah-Floer}\\\text{correspondence}}", "\textbf{d}"']
\\
&\boxed{\substack{\text{VW Floer}\\\text{homology of}\\M_3 \text{, with } \tau,\, G}}  
\arrow[rrr, leftrightarrow, dashed,"\substack{\text{VW Floer}\\\text{Langlands duality}}"]
\arrow[dl, leftrightarrow, "\textbf{b: }M_4=M_3 \times \mathbb{R^+}"]
&&&\boxed{\substack{\text{VW Floer}\\\text{homology of}\\M_3  \text{, with } -1/n_{\mathfrak{g}}\tau,\, ^LG}} 
\arrow[dl, leftrightarrow, "\textbf{b }"]\\
\boxed{\substack{\text{\textbf{VW invariants of} }\bf{M_4}\\\textbf{with }\bf{\boldsymbol{\tau}, \,G }}} 
\arrow[rrr, leftrightarrow, dashed, "\textbf{Langlands dual}"]
\arrow[dr, leftrightarrow, "\substack{\textbf{a: }\text{$M_4=\Sigma \times C$,}\\ \text{$C\to 0$}}"]
&&&\boxed{\substack{\textbf{VW invariants of }\bf{M_4}\\\textbf{with }\boldsymbol{-1/n_{\mathfrak{g}}\tau, \,^LG }}} 
\arrow[dr, leftrightarrow, "\textbf{a }"]\\
&\boxed{\substack{\text{GW invariants of}\\\mathcal{M}^G_{\text{Higgs}}(C)\text{, with }  \tau} }
\arrow[rrr, leftrightarrow, dashed, "\substack{\text{Mirror symmetry}\\\text{of Higgs bundles}}", crossing over]
&&&\boxed{\substack{\text{GW invariants of}\\ \mathcal{M}^G_{\text{Higgs}}(C)\text{, with }  -1/n_{\mathfrak{g}}\tau} }  \\
\boxed{\substack{\text{Category of $A$-branes}\\\text{on $\mathcal{M}^G_{\text{Higgs}}(C)$ with $\tau$}}}
\arrow[dd,leftrightarrow,  "\text{Re}(\tau) = 0"']
\arrow[dr,leftrightarrow,  " \substack{\textbf{e: } C = I' \times S^1,\\ S^1 \to 0,\, \text{abelian $G$, }\,\text{Re}(\tau) = 0}"]
\arrow[uu, leftrightarrow,  " \substack{{\textbf{c: }M_4=I\times\mathbb{R}^+\times C,}\\{ C\to 0}}"] 
\arrow[rrr, leftrightarrow, dashed, "\substack{\text{Homological mirror}\\\text{symmetry of}\\\text{ Higgs bundles}}", crossing over]
&&&\boxed{\substack{\text{Category of $A$-branes}\\\text{on $\mathcal{M}^G_{\text{Higgs}}(C)$ with $-1/n_{\mathfrak{g}}\tau$}}}
\arrow[uu, leftrightarrow, "\substack{\\\\\\\\\\\textbf{c }}"]
\arrow[dd, leftrightarrow, "\substack{\\\\\\\\\\\\\\\\\text{Re}(\tau) = 0}"']
\arrow[dr,leftrightarrow,  "\textbf{e}"]\\
&\boxed{\substack{\text{ 2-category of module}\\\text{ categories of the Fukaya-Floer}\\\text{category on $T^2$}\\}}
\arrow[rrr, leftrightarrow, dashed, "\substack{\text{Langlands duality of}\\\text{ 2-categories}}", crossing over]
&&&\boxed{\substack{\text{ 2-category of module}\\\text{ categories of the Fukaya-Floer}\\\text{category on dual $T^2$}\\}} \\
\boxed{\substack{\text{Category of}\\\text{Twisted $D$-modules}\\\text{on }\text{Bun}_{G_\mathbb C} (C)\text{with }\tau}}
\arrow[rrr, leftrightarrow, dashed, "\substack{{\text{Quantum geometric}}\\{\text{ Langlands correspondence}}}"]
\arrow[d,leftrightarrow,  "\tau\to \infty"']
&&&\boxed{\substack{\text{Category of}\\\text{Twisted $D$-modules }\\\text{on }\text{Bun}_{^LG_\mathbb C} (C) \text{with }-1/n_{\mathfrak{g}}\tau}}
\arrow[d,leftrightarrow,  "\tau\to \infty"']\\
\boxed{\substack{\text{Category of}\\\text{coherent sheaves}\\\text{on }\mathcal{M}^{G_{\mathbb{C}}}_{\text{flat}} (C)\text{ with }\tau}}
\arrow[rrr, leftrightarrow, dashed, "\substack{{\text{Classical geometric}}\\{\text{ Langlands correspondence}}}"]
&&&\boxed{\substack{\text{Category of}\\\text{critically twisted $D$-modules }\\\text{on }\text{Bun}_{^LG_{\mathbb{C}}} (C) \text{with }-1/n_{\mathfrak{g}}\tau}}
\end{tikzcd}
\caption{A novel web of mathematical relations stemming from Vafa-Witten Theory.}
\label{fig:equimath0}
\end{figure}
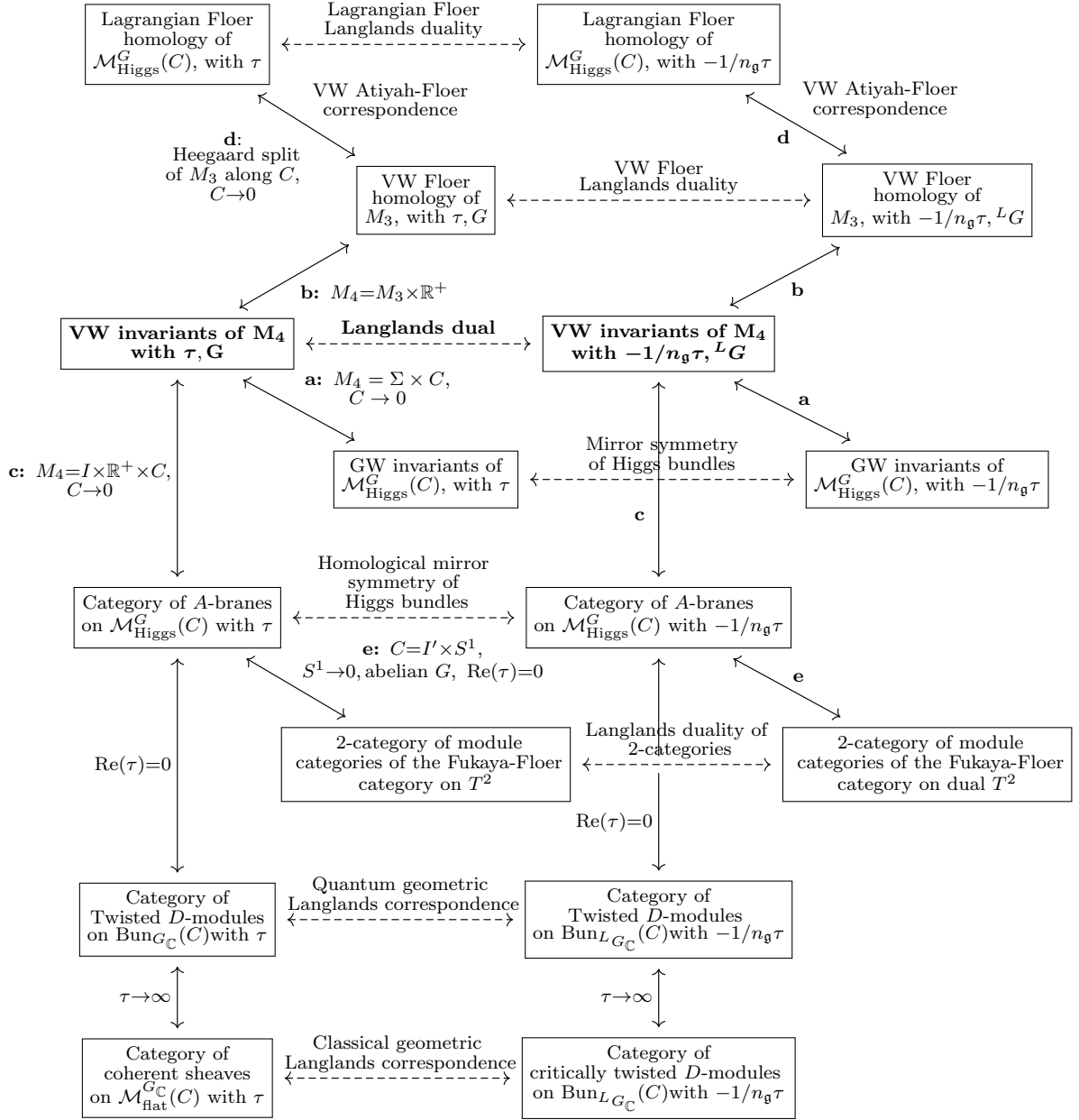

\bigskip
In $\S$\ref{sec:web of dualities}, we will present a novel web of mathematical relations, summarizing the dualities, correspondences, and identifications between the various mathematical objects we physically derived in $\S$\ref{vwgeneral}--\ref{sec:geometric langlands} starting from VW theory, in Fig.~\ref{fig:equimath}. We will go on to explain how the VW invariant will be systematically categorified in our framework as depicted in \eqref{categorifying VW - extend}:
\begin{equation} \label{categorifying VW - extend 0}
\boxed{   {{{\cal Z}_{\text{VW}}} \xrightarrow{\text{categorification}} {{\text {HF}}^{\text {VW}}_*} \xrightarrow{\text{categorification}} {\text{Cat}_{A\text{-branes}}}}
\xrightarrow{\text{categorification}} {\text{2-Cat}_{\text{mod-cat}} \big({\text{FF-cat}}(T^2)\big)} }
\end{equation}
where Fig.~\ref{fig:equimath} will be enhanced to Fig.~\ref{fig:equimath - cat}.


In summary, the physical approach that we have taken in this paper is given in Fig.~\ref{fig:ideaflow}, where it will lead us to the novel mathematical relations in Fig.~\ref{fig:equimath0}.




\bigskip\noindent\textit{Acknowledgements}
\vspace*{0.5em}\\
We would like to thank M.~Ashwinkumar for many useful discussions, and S.~Gukov for his comments on v1 of this paper. We would also like to thank A.~Haydys for his assistance on various mathematical issues, and R.P. Thomas for his interest and questions on our work. Last but not least, we would like to thank the ATMP referee for pointing out our oversight regarding the dimension of VW moduli space, thereby enabling us to make the necessary edits. This work is supported in part by the MOE AcRF Tier 1 grant R-144-000-470-114.

\section{Vafa-Witten Twist of \texorpdfstring{$\mathcal{N}=4$}{N4} Gauge Theory, and a Vafa-Witten Invariant}\label{vwgeneral}

In this section, we start by reviewing aspects of VW theory on $M_4$ with gauge group $G$ necessary for this paper, referring to ~\cite{vafa1994strong, labastida1997mathai}. 
We then physically derive a novel VW invariant of $M_4$.  

\subsection{Vafa-Witten Theory} \label{VW twist discussion}

First, note that in Euclidean signature (the case natural to TQFT's), we can express the 4d spacetime group as $SO(4)=SU(2)_L \otimes SU(2)_R$. Next, note that for ${\cal N} = 4$ supersymmetry in 4d, we have an $SU(4)_{\cal R}$ $R$-symmetry group that can be broken down and expressed as $SO(4)_{\cal R} =SU(2)_{A}\otimes SU(2)_{B}$. Then, in order to obtain the VW-twist of \cite{vafa1994strong}, we just need to replace the $SU(2)_L \subset SO(4)$ with $SU(2)_{L'}$, the diagonal subgroup of $SU(2)_L$ and $SU(2)_A$. The resulting fields will consequently have quantum numbers corresponding to the total group $SU(2)_{L'}\otimes SU(2)_{R} \otimes SU(2)_B$. The field content of the $\mathcal{N}=4$ theory is then modified as follows:
\begin{equation}\label{fieldcontent}
    \begin{aligned}
        A_{\alpha \dot{\alpha}} &\to A_{\alpha \dot{\alpha}}\: (\textbf{2,2,1})\: ,\\
            \phi_{ij} &\to B_{\alpha\beta}\:(\textbf{3,1,1})\:,\quad C_{ab}\:(\textbf{1,1,3})\:,\\
        \lambda_{\dot{\alpha}}^{i} &\to \psi_{\alpha \dot{\alpha}}^{a}\:(\textbf{2,2,2})\:,\\
        \lambda_{\alpha}^{i} &\to \chi_{\alpha\beta}^{a}\:(\textbf{3,1,2})\:,\quad \eta^{a}\:(\textbf{1,1,2}) \:.
    \end{aligned}
\end{equation}
In \eqref{fieldcontent}, the labels $a,\,b$ represent indices for $SU(2)_B$; the labels $\alpha, \dot \alpha$ represent spinor indices. The supercharges $\mathcal{Q}_{\alpha}^{i}$ and $\mathcal{Q}_{\dot{\alpha}}^{i}$,  being fermions, are modified in the same way as the gauge fermions $\lambda_{\alpha}^{i}$ and $ \lambda_{\dot{\alpha}}^{i}$ to
\begin{equation}
    \begin{aligned}
        \mathcal{Q}_{\alpha}^{i} &\to \mathcal{Q}_{\alpha\beta}^{a}\:(\textbf{3,1,2}), \quad \mathcal{Q}^{a}\:(\textbf{1,1,2}),\\
        \mathcal{Q}_{\dot{\alpha}}^{i} &\to \mathcal{Q}_{\alpha \dot{\alpha}}^{a}\:(\textbf{2,2,2}).
    \end{aligned}
\end{equation}

The VW twist thus produces a scalar supercharge $\mathcal{Q}^a$ within an $SU(2)_B$ doublet. We now split the fields along their $SU(2)_B$ representation. There are 3 independent components for $C_{ab}$ (being in the \textbf{3} of $SU(2)_B$), and we will label them as separate scalar fields $C(0)$, $\phi(+2)$ and $\bar{\phi}(-2)$. Here $C(0)$ represents the field $C$ with a ghost number of 0. Similarly, $\psi_{\alpha \dot{\alpha}}^{a}$ will be labelled as $\psi_{\alpha \dot{\alpha}}(+1)$ and $\tilde{\chi}_{\alpha \dot{\alpha}}(-1)$; $\chi_{\alpha\beta}^{a}$ will be labelled as $\chi_{\alpha\beta}(-1)$ and $\tilde{\psi}_{\alpha\beta}(+1)$; and $\eta^a$ will be labelled as $\eta(-1)$ and $\zeta(+1)$. The two bosonic fields $A_\mu(0)$ and $B_{\alpha\beta}(0)$ remain unchanged since they are singlets of $SU(2)_B$. We can also split $\mathcal{Q}^{a}$ into $\mathcal{Q}^{\pm}$.

The supersymmetry transformations are then 
\begin{equation}\label{susytransform}
    \begin{split}
        [\mathcal{Q}^+, A ] &= \psi\;, \quad \{\mathcal{Q}^+, \psi \} = -\mathcal{D}\phi\;,\\
        [\mathcal{Q}^+, B] &= \tilde{\psi}\;, \quad \{\mathcal{Q}^+, \tilde{\psi} \} = [\phi, B]\;,\\
        [\mathcal{Q}^+, C] &= \zeta\;, \quad \{\mathcal{Q}^+, \zeta \} = [\phi, C]\;,\\
        [\mathcal{Q}^+, \phi] &= 0\;, \\
        [\mathcal{Q}^+, \bar{\phi}] &= \eta\;, \quad \{\mathcal{Q}^+, \eta \} = [\phi, \bar{\phi}]\;,\\
        \{\mathcal{Q}^+, \tilde{\chi}\} &= \tilde{H}\;, \quad [\mathcal{Q}^+, \tilde{H} ] = [\phi, \tilde{\chi}]\;,\\
        \{\mathcal{Q}^+, \chi\} &=  H\;, \quad [\mathcal{Q}^+, H ] = [\phi, \chi]\;,\\
    \end{split}
    \quad\quad\quad\quad
    \begin{split}
        [\mathcal{Q}^-, A ] &= -\tilde{\chi}\;, \quad \{\mathcal{Q}^-, \tilde{\chi} \} = \mathcal{D}\bar{\phi}\;,\\
        [\mathcal{Q}^-, B] &= -\chi\;, \quad \{\mathcal{Q}^-, \chi \} = [\bar{\phi}, B]\;,\\
        [\mathcal{Q}^-, C] &= -\eta\;, \quad \{\mathcal{Q}^-, \eta \} = [\bar{\phi}, C]\;,\\
        [\mathcal{Q}^-, \bar{\phi}] &= 0\;, \\
        [\mathcal{Q}^-, \phi] &= \zeta\;, \quad \{\mathcal{Q}^-, \zeta \} = [\bar{\phi}, \phi]\;,\\
        \{\mathcal{Q}^-, \psi\} &= -\tilde{H}\;, \quad [\mathcal{Q}^-, \tilde{H} ] = [\psi,\bar{\phi}]\;,\\
        \{\mathcal{Q}^-, \tilde{\psi}\} &=  H\;, \quad [\mathcal{Q}^-, H ] = [\bar{\phi}, \tilde{\psi}]\;,\\
    \end{split}
\end{equation}
satisfying the algebra 
\begin{equation}\label{nilpotent}
    \begin{aligned}
        \{\mathcal{Q}^+,\mathcal{Q}^+\} &= \delta_{g}(\phi),\\
        \{\mathcal{Q}^-,\mathcal{Q}^-\} &= \delta_{g}(\bar{\phi}),\\
        \{\mathcal{Q}^+,\mathcal{Q}^-\} &= \delta_{g}(C),
    \end{aligned}
\end{equation}
where $\delta_g$ represents a gauge transformation. From \eqref{susytransform}, we see that $\mathcal{Q}^{\pm}$ is nilpotent up to a gauge transformation. The auxiliary fields $H$ and $\tilde{H}$ have been included in \eqref{susytransform} for \eqref{nilpotent} to hold off-shell.

We note that despite the existence of two supercharges $\mathcal{Q}^{\pm}$, linear combinations of $\mathcal{Q}^{\pm}$ are equivalent up to an $SU(2)_B$ symmetry transformation~\cite{vafa1994strong}. Hence, it does not matter which of $\mathcal{Q}^{\pm}$ we consider. Therefore, let us use ${\cal Q}^+$ for our construction of VW theory. 

With the complex coupling parameter 
\begin{equation}
    \tau = \frac{\theta}{2\pi}+i\frac{4\pi}{e^2},
\end{equation}
the action can be written as the sum of a $\mathcal{Q}^+$-exact term and a topological term:\footnote{One can also write the action as a $\mathcal{Q}^-$-exact term, but we will only show the case for it being $\mathcal{Q}^+$-exact.} 
\begin{equation}\label{Qplusexact}
    \begin{aligned}
        S_{\text{VW}} = \frac{1}{e^2}\int_{M_4}d^{4}x\sqrt{g}\,\text{Tr}\{\mathcal{Q}^+, \mathcal{V} \}-\frac{i\tau}{4\pi}\int_{M_4}\text{Tr}F\wedge F,
    \end{aligned}
\end{equation}
where\footnote{Clebsh-Gordan coefficients $(\sigma)^{\mu}_{\alpha\dot{\alpha}}$ and  $(\bar{\sigma}^{\mu})^{\alpha\dot{\alpha}}$ allow us to express $A_{\alpha\dot{\alpha}}(\bar{\sigma}^{\mu})^{\alpha\dot{\alpha}}=A^{\mu}$, and $(\sigma^{\mu\nu})^{\alpha\beta}B_{\alpha\beta}=B^{\mu\nu}$, where $(\sigma^{\mu\nu})^{\alpha\beta}=\frac{1}{4}\big[\sigma^{\mu}\bar{\sigma}^{\nu}-\sigma^{\nu}\bar{\sigma}^{\mu}\big]^{\alpha\beta}$. }
\begin{equation}
    \begin{aligned}
        \mathcal{V} &=\chi_{\mu\nu}\bigg(H^{\mu\nu}-2F^{+\mu\nu}\bigg)+2\bar{\phi}\mathcal{D}_{\mu}\psi^{\mu}+\tilde{\chi_{\mu}}\bigg(\tilde{H}^{\mu}-2\mathcal{D}^{\mu}C-2\mathcal{D}_{\nu}B^{\nu\mu}\bigg)\\
        &-\chi_{\mu\nu}\bigg([B^{\mu\nu}, C]+\frac{1}{2}[B^{\mu\tau}, B_{\tau}^{\nu}] \bigg)-\bar{\phi}\bigg(\frac{1}{2}[\tilde{\psi_{\mu\nu}},B^{\mu\nu}]+2[\zeta,C] \bigg)+\eta[\phi,\bar{\phi}].
    \end{aligned}
\end{equation}

Upon integrating out the auxiliary fields $H$ and $\tilde{H}$, we obtain the localization equations by setting to zero $\{\mathcal{Q}^{+},\text{fermion}\}$ in \eqref{susytransform}: 
\begin{subequations}\label{4dbps0}
\begin{empheq}{align}
  F^{+}_{\mu\nu}+\frac{1}{2}[B_{\mu\nu},C]+\frac{1}{4}[B_{\mu\rho},B_{\lambda\nu}]g^{\rho\lambda}=0,\\
    \mathcal{D}_{\mu}C+\mathcal{D}_{\nu}B^{\nu\mu}=0, \\
    \mathcal{D}\phi = [\phi, B] = [\phi, C] = [\phi, \bar \phi] =0.
\end{empheq}
\end{subequations}
These constitute the BPS equations for the theory, with the zero modes of $A_{\mu}$, $B_{\mu\nu}$, $C$, $\phi$ and $\bar \phi$ that satisfy \eqref{4dbps0} defining a moduli space which the path integral localizes on.
Like in~\cite{witten1988topological}, we can set (the zero modes of) $C$, $\phi$ and $\bar \phi$ to vanish in \eqref{4dbps0} if we wish to consider only irreducible connections.
In short, we will henceforth concern ourselves with the following localization equations:\footnote{The 2-form $B_{\mu\nu}$ need not vanish if the scalar curvature of K\"ahler $M_4$ and the gauge group $G$ are not simultaneously non-negative and locally a product of $SU(2)$’s \cite{vafa1994strong}, and we have assumed this to be the case here. }
\begin{subequations}\label{4dbps}
\begin{empheq}[box=\widefbox]{align}
  F^{+}_{\mu\nu}+\frac{1}{4}[B_{\mu\rho},B_{\lambda\nu}]g^{\rho\lambda}=0\\
    \mathcal{D}_{\nu}B^{\nu\mu}=0
\end{empheq}
\end{subequations}
This set of equations in \eqref{4dbps} have also been studied in \cite{haydys2010fukaya, tanaka2017vafa, huang2017compactness}.
Upon evaluating \eqref{Qplusexact}, the part of $S_{\text{VW}}$ involving only 
$A$ and $B$ 
is
\begin{equation} \label{bosonaction}
\begin{aligned}
    S^{A,B}_{\text{VW}}&=\frac{1}{e^2}\int_{M_4} d^4 x \sqrt{g}\text{Tr}\bigg( \big(F^{+}_{\mu\nu}+\frac{1}{4}\big[B_{\mu\rho}, B^{\rho}_{\nu}\big] \big)^2+ (\mathcal{D}^{\mu}B_{\mu\nu})^2  \bigg) \\
    &- \frac{ i \tau }{4\pi}\int_{M_4} \text{Tr}\, \bigg(F\wedge F  
    +  dB\wedge \star DB + B\wedge d(\star DB) \bigg),
\end{aligned}
\end{equation}
where we have taken the liberty to add the term $\{\mathcal{Q}^+, d (B \wedge \tilde\chi) \}$ (that is null in the spectrum of VW theory given by the $\mathcal {Q}^+$-cohomology), for later convenience. Also, we have used the fact that $B$ is self-dual, whence $\star B = B$, and here, $D=d+A$ where $\star DB$ is a one-form on $M_4$.

With $\mathcal{N}=4$ supersymmetry, VW theory posseses an $SL(2,\mathbb{Z})$ symmetry, having both $S$-duality and $T$-duality. On a generic $M_4$, $T$-duality corresponds to shifting $\tau\to\tau+1$, generating a $2\pi$ shift of $\theta$, which is a symmetry. Less obvious is $S$-duality, which, at the quantum level, says that a theory with coupling $\tau$ and simply-laced gauge group $G$ is isomorphic to a dual theory with the Langlands dual group $^L G$ and coupling 
\begin{equation}
    ^L \tau = -\frac{1}{\tau}.
\end{equation}
For non-simply-laced gauge groups, the coupling transforms as 
\begin{equation}
    ^L \tau = -\frac{1}{n_{\mathfrak{g}}\tau}
\end{equation}
instead, where $n_{\mathfrak{g}}$ is the lacing number of the group.

\subsection{A Vafa-Witten Invariant}



The VW equations in (\ref{4dbps}), and its moduli space, $\mathcal{M}_{\text {VW}}$, will now enable us to furnish a purely physical derivation of a novel Vafa-Witten invariant of $M_4$.

To this end, first note that an examination of the supersymmetry transformations \eqref{susytransform} indicates that the observables for VW theory ought to be similar to that for Donaldson-Witten (DW) theory. Insertion of these operator observables $\mathcal{O}_i$ into the path integral amounts to computing the correlation function
\begin{equation}\label{vwinvariant4d}
    \langle \prod_i \mathcal{O}_i \rangle_{\text{VW}} = \int_{{\cal M}_{\text{VW}}} \prod_i\, \mathcal{O}_i e^{-S_{\text{VW}}},
\end{equation}
where the subscript `$\mathcal {M}_{\text{VW}}$' means that the zero modes of $A$ and $B$ in the path integral measure lie along $\mathcal {M}_{\text{VW}}$. Enforcing $R$-charge  anomaly  cancellation,  one can interpret the correlation function as an integral of a top-form  on ${\cal M}_{\text{VW}}$.  

That said, note that unlike DW theory, VW theory belongs to a class of TQFT's called `balanced TQFT', where there is never an $R$-charge anomaly~\cite{dijkgraaf1997balanced}, whence the virtual dimension of  $\mathcal{M}_{\text {VW}}$ is always zero.\footnote{The virtual dimension of ${\cal M}_{\text{VW}}$ was first computed in \cite{labastida1997mathai} by analyzing the deformation complex corresponding to the moduli space of solutions to \eqref{4dbps0}, and it was shown to be zero for any $M_4$. Specifically, it is given by the number of 1-form fermion zero modes minus the total number of 2-form and 0-form fermion zero modes. In our case of \eqref{4dbps} where we consider (the zero modes of) $C$, $\phi$ and $\bar \phi$ to vanish in \eqref{4dbps0} whence there are no 0-form fermion zero modes by supersymmetry, the virtual dimension of ${\cal M}_{\text{VW}}$ continues to be zero~\cite{haydys2010fukaya, tanaka2017vafa}, reflecting the physical fact that VW is a balanced TQFT such that there continues to be an equal number of 1-form and 2-form fermion zero modes.}  Hence, the nonvanishing correlation function of VW theory will be
\begin{equation}\label{4dpartition}
    \langle 1\rangle_{\text{VW}} =  \int_{{\cal M}_{\text{VW}}} \, e^{-S_{\text{VW}}}= \mathcal{Z}_{\text{VW},M_4}(\tau, G),
\end{equation}
where $\mathcal{Z}_{\text{VW},M_4}(\tau, G)$ is the VW partition function 
 that can be interpreted as an integral of a virtual zero-form  on virtually zero-dimensional ${\cal M}_{\text{VW}}$,\footnote{Although the virtual dimension is zero, there are still fermion zero modes in the path integral measure that need to be absorbed for the path integral to be nonvanishing. One then needs to ``pull down''  the interaction terms of $S_{\text{VW}}$ in the path integral to absorb these fermion zero modes. These terms can then be interpreted as a virtual zero-form, where their total contribution to the path integral would be given by an integral of this virtual zero-form on virtually zero-dimensional ${\cal M}_{\text{VW}}$.}
 whence it can be evaluated as
\begin{equation} \label{vwk}
  \boxed {  \mathcal{Z}_{\text{VW},M_4}(\tau, G) = \sum_{k}  a_{k} q^{m_k} }
\end{equation}
Here, $q = e^{2\pi i \tau}$, $k$ denotes the $k^{\text{th}}$ sector of ${\cal M}_{\text{VW}}$, the number $a_k$ is given by
\begin{equation} \label{ak}
\boxed{a_k = \int_{\mathcal{M}^k_{\text{VW}}} \Omega^0 \wedge e({T_{\mathcal{M}^k_{\text{VW}}}})\,, \quad \text{where \, $\Omega^0(\mathcal{M}^k_{\text{VW}})=(1 + B^4)^{dim_{\mathbb{C}}\mathcal{M}^k_{\text{VW}}}$}
}
\end{equation}
$B$ is a coordinate on $\mathcal{M}^k_{\text{VW}}(A,B)$, $e$ is the signed Euler class of the tangent bundle ${T_{\mathcal{M}^k_{\text{VW}}}}$, and $m_k$ is the corresponding VW number given by
\begin{equation} \label{m_k}
  \boxed{  m_k =  \frac{ 1}{8\pi^2}\int_{M_4} \text{Tr}\, \bigg(F_{(k)}\wedge F_{(k)}  +  dB_{(k)}\wedge \star DB_{(k)} + B_{(k)}\wedge d(\star DB_{(k)})\bigg) }
\end{equation} 
 
Notice that $\mathcal{Z}_{\text{VW},M_4}$ is a topological invariant of $M_4$ which is an algebraic count of VW solutions with corresponding weight given by  $a_{k} q^{m_k}$ that we elaborated on above. This defines a novel $\tau$-dependent Vafa-Witten invariant of $M_4$.\footnote{A purely algebro-geometric definition of $\mathcal{Z}_{\text{VW},M_4}$, in particular the $a_k$'s, was first given by Tanaka-Thomas in in~\cite{tanaka2017vafa}, albeit for projective algebraic surfaces only. The novelty here is that we provide a purely differentio-geometric definition of the $a_k$'s for a more general $M_4$. }

When $B = 0$, $a_k$ will become the Euler characteristic $\chi(\mathcal{M}^k_{\text{inst}})$, while $m_k$ will become the instanton number. Then, $\mathcal{Z}_{\text{VW},M_4}$ will just become the usual partition function for instantons first derived in~\cite{vafa1994strong}, as expected.





\section{An \texorpdfstring{$\mathcal{N}=(4,4)$}{N} \texorpdfstring{$A$}{A}-model, Higgs Bundles and Gromov-Witten Theory}\label{sigmareduction}

In this section, we will perform dimensional reduction of the 4d VW theory with action \eqref{Qplusexact} down to 2d. The four-manifold $M_4$ will be taken to be $M_4 = \Sigma \times C$, where $\Sigma$ and $C$ are both closed Riemann surfaces, and $C$ is of genus $g\geq 2$.  
Dimensional reduction is then performed by shrinking $C$, whence we will show that we obtain an $\mathcal{N}=(4,4)$ theory in 2d that is a topological $A$-model on $\Sigma$. In turn, we obtain a correspondence between the VW invariant of $M_4$ and the GW
invariant of Higgs bundles.
The method employed for dimensional reduction will be the one in \cite{bershadsky1995topological}. 


\subsection{Reduction of 4d Terms}

We consider a block diagonal metric $g$ for $M_4 = \Sigma \times C$, 
\begin{equation}\label{4d metric}
    g = \text{diag}\big(g_{\Sigma}, \epsilon g_C \big),
\end{equation}
where $\epsilon$ is a small parameter to deform $g_C$. We shall use capital letters $A,B=x^1,x^2$ to denote coordinates on $\Sigma$, and small letters $a,b=x^3,x^4$ to denote coordinates on $C$. Taking the limit $\epsilon\to0$ then gives us a 2d theory on $\Sigma$ with ${\cal N} = (4,4)$ supersymmetry.\footnote{Compactification of an ${\cal N}=4$ theory in 4d on a Riemann surface $C$ breaks half of the 16 supersymmetries to give an ${\cal N} = (4,4)$ theory in 2d.} 

Deforming the metric inevitably affects the fields in the action, since they involve contraction of indices by the metric tensor. With the introduction of the $\epsilon$ paramter, the determinant changes by $\sqrt{g}\to \epsilon\sqrt{g}$. Thus, fields that survive after taking the limit $\epsilon\to0$ require one contraction of indices $a,b$ on $C$, giving a factor of $\epsilon^{-1}$.

On the other hand, terms with higher negative powers of $\epsilon$ will blow up, and we are forced to set to zero these terms to ensure finiteness of the action. The topological term aside, terms in \eqref{bosonaction} with $\mu,\nu,\rho=A,B$ vanish as $\epsilon\to0$. For $\mu,\nu,\rho=a,b$, each term must be set to zero individually since the action \eqref{bosonaction} is a sum of squares . Using $F^{+}_{\mu\nu}=\frac{1}{2}(F_{\mu\nu}+\frac{1}{2}\epsilon_{\mu\nu\rho\lambda}F^{\rho\lambda})$, 
we obtain a finiteness condition for the first squared term:
\begin{equation}\label{finite0}
    F_{34} + \frac{1}{4}\big[ B_{3\rho}, B^{\rho}_{4} \big]=0 .
\end{equation}

Since $B_{\mu\nu}$ is an anti-symmetric and self-dual 2-form ($B_{\mu\nu}=\frac{1}{2}\epsilon_{\mu\nu\rho\lambda}B^{\rho\lambda}$), there are only 3 independent components which we can take to be $B_{12}$, $B_{13}$ and $B_{14}$. We then perform a final contraction of indices in \eqref{finite0} with $g^{AB}$ to obtain
\begin{equation}\label{finite1}
    \boxed{F_{34} - \big[ B_{13}, B_{14} \big]=0}
\end{equation}
after a rescaling of the metric on $\Sigma$ and using the self-dual properties of $B_{\mu\nu}$. This is our first finiteness condition. 

Similarly, the second finiteness condition comes from the other squared term as
\begin{subequations}\label{finite2}
\begin{empheq}[box=\widefbox]{align}
  D_{3}B_{13}+D_{4}B_{14} & = 0 \\
  D_{3}B_{14}-D_{4}B_{13} & = 0
\end{empheq}
\end{subequations}
where we have used the self-dual properties of $B_{\mu\nu}$. 

Identifying $B_{13}$ and $B_{14}$ as the two components of a 1-form $\varphi$ on $C$, equations \eqref{finite1} and \eqref{finite2} are in fact Hitchin's equations on $C$~\cite{hitchin1987self} given by\footnote{$D^{*}\varphi = \star D \star \varphi = D_{\mu}\varphi^{\mu}$, where $\star$ is the Hodge star operator.}
\begin{subequations}\label{Hitchin eqns}
\begin{empheq}[box=\widefbox]{align}
   F - \varphi\wedge\varphi & = 0 \\
  D\varphi = D^{*}\varphi & = 0
\end{empheq}
\end{subequations}
where
\begin{equation} \label{varphi}
 \boxed{   \varphi = B_{13}dx^3 + B_{14}dx^4 =\varphi_{3}dx^3+\varphi_{4}dx^4 }
\end{equation}

The space of solutions of $(A_C, \varphi)$ to \eqref{finite1} and \eqref{finite2} modulo gauge transformations then span Hitchin's moduli space $\mathcal{M}^G_H(C)$ for a connection $A_C$ on a principal $G$-bundle $P$ over the Riemann surface $C$, and a section $\varphi\in\Omega^1(C)$. The above equations leave the $(x^1, x^2)$ dependence of $A_C$ and $\varphi$ arbitrary, and thus the fields $(A_C, \varphi)$ define a map $\Phi: \Sigma \to \mathcal{M}^G_H(C)$.

Another finiteness condition we obtain is
\begin{equation}
    D_{3}B_{34}=D_{4}B_{43}=0,
\end{equation}
which again using the self-duality properties of $B_{\mu\nu}$, we obtain
\begin{equation}\label{finite3}
    \boxed{D_{3}B_{12}=-D_{4}B_{12}=0}
\end{equation}
The field $B_{12}$ is a 0-form w.r.t rotations on both $C$ and $\Sigma$, so \eqref{finite3} tells us that the 0-form $B_{12}$ is covariantly constant on $C$, which means $B_{12}$ generates infinitesimal gauge transformations while leaving $A_C$ fixed. We can however set $B_{12}=0$, since we require gauge connections to be irreducible to avoid complications on $\mathcal{M}^G_H(C)$. 

For the 2d action on $\Sigma$, we require terms with at most one contraction of indices of $C$. These are (excluding the topological term for now)
\begin{equation} \label{effaction}
    \begin{aligned}
    S_{\text{eff}}&=\frac{1}{e^2}\int d^4 x \sqrt{g}\text{Tr}\bigg((F^{+}_{Aa})^2 + (D_{A}B^{Aa})^2 \bigg)+ \text{fermions}\\
    &= \frac{1}{e^2}\int d^4 x \text{Tr}\bigg( \frac{1}{2}(F_{13}-F_{24})^2+\frac{1}{2}(F_{14}+F_{23})^2 \\
    &+ (\partial_{1}B_{13}+\partial_{2}B_{14})^2 + (\partial_{1}B_{14}-\partial_{2}B_{13})^2 \bigg)+ \text{fermions} .\\
    \end{aligned}
\end{equation}
We can take $F_{13}=\partial_{1}A_{3}$, since $A_1$ does not have derivatives on $\Sigma$ and are thus non-dynamical fields which can be integrated out in the 2d action on $\Sigma$. $A_1$ will then be equal to a combination of fermionic fields (and the same goes for $A_2$). 
Switching to complex coordinates $z=x^1+ix^2$ and $w=x^3+ix^4$, we obtain
\begin{equation}
    \begin{aligned} \label{reducedaction}
    S_{\text{eff}}&=\frac{1}{e^2}\int_{\Sigma}(idz\wedge d\bar{z})\int_{C}(idw\wedge d\bar{w})\text{Tr}\bigg(2\partial_{z}A_{w}\partial_{\bar{z}}A_{\bar{w}} + 4\partial_{z}\varphi_{\bar{w}}\partial_{\bar{z}}\varphi_{w}  \bigg)+ \text{fermions}.\\
    \end{aligned}
\end{equation}

After suitable rescalings, we can then rewrite \eqref{reducedaction} (with $idz\wedge d\bar{z}=|dz^2|)$ as 
\begin{equation}\label{2dfinalaction}
    S_{\text{2d}}=\frac{1}{e^2}\int_{\Sigma}|dz^2| g_{i\bar{j}}\bigg(\partial_{z}X^{\bar{i}}\partial_{\bar{z}}X^{j}+\partial_{z}X^{i}\partial_{\bar{z}}X^{\bar{j}}+\partial_{z}Y^{\bar{i}}\partial_{\bar{z}}Y^{j}+\partial_{z}Y^{i}\partial_{\bar{z}}Y^{\bar{j}} \bigg)+ \text{fermions},
\end{equation}
where $X$ corresponds to $A_C$, and $Y$ to $\varphi_C$. Thus, we have an ${\cal N} = (4,4)$ sigma model on $\Sigma$,\footnote{Even though we have, for brevity, only demonstrated the reduction of the 4d bosonic terms to 2d bosonic ones, the rest of the  4d fermionic terms can also be shown to reduce to 2d fermionic ones consistently, a fact that is also guaranteed by the surviving ${\cal N} = (4,4)$ supersymmetry in 2d.} which hyper-K\"{a}hler target $\mathcal{M}^G_H(C)$ is split into two halves, each parameterized by coordinates $(X^{i}, X^{\bar{i}})$ and $(Y^{i}, Y^{\bar{i}})$ with basis $(\alpha_{\bar{w}i}, \alpha_{w\bar{i}})$ and $(\beta_{wi}, \beta_{\bar{w}\bar{i}})$, respectively. The cotangent space to $\mathcal{M}^G_H(C)$ are spanned by the one-form fermions $\psi_C$ and $\tilde{\chi}_C$, and from \eqref{susytransform}, we see that these will be cotangent to $A_C$ and $\varphi_C$, respectively. More details about $\mathcal{M}^G_H(C)$ will be discussed shortly.


\subsection{BPS Equations in 2d and an $\mathcal{N} = (4,4)$ $A$-model} \label{4d bps to 2d bps}

To obtain the corresponding 2d BPS equations of the ${\mathcal N} = (4,4)$ sigma model on $\Sigma$, we start with the 4d, ${\cal N} = 4$ action \eqref{bosonaction}. 
It is of the form 
\begin{equation} \label{4dsk}
    \frac{1}{e^2}\int_{M_4} \text{Tr}(|s|^2+ |k|^2) + \text{topological term},
\end{equation}
with
\begin{equation}\label{sandk}
    \begin{aligned}
        s_{\mu\nu} &= F^{+}_{\mu\nu}+\frac{1}{4}\big[B_{\mu\rho}, B^{\rho}_{\nu}\big],\\
        k_{\nu} &= \mathcal{D}^{\mu}B_{\mu\nu}.
    \end{aligned}
\end{equation}
By taking $s=k=0$, we can also obtain \eqref{4dbps}, the 4d BPS equations of VW theory, i.e., the equations which the 4d VW path integral localizes on. 
By performing dimensional reduction of \eqref{sandk} on $C$ with $s=k=0$, we can directly obtain the corresponding 2d BPS equations. 

Noting the fact that only terms with mixed indices on $\Sigma \times C$ survive the reduction on $C$, together with the self-duality properties of $B_{\mu\nu}$, we obtain, from \eqref{sandk} and $s=k=0$,
\begin{equation}\label{effbps}
    \begin{aligned}
       F_{Aa}^{+} &= 0, \\
        \mathcal{D}_{A}B^{Aa} &=0.
    \end{aligned}
\end{equation}
Via the first equation, $F^{+}=0$, and its implied anti-self-duality of $F$, we get
\begin{equation}\label{2dbpsa}
    \begin{aligned}
        \partial_{1}A_{3} &= \partial_{2}A_{4},\\
        \partial_{1}A_{4} &= -\partial_{2}A_{3}.
    \end{aligned}
\end{equation}
These are Cauchy-Riemann equations for $A_{\bar{w}}=\frac{1}{2}(A_3 + iA_4)$. Switching to complex coordinates as before, \eqref{2dbpsa} can be written as $\partial_{\bar{z}}A_{\bar{w}}=0$. A similar computation can be performed for $\mathcal{D}_{A}B^{Aa} =0$, where instead, we obtain the Cauchy-Riemann equations\footnote{$A_{1}, A_{2}$ in the covariant derivative can be ignored since we are only considering bosonic terms for BPS equations.} 
\begin{equation}\label{2dbpsb}
    \begin{aligned}
        \partial_{1}B_{13} &= -\partial_{2}B_{14},\\
        \partial_{1}B_{14} &= \partial_{2}B_{13},
    \end{aligned}
\end{equation}
for an anti-holomorphic field $\varphi_{\bar{w}}=\frac{1}{2}(B_{13}+i B_{14})$. In complex coordinates, \eqref{2dbpsb} becomes $\partial_{z}\varphi_{\bar{w}}=0$. Alternatively, we can also express \eqref{2dbpsb} as $\partial_{\bar{z}}\varphi_{w}=0$, with $\varphi_{w}=\frac{1}{2}(B_{13}-i B_{14})$. Seeing that $A_{\bar{w}}$ corresponds to $X^{i}$ and $\varphi_{w}$ to $Y^{i}$, we get the 2d BPS equations as
\begin{subequations}\label{2dbpsfull}
    \begin{empheq}[box=\widefbox]{align}
        \partial_{\bar{z}}X^{i} &= 0\\
        \partial_{\bar{z}}Y^{i} &= 0
    \end{empheq}
\end{subequations}

Hence, the path integral of the 2d, ${\cal N} = (4,4)$ sigma model on $\Sigma$ with action \eqref{2dfinalaction}, localizes on the moduli space of holomorphic maps $\Phi(X^i, Y^i):\Sigma\to\mathcal{M}^G_H(C)$:
\begin{equation}\label{mmaps2d}
    \mathcal{M}_{\text{maps}} = \{ \Phi(X^i, Y^i):\Sigma\to\mathcal{M}^G_H(C) \;|\; \partial_{\bar{z}}X^{i}=\partial_{\bar{z}}Y^{i}=0 \}.
\end{equation}
In other words, we have a 2d, ${\cal N} = (4,4)$ $A$-model on $\Sigma$ with target $\mathcal{M}^G_H(C)$. This conclusion has also been anticipated in~\cite{bershadsky1995topological}.


\subsection{An $A$-model in Complex Structure $I$}

The space of fields $(A_C$, $\varphi)$ span an infinite-dimensional affine space $\mathcal{W}$. The cotangent vectors $\delta A_C$ and $\delta \varphi$ to $\mathcal{M}^G_H(C)$ are solutions to the variations of equations  \eqref{finite1} and \eqref{finite2}. We can then introduce a basis $(\delta A_w, \delta \varphi_{\bar{w}})$ and $(\delta A_{\bar{w}}, \delta \varphi_{w})$ in $\cal W$, where the (flat) metric on $\mathcal{M}^G_H(C)$ is given by 
\begin{equation}\label{hitchinmetric}
    ds^2 = -\frac{1}{2\pi}\int_{C} \text{Tr}\bigg(\delta A_w \wedge \star \delta A_{\bar{w}} + \delta\varphi_{w} \wedge \star \delta\varphi_{\bar{w}} \bigg).
\end{equation}
Note that $\mathcal{M}^G_H(C)$ is necessarily hyper-K\"{a}hler~\cite{alvarez1981geometrical}. 
As a hyper-K\"{a}hler manifold, the metric \eqref{hitchinmetric} has three independent complex structures $I$, $J$ and $K$, satisfying quarternion relations $I^2=J^2=K^2=-1$. 

From the BPS equations \eqref{2dbpsfull}, which are $\del_{\bar z} A_{\bar w} =0$ and $\del_{\bar z} \varphi_w = 0$, one can see that the complex structure relevant to the $A$-model is $I$, with linear holomorphic functions consisting of $A_{\bar{w}}$ and $\varphi_w$.\footnote{The holomorphic functions for $J$ are $A_{\bar{w}}+i\varphi_{\bar{w}}$ and $A_w+i\varphi_w$, and for $K$ are $A_{\bar{w}}-\varphi_{\bar{w}}$ and $A_w+\varphi_w$.}
In complex structure $I$, $\mathcal{M}^G_H(C)$ can be identified as the moduli space of stable Higgs $G$-bundles on $C$,  $\mathcal{M}^G_{\text{Higgs}}(C)$. One can write the corresponding symplectic form as $\omega_I = \omega'_I - \delta \lambda_I$, where
\begin{equation}\label{omegaexact}
    \omega'_I = -\frac{1}{4\pi}\int_C\,\text{Tr}\,\delta A_C \wedge\delta A_C \quad \text{and} \quad \lambda_I = \frac{1}{4\pi}\int_C\,\text{Tr}\,\varphi\wedge\delta\varphi,
\end{equation}
and $\omega_I$ is cohomologous to $\omega'_I$. 


Comparing the 4d topological term in \eqref{bosonaction} to \eqref{omegaexact}, we see that the topological term can be written as 
\begin{equation}
    i\tau\int_{\Sigma}\,\Phi^{*}(\omega_I).
\end{equation}
Clearly, $\Phi^\ast$ represents a pullback  from $\mathcal{M}^G_{\text{Higgs}}(C)$ onto $\Sigma$. 
The 2d action \eqref{2dfinalaction}, including the topological term, is then 
\begin{equation}\label{2dfinalaction2}
    S_{\text{2d}}=\frac{1}{e^2}\int_{\Sigma}|dz^2| g_{i\bar{j}}\bigg(\partial_{z}X^{\bar{i}}\partial_{\bar{z}}X^{j}+\partial_{z}X^{i}\partial_{\bar{z}}X^{\bar{j}}+\partial_{z}Y^{\bar{i}}\partial_{\bar{z}}Y^{j}+\partial_{z}Y^{i}\partial_{\bar{z}}Y^{\bar{j}} \bigg)+i\tau\int_{\Sigma}\Phi^{*}(\omega_I)+ \dots
\end{equation}
where ``\dots" represent fermionic terms. \eqref{mmaps2d} then becomes
\begin{equation} \label{Mmaps final}
    \mathcal{M}_{\text{maps}} = \{ \Phi(X^i, Y^i):\Sigma\to\mathcal{M}^G_{\text{Higgs}}(C) \;|\; \partial_{\bar{z}}X^{i}=\partial_{\bar{z}}Y^{i}=0 \},
\end{equation}
the moduli space of holomorphic maps $\Phi:\Sigma\to \mathcal{M}^G_{\text{Higgs}}(C)$. 

In short, we have a 2d, ${\cal N} = (4,4)$ $A$-model on $\Sigma$ with target $\mathcal{M}^G_{\text{Higgs}}(C)$.

\subsection{Vafa-Witten Invariants as Gromov-Witten Invariants of Higgs Bundles}

The virtual dimension of $\mathcal{M}_{\text{maps}}$, like that of ${\cal M}_{\text {VW}}$, ought to also be zero. This is because the 2d $A$-model is obtained via a topological deformation that sets $C\to0$ in the original 4d VW theory, whence the relevant index of kinetic operators counting the virtual dimension of moduli space remains the same. 
Thus, as in the 4d case, the nonvanishing correlation function here is the partition function
\begin{equation} \label{ZAclosed}
    \langle 1 \rangle_{{A,\Sigma}} = \int_{\mathcal{M}_{\text{maps}}} e^{-S_{\text{2d}}}= \mathcal{Z}^{\text{closed}}_{A,\Sigma}(\tau, \mathcal{M}^G_{\text{Higgs}}(C)), 
\end{equation}
where the subscript `$\mathcal{M}_{\text{maps}}$' means that the zero modes of $X$ and $Y$ in the path integral measure lie along $\mathcal{M}_{\text{maps}}$. Like ${\cal Z}_{\text{VW}, M_4}$ in 4d, $\mathcal{Z}^{\text{closed}}_{A,\Sigma}$ can be interpreted as an integral of a virtual zero-form on virtually zero-dimensional ${\cal M}_{\text {maps}}$, whence it can be evaluated as 
\begin{equation}
   \mathcal{Z}^{\text{closed}}_{A,\Sigma}(\tau, \mathcal{M}^G_{\text{Higgs}}(C)) =\sum_{l} {\tilde a}_l q^{{\tilde m}_l}.
\end{equation}
Here, $l$ denotes the $l^{\text{th}}$ sector of  ${\mathcal M}_{\text {maps}}$ defined in \eqref{Mmaps final} for \emph{genus one} $\Sigma$, the rational number ${\tilde a}_l$ is given by~\cite{hori2003mirror}
 \begin{equation} \label{al}
 \boxed{{\tilde a}_l = \int_{{\cal M}^l_{\text {maps}}} e(\mathcal{V}) }
 \end{equation}
where $e$ is the signed Euler class of the vector bundle $\mathcal{V}$ with fiber $H^0(\Sigma, K \otimes \Phi^*T^*{{\cal M}^l_{\text {maps}}})$ and canonical bundle $K$ on $\Sigma$, and ${\tilde m}_l$ is the corresponding worldsheet instanton number given by
\begin{equation} \label{q_l}
    \boxed{{\tilde m}_l = \frac{1}{2\pi}\int_{\Sigma}\,\Phi^{*}_l(\omega_I)}
\end{equation}


Notice that $\mathcal{Z}^{\text{closed}}_{A,\Sigma}$ is an enumerative invariant which is an algebraic count of holomorphic maps  with corresponding weight given by ${\tilde a}_l q^{{\tilde m}_l}$ that we elaborated on above. This coincides with the definition of the GW invariant, which then means that one can identify $\mathcal{Z}^{\text{closed}}_{A,\Sigma}$ as 
\begin{equation}\label{GW invariant}
    \boxed{\mathcal{Z}_{\text{GW},\Sigma}(\tau, \mathcal{M}^G_{\text{Higgs}}(C)) = \sum_l {\tilde a}_l q^{{\tilde m}_l}}  
\end{equation}
where $\mathcal{Z}_{\text{GW},\Sigma}$ is a $\tau$-dependent GW invariant of $\mathcal{M}^G_{\text{Higgs}}(C)$.



From the topological invariance of the 4d theory, we have a 4d-2d correspondence of partition functions 
\begin{equation}\label{4d2dpartition}
    {\mathcal{Z}_{\text{VW},M_4}(\tau, G) = \mathcal{Z}^{\text{closed}}_{A,\Sigma}(\tau, \mathcal{M}^G_{\text{Higgs}}(C)) },  
\end{equation}
whence from our above discussion, it will mean that  
\begin{equation}\label{VW=GW}
    \boxed{\mathcal{Z}_{\text{VW},M_4}(\tau, G) = \mathcal{Z}_{\text{GW},\Sigma}(\tau, \mathcal{M}^G_{\text{Higgs}}(C)) }  
\end{equation}
In other words, we have a correspondence between the VW invariant of $M_4 = \Sigma \times C$ and the GW invariant of $\mathcal{M}^G_{\text{Higgs}}(C))$.


In fact, recall that the numbers ${\tilde m}_l$ (in \eqref{GW invariant}) correspond to the numbers $m_k$ (in \eqref{vwk}). Hence, \eqref{VW=GW} means that we have 
\begin{equation}
\label{ak=al}
 \boxed{   a_k = \tilde{a}_l}
\end{equation}
where $a_k$ and $\tilde{a}_l$ are given in \eqref{ak} and \eqref{al}, respectively. Thus, one can also determine the $a_k$'s, the VW invariants of $T^2 \times C$, via the signed Euler class of a bundle $\mathcal{V}$ over $\mathcal{M}^l_\text{maps}$.\footnote{Computing the ${\tilde a}_l$'s and thus $a_k$'s for $T^2 \times C$ explicitly is a purely mathematical endeavour that is beyond the scope of this physical mathematics paper which main objective is to furnish their fundamental definitions via the expressions \eqref{al} and \eqref{ak}, respectively. The reader who seeks an explicit computation of these invariants may be happy to know that after our work appeared, this was done purely mathematically in~\cite{nesterov2023enumerative}.}


\section{A Novel Floer Homology from Boundary Vafa-Witten Theory}\label{vwsqm}

In this section, we will show how we can physically derive a novel Floer homology by considering boundary VW theory on $M_4 = M_3 \times \mathbb{R}^+$.\footnote{To be precise, VW theory is still being defined on an $M_4$ with no boundary. However, to make contact with Floer theory, we will need to examine a hyper-slice of $M_4$, which we can topologically regard as $ M_3 \times I \cong M_3 \times \mathbb{R}^- \cup_{M_3} M_3 \times \mathbb{R}^+$. 
As there is no evolution in the $\mathbb{R}^\pm$ time-direction in our topological theory, it suffices to examine only $M_3 \times \mathbb{R}^+$, where $M_3$ can then be viewed as a boundary. 
This is consistent with the idea that categorification of topological invariants can be achieved via successive introductions of boundaries to $M_4$, which we will elaborate upon in \S\ref{sec:web of dualities}.}
We will first give a relevant summary of supersymmetric quantum mechanics (SQM). After which,
we will recast the 4d $\mathcal{N}=4$ boundary VW theory into an  SQM model, which will in turn allow us to physically derive a VW Floer homology assigned to $M_3$.

\subsection{A Summary of Supersymmetric Quantum Mechanics}

Supersymmetric quantum mechanics is a one-dimensional topological sigma model with a map $\phi: t \to \mathcal{M}$, where time $t$ parameterizes the worldline, and $\mathcal{M}$ represents a generic target manifold. The worldline can either be closed or open, i.e., either $S^1$ or $\mathbb{R}^+$, but for our purposes, we shall take it to be open, i.e., $\mathbb{R}^+$. 
For a comprehensive review of SQM, the reader can refer to \cite{birmingham1991topological, hori2003mirror}. 

The action for SQM is of the form
\begin{equation}\label{sqmaction0}
    S_{\text{SQM}} = \int dt \bigg[i \frac{d\phi^{i}}{dt} H_{i}+\frac{1}{2}g^{ij}H_{i}H_{j}+ \frac{1}{4}R_{kl}^{ij}\bar{\psi}_{i}\psi^{k}\bar{\psi}_{j}\psi^{l} -i\bar{\psi_{i}}\nabla_{t}\psi^{i}\bigg].
\end{equation}
Indices $i,j$ belong to $\mathcal{M}$, with the $\phi^{i}$'s being coordinates on $\mathcal{M}$. The $\bar{\psi}_{i}, \psi^{i}$'s are Grassmann odd coordinates (that are the supersymmetric partners to the $\phi^i$'s), and $g^{ij}$ is the metric on $\mathcal{M}$. The field $H_i$ is an auxiliary field which can be integrated out from the action. The covariant derivative $\nabla_{t}$ is the pull-back of the covariant derivative on $\mathcal{M}$ to the worldine (parameterized by) $t$, and $R_{kl}^{ij}$ is the Riemann curvature tensor on $\mathcal{M}$. 

There is only one nilpotent scalar supersymmetry generator $\mathcal{Q}$, generating the transformations 
\begin{equation}\label{sqmsusy}
    \begin{aligned}
        \{ \mathcal{Q}, \phi^i\}&=\psi^i,\\
        \{ \mathcal{Q}, \psi^i\}&=0,\\
        \{ \mathcal{Q}, \bar{\psi}_i\}&= H_i-\bar{\psi}_j\Gamma^j_{ik}\psi^k,\\
        \{ \mathcal{Q}, H_i\}&= H_j\Gamma^j_{ik}\psi^k-\frac{1}{2}\bar{\psi}_{j}R^j_{ilk}\psi^{l}\psi^{k},
    \end{aligned}
\end{equation}
where $\Gamma^j_{ik}$ is the Riemannian connection on $\mathcal{M}$.

One can always generalize the action \eqref{sqmaction0} by including a potential $V(\phi)$. The action then becomes 
\begin{equation}\label{sqmaction0}
    S_{\text{SQM}} = \int dt \bigg[i \bigg(\frac{d\phi^{i}}{dt}+s g^{ij}\frac{\partial V(\phi)}{\partial \phi^{j}}  \bigg)H_{i}+\frac{1}{2}g^{ij}H_{i}H_{j}+ \frac{1}{4}R_{kl}^{ij}\bar{\psi}_{i}\psi^{k}\bar{\psi}_{j}\psi^{l} -i\bar{\psi_{i}}\bigg(\delta^{i}_{j}\nabla_{t}+sg^{ik}\nabla_{k}\partial_{j}V(\phi)\bigg)\psi^{j} \bigg],
\end{equation}
where $V(\phi)$ is some functional on $\mathcal{M}$, and $s$ is a parameter. Upon integrating out $H_i$ via its equation of motion, \eqref{sqmaction0} becomes
\begin{equation}\label{sqmaction}
    S_{\text{SQM}} = \int dt \bigg[\frac{1}{2} \bigg(\frac{d\phi^{i}}{dt}+s g^{ij}\frac{\partial V(\phi)}{\partial \phi^{j}}  \bigg)^2+ \frac{1}{4}R_{kl}^{ij}\bar{\psi}_{i}\psi^{k}\bar{\psi}_{j}\psi^{l} -i\bar{\psi_{i}}\bigg(\delta^{i}_{j}\nabla_{t}+sg^{ik}\nabla_{k}\partial_{j}V(\phi)\bigg)\psi^{j} \bigg].
\end{equation}

The resulting action \eqref{sqmaction} is minimized by the gradient flow equation
\begin{equation}\label{sqminstanton}
    \frac{d\phi^{i}}{dt}+s g^{ij}\frac{\partial V}{\partial \phi^{j}}=0.
\end{equation}
We thus have \eqref{sqminstanton} as the BPS equation for this theory. One can see that a non-fixed $\phi^{i}$ satisfying \eqref{sqminstanton} flows along the $t$-direction between boundary configurations where $\dot {\phi^i} = 0$, i.e., it is fixed. Notice that \eqref{sqminstanton} tells us that these boundary configurations are also critical points of $V(\phi)$. Thus, this is similar to how an instanton tunnels between the ground states of a potential. 



In non-topological theories, minimisation of the action only gives a semiclassical approximation to the theory. In supersymmetric topological theories, which is the case here, the semiclassical approximation is in fact, exact, as pointed out in the introduction. Specifically, the path integral of the theory localizes on a moduli space defined by \eqref{sqminstanton}, whence one can compute the path integral exactly. A relevant fact at this point is that the `squaring argument' (see \cite{blau1993n}) tells us that for \eqref{sqminstanton} to hold identically whence the path integral localizes, it must be that $\dot{\phi^{i}}=s g^{ij}\partial V/\partial \phi^{j}=0$. In other words, the path integral of the theory localizes on the \emph{fixed} critical points of $V(\phi)$. Indeed, these fixed points are also time-invariant points that therefore correspond to the $\mathcal{Q}$-cohomology (since its Hamiltonian is necessarily zero), and the path integral is expected to count just that.

Assuming that the fixed critical points of $V(\phi)$ are isolated and non-degenerate, and, for $s\neq 0$, each fixed critical point contributes $\pm$1 to the partition function $\mathcal{Z}_{SQM}$, then 
\begin{equation}\label{sqmpartitionfunction}
    \mathcal{Z}_{SQM}\, = \sum_{\phi^{i}: \ \dot \phi^i =dV(\phi^i)= 0} \pm 1 
\end{equation}
exactly. Notice that $\mathcal{Z}_{SQM}$ is just an algebraic count of the fixed critical points of $V(\phi)$, where there are BPS flow lines between these fixed critical points.  

\subsection{SQM Interpretation of Boundary Vafa-Witten Theory} \label{SQM interpret of Bd VW}

Let the manifold of the 4d theory in \eqref{bosonaction} be $M_4 = M_3 \times \mathbb{R}^+$, where the $M_3$ boundary is a closed three-manifold, and $\mathbb{R}^+$ is the `time' coordinate. We also let spacetime indices take the values $\mu = 0,1,2,3$, with $\mu=0$ being the time direction, while $\mu = i,j,k = 1,2,3$ being the spatial directions. We shall first review the method where boundary DW theory can be recast as an SQM model. Then, we will apply this same method to boundary VW theory.

\bigskip\noindent\textit{Review of SQM Interpretation of Boundary DW Theory}
\vspace*{0.5em}\\
We first consider a 4d $\mathcal{N}=2$ topologically twisted boundary DW theory of gauge group $G$ with a principal $G$-bundle $P\to M_4$ and nilpotent scalar supercharge $\cal Q$. Our aim is to review how this theory can be recast as an SQM model, as was first done in~\cite{blau1993n, blau1993topological}.


 
Of central importance in DW theory is the BPS equation
\begin{equation}\label{dwbps}
    F^+ = 0,
\end{equation}
which characterises instantons. The path integral of the 4d theory localizes on the moduli space of this equation, i.e., instantons.  Using $F^{+}_{\mu\nu}=\frac{1}{2}(F_{\mu\nu}+\frac{1}{2}\epsilon_{\mu\nu\rho\lambda}F^{\rho\lambda})$, \eqref{dwbps} can be written as
\begin{equation}\label{dwinstantonflow}
    \dot{A}^i +\frac{1}{2}\epsilon^{ijk}F_{jk}=0,
\end{equation}
where the temporal gauge $A^0=0$ is taken, and $\dot{A}^i = F^{0i}$. The boundary DW action can then be written as 
\begin{equation}\label{dwsqm}
    \begin{aligned}
        S^{\text{bdry}}_{\text{DW}} &= \frac{1}{e^2}\int_{M_4} \text{Tr}\big(F^+ \big)^2 - \frac{i\tau}{4\pi}\int_{M_4}\text{Tr}F \wedge F + \dots \\
        &= \frac{1}{e^2}\int dt \int_{M_3}\text{Tr}\big(\dot{A}^i+\frac{1}{2}\epsilon^{ijk}F_{jk} \big)^2- \frac{i\tau}{4\pi}\int_{M_3}\text{Tr}\big(A\wedge dA + \frac{2}{3} A \wedge A \wedge A \big)+\dots,
    \end{aligned}
\end{equation}
where ``$\dots$" refers to fermionic terms and scalar fields in the $\mathcal{N}=2$ multiplet. Note that $A\in\Omega^1(M_3)$ in the final expression of the topological term, i.e., it is a one-form on $M_3$. 


Next, let $\mathscr{A}$ be the space of irreducible connections $A$ on $P$, where the cotangent space $T_A^*{\mathscr{A}}$ to $\mathscr{A}$ is spanned by $\delta A$.
The metric $g_{\mathscr{A}}$ on $\mathscr{A}$ can then be defined as 
\begin{equation}\label{sqmmetric} 
    g_{\mathscr{A}}=\int_{M_3} \text{Tr}\big( \delta A \wedge \star  \delta A\big). 
\end{equation}
With the metric on $\mathscr{A}$ defined as such, one can see that the first term in \eqref{dwsqm} resembles the bosonic kinetic term of the SQM action in \eqref{sqmaction}, where $\epsilon^{ijk}F_{jk}$, being the gradient vector field of a Chern-Simons functional, means that $V(\phi)$ can be interpreted as the Chern-Simons functional itself, while $\dot{A}^i = d A^i/dt$ can be identified with $d\phi^i/d t$. 
The terms indicated by ``$\dots$" then give,  via equations of motion, the Riemann curvature terms and the fermion kinetic terms in \eqref{sqmaction}. Altogether, this means that we can interpret \eqref{dwsqm} as the action of an SQM model with target ${\cal M} = \mathscr{A}$ that also has a single nilpotent topological scalar supercharge $\mathcal{Q}$. 



Thus, with the potential on $\mathscr{A}$ being the Chern-Simons functional, and the identification of \eqref{dwinstantonflow} with \eqref{sqminstanton}, we conclude that \eqref{dwinstantonflow}, which is the instanton equation, can be interpreted as a gradient flow equation between fixed critical points of the Chern-Simons functional. Hence, just like \eqref{sqmpartitionfunction}, assuming that the fixed critical points are isolated and nondegenerate in $\mathscr A$, the partition function of boundary DW theory will be an algebraic count of {fixed} critical points of the Chern-Simons functional, i.e., fixed flat $G$-connections on $M_3$, where there are instanton flow lines between these fixed critical points.

The second term in \eqref{dwsqm} is a topological term that only contributes to an overall factor in the path integral. The $\tau$-dependence of this term will not be important for boundary DW theory. It will, however, play a significant role in boundary VW theory, as we will explain shortly.

\bigskip\noindent\textit{SQM Interpretation of Boundary VW Theory}
\vspace*{0.5em}\\
Likewise, let us turn to the BPS equations \eqref{4dbps} of boundary VW theory, and split the indices into space and time directions. Using $F^{+}_{\mu\nu}=\frac{1}{2}(F_{\mu\nu}+\frac{1}{2}\epsilon_{\mu\nu\rho\lambda}F^{\rho\lambda})$ and $B_{\mu\nu}=\frac{1}{2}\epsilon_{\mu\nu\rho\lambda}B^{\rho\lambda}$, we can reexpress the VW equations \eqref{4dbps} as
\begin{equation}\label{vwflow}
    \begin{aligned}
        \dot{A}^{i}+\frac{1}{2}\epsilon^{ijk}\big(F_{jk}-[B_{j}, B_{k}] \big) &= 0,\\
        \dot{B}^{i} + \epsilon^{ijk}\big(\partial_j B_k + [A_j, B_k]\big) &= 0,
    \end{aligned}
\end{equation}
where the temporal gauge $A^0=0$ is taken, $B^i = B^{0i}$, $\epsilon^{ijk}=\epsilon^{0ijk}$, and $A^i, B^{i}\in \Omega^1(M_3)$.\footnote{Using self-duality properties, we have $B^{0i} = B^i = \epsilon^{ijk}B_{jk}$. }

Our aim is to recast boundary VW theory into an SQM model, in the same way that was done for boundary DW theory above. To this end, let us introduce a complexified connection $\mathcal{A}=A+iB \in \Omega^1(M_3)$, of a $G_\mathbb{C}$-bundle on $M_3$. We then find that \eqref{vwflow} can be expressed  as 
\begin{equation}\label{flatcomplex}
  {  \dot{\mathcal{A}}^i+\frac{1}{2}\epsilon^{ijk}\mathcal{F}_{jk}=0, }
\end{equation}
where $\mathcal{F}\in \Omega^2(M_3)$ is the complexified field strength. This is just a complexified gauge field version of \eqref{dwinstantonflow}. 

As in the boundary DW theory case, we can write the action for boundary VW theory 
as 
\begin{equation}\label{vwsqmcomplex}
    {S_{\text{VW}}^{\text{bdry}} = \frac{1}{e^2}\int dt \int_{M_3}\text{Tr}\bigg( \dot{\mathcal{A}}^i +  \frac{1}{2}\epsilon^{ijk}\mathcal{F}_{jk}\bigg)^2 
    -\frac{i\tau}{4\pi}\int_{M_3}\text{Tr}\bigg(A\wedge dA + \frac{2}{3}A\wedge A\wedge A + B\wedge \star D B\bigg)+\dots}
\end{equation}
where ``\dots" refers to fermionic terms and scalar fields in the $\mathcal{N}=4$ multiplet. 

Now, let $\mathfrak{A}$ denote the space of complexified connections $\mathcal{A}$. Then, we can define a metric $g_{\mathfrak{A}}$ on $\mathfrak{A}$ in similar fashion to \eqref{sqmmetric} as
\begin{equation}\label{sqmmetric-complex}
    g_{\mathfrak{A}}=\int_{M_3} \text{Tr}\big( \delta {\cal A} \wedge \star  \delta {\cal A}\big). 
\end{equation}
Noticing also that $\epsilon^{ijk}\mathcal{F}_{jk}$ is a gradient vector field of a complex Chern-Simons functional, it will then mean that we can rewrite \eqref{vwsqmcomplex} as 
\begin{equation}\label{vwsqmcomplex-final}
 \boxed   {S_{\text{VW}}^{\text{bdry}} = \frac{1}{e^2}\int dt \bigg( {d{\mathcal{A}}^i \over dt} +  s g_{\frak A}^{ij}\frac{\partial V({\cal A})}{\partial {\cal A}^{j}}\bigg)^2 
    -\frac{i\tau}{4\pi}\int_{M_3}\text{Tr}\bigg(A\wedge dA + \frac{2}{3}A\wedge A\wedge A + B\wedge \star D B\bigg)+\dots}
\end{equation}
where
\begin{equation} \label{complex CS fn}
 \boxed{  V(\mathcal{A})= -\frac{1}{4\pi^2}\int_{M_3}\text{Tr}\bigg(\mathcal{A}\wedge d\mathcal{A} + \frac{2}{3}\mathcal{A}\wedge \mathcal{A}\wedge \mathcal{A}\bigg) }
\end{equation}
and from \eqref{flatcomplex}, 
\begin{equation}\label{flatcomplex - final}
\boxed  {  {d{\mathcal{A}}^i \over dt}+ s g_{\frak A}^{ij}\frac{\partial V({\cal A})}{\partial {\cal A}^{j}}=0 }
\end{equation}

One can see that \eqref{vwsqmcomplex-final} and \eqref{flatcomplex - final} resemble  \eqref{sqmaction} and \eqref{sqminstanton}, respectively, with $\cal A$ corresponding to $\phi$. In fact, the terms in \eqref{vwsqmcomplex-final} indicated by ``\dots" give, via equations of motion, the Riemann curvature terms and the fermion kinetic terms in (an ${\cal N} = 4$ generalization of) \eqref{sqmaction}. Altogether, this means that we can interpret \eqref{vwsqmcomplex-final} as the action of an SQM model with target $\frak A$ and a single nilpotent topological scalar supercharge ${\cal Q}^+$, where \eqref{flatcomplex - final}, which describes the VW equations, can be interpreted as a gradient flow equation between fixed critical points ($\dot{{\mathcal A}^i} = 0$) of the potential on $\frak A$ given by \eqref{complex CS fn}. Hence, just like \eqref{sqmpartitionfunction}, assuming that the fixed critical points are isolated and nondegenerate in $\mathfrak A$,\footnote{This is guaranteed (though not necessary) when all critical points are isolated and nondegenerate.  This can be the case for an appropriate choice of $G$ and $M_3$. For example, one could choose (1) $G$ compact and $M_3$ of nonnegative Ricci curvature such as a three-sphere or its quotient, or (2) an $M_3$ with a finite $G_{\mathbb C}$ representation variety, and introduce physically-trivial $\mathcal Q$-exact terms to the action to perturb $V({\mathcal A})$. We would like to thank A.~Haydys for discussions on this. \label{isolated}} \emph{the partition function of boundary VW theory will be an algebraic count of fixed critical points of the complex Chern-Simons functional, i.e., fixed flat $G_\mathbb{C}$-connections on $M_3$, where there are VW flow lines between these fixed critical points.}



The second term in \eqref{vwsqmcomplex-final} is a $\tau$-dependent topological term that contributes to an overall factor in the path integral. Contrary to the situation in boundary DW theory, $\tau$ is now scale-invariant, and will thus play a significant role in the $S$-duality of the path integral later.

\subsection{A Novel Vafa-Witten Floer Homology}




\bigskip\noindent\textit{The Spectrum of States of Boundary VW Theory as States on $M_3$}

Recall from the introduction that for a TQFT such as VW theory, the Hamiltonian $H$ vanishes in the ${\cal Q}^+$-cohomology, whence this means that for any state $|\mathcal{O}\rangle$ that is nonvanishing in the ${\cal Q}^+$-cohomology, we have
\begin{equation}
    H |\mathcal{O}\rangle = \{{\cal Q}^+, \cdots \} |\mathcal{O}\rangle = {\cal Q}^+ (\cdots |\mathcal{O}\rangle ) = {\cal Q}^+ |\mathcal{O}'\rangle = \{ {\cal Q}^+, \mathcal{O}' \} |0 \rangle = |\{{\cal Q}^+, \mathcal{O}' \} \rangle \sim 0.
\end{equation}
In other words, the $|\mathcal{O}\rangle$'s which span the spectrum of states in VW theory are actually ground states that are therefore time-invariant. In particular, for boundary VW theory on $M_4 = M_3 \times \mathbb R^+$, where $\mathbb R^+$ is the `time' coordinate, its spectrum of states is associated only with $M_3$. This will indeed be the case, as we will see shortly.  

Now, for an $M_4$ with boundary $\partial M_4=M_3$, one needs to specify ``boundary conditions'' on $M_3$ to compute the path integral. We can do this by first defining a restriction of the fields to $M_3$, which we shall denote as $\Psi_{M_3}$, and then specifying boundary values for these restrictions.  Doing this is equivalent to inserting in the path integral, an operator  functional ${F}(\Psi_{M_3})$ that is nonvanishing in the ${\cal Q}^+$-cohomology (so that the path integral will continue to be topological). This means that the partition functions in boundary VW theory can be computed as \cite[eqn.~(4.12)]{witten1988topological}
\begin{equation}\label{floerfunctional}
    \langle 1 \rangle_{{ F}(\Psi_{M_3})} = \int_{\mathcal{M}_{\text{VW}}} {F}(\Psi_{M_3}) \,    e^{-S^{\text{bdry}}_{\text{VW}}}.
\end{equation}

In other words, we can write the partition function on $M_4$ as
\begin{equation}\label{4d3dpartition}
    {\mathcal{Z}_{\text{VW},M_4}(\tau, G) = \langle 1 \rangle_{{F}(\Psi_{M_3})}  = \sum_k  {\cal F}^{G,\tau}_{\text{VW}}(\Psi_{M_3}^k)}.
\end{equation}
Here, the summation in `$k$' is over all sectors of $\mathcal{M}_{\text{VW}}$ labeled by the VW number $m_k$, and 
${\cal F}^{G,\tau}_{\text{VW}}(\Psi_{M_3}^k)$ is the $k^{\text{th}}$ contribution to the partition function that depends on the expression of ${{F}(\Psi_{M_3})}$ in the bosonic fields on $M_3$ evaluated over the corresponding solutions of the VW equations restricted to $M_3$.  




What else can we say about ${\cal F}^{G,\tau}_{\text{VW}}(\Psi_{M_3}^k)$? 

\bigskip\noindent\textit{A Novel Vafa-Witten Floer Homology Assigned to $M_3$}

To this question, first note that in the previous subsection, we showed that boundary VW theory on $M_3 \times \mathbb R^+$ can also be interpreted as an SQM model on $\frak A$, the space of complexifed connections $\cal A$ on $M_3$, and the partition function can be expressed as an algebraic count of fixed critical points of the complex Chern-Simons functional \eqref{complex CS fn}, i.e., fixed flat $G_\mathbb{C}$-connections on $M_3$, where there are VW flow lines between these fixed critical points described by the gradient flow equation \eqref{flatcomplex - final}. 

Next, note that according to~\cite{floer1988instanton}, 
the fixed critical points as described above, just generate a Floer complex with Morse functional
\begin{equation} \label{HFvw functional}
 \boxed{  CS(\mathcal{A})= -\frac{1}{4\pi^2}\int_{M_3}\text{Tr}\bigg(\mathcal{A}\wedge d\mathcal{A} + \frac{2}{3}\mathcal{A}\wedge \mathcal{A}\wedge \mathcal{A}\bigg) }
\end{equation}
the complex Chern-Simons functional, where the VW flow lines, described by the gradient flow equation
\begin{equation}\label{HFvw grad flow}
\boxed  {  {d{\mathcal{A}}^i \over dt} = - s g_{\frak A}^{ij}\frac{\partial CS({\cal A})}{\partial {\cal A}^{j}}}
\end{equation}
can be interpreted as the Floer differential, whence the number of outgoing flow lines at each fixed critical point would be the degree of the corresponding chain in the complex. 

In other words, we can also write \eqref{4d3dpartition} as
\begin{equation} \label{4d3dpartition - 2nd}
     \mathcal{Z}_{\text{VW},M_4}(\tau, G) =   \sum_k  {\cal F}^k_{\text{VW-Floer}} (M_3, G, \tau), 
\end{equation}
where 
 each ${\cal F}^k_{\text{VW-Floer}} (M_3, G, \tau)$ can be identified with a class in what we shall henceforth call a Vafa-Witten Floer homology $\text{HF}^{\text{VW}}_{d_k}(M_3, G, \tau)$ assigned to $M_3$ of degree $d_k$, defined by \eqref{HFvw functional} and \eqref{HFvw grad flow}.

In summary, from \eqref{4d3dpartition} and \eqref{4d3dpartition - 2nd}, we can write 
\begin{equation}\label{4d3dpartitionfinal}
    \boxed{\mathcal{Z}_{\text{VW},M_4}(\tau, G)  = \sum_k {\cal F}^{G,\tau}_{\text{VW}}(\Psi_{M_3}^k)
    =\sum_k \text{HF}_{d_k}^{\text{VW}}(M_3, G, \tau) = \mathcal{Z}^{\text{Floer}}_{\text{VW},M_3}(\tau,G)}
\end{equation}
where 
`$k$' sums from zero to the maximum number of fixed VW solutions on $M_3 \times \mathbb R^+$ that correspond to isolated and non-degenerate fixed critical points of $CS(\cal A)$.\footnote{See footnote~\ref{isolated}.}


\bigskip\noindent\textit{About the $\tau$-dependence}

Notice the $\tau$-dependence of ${\cal F}^{G,\tau}_{\text{VW}}(\Psi_{M_3}^k)$ and therefore $\text{HF}_{d_k}^{\text{VW}}(M_3, G, \tau)$ that we have yet to explain. 
This arises because in evaluating \eqref{4d3dpartition}, there will be a factor of $q^{s_k}$ for the $k^{\text{th}}$ term, where from the action $S^{\text{bdry}}_{\text{VW}}$ in \eqref{vwsqmcomplex}, we have a number   
\begin{equation} \label{sk}
    \boxed{s_k = \frac{1}{8\pi^2}\int_{M_3} \text{Tr}\bigg(A_{(k)}\wedge dA_{(k)} + \frac{2}{3}A_{(k)}\wedge A_{(k)}\wedge A_{(k)} + B_{(k)}\wedge \star D B_{(k)}\bigg)}
\end{equation}
Here, the subscript `$(k)$' denotes that they are the $k^{\text{th}}$ fixed solution to the VW equations on $M_3 \times \mathbb R^+$ restricted to $M_3$.  



\section{A Vafa-Witten Atiyah-Floer Correspondence}\label{vwafconj}

In this section, we consider a four-manifold of the form $M_4=M_3 \times \mathbb{R}^+$, where a Heegaard split of $M_3$ into $M'_3$ and $M''_3$ along a Riemann surface $C$ is performed. This will allow us to relate Vafa-Witten Floer homology obtained in the previous section to Lagrangian Floer homology, in what is a novel Vafa-Witten version of the Atiyah-Floer correspondence~\cite{Atiyah1987NewIO} based on instantons. In doing so, we would be able to physically prove and generalize a conjecture by mathematicians Abouzaid-Manolescu about the hypercohomology of a perverse sheaf of vanishing cycles in the moduli space of irreducible flat $SL(2, \mathbb{C})$-connections on $M_3$.  

\subsection{Heegaard Splitting} \label{subsec: Heegaard splitting}

\begin{figure}[hbt!]
    \centering
    \includegraphics[width=0.6\textwidth]{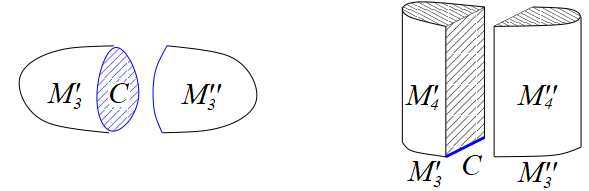} 
    \caption{\textbf{Left:} $M_3$ as a connected sum of three-manifolds $M'_3$ and $M''_3$ along a common Riemann surface $C$. \textbf{Right:} $M_4$ split along four-manifolds $M'_4$ and $M''_4$ with corners. }
    \label{fig:halfeggandblock}
\end{figure}

We perform a Heegaard split of $M_3=M'_3 \cup_{C}M''_3$ along $C$, as shown in Fig.~\ref{fig:halfeggandblock} (left), whence we can view  $M'_3$ and $M''_3$ as nontrivial fibrations of $C$ over intervals $I'$ and $I''$, respectively, where $C$ goes to zero size at one end of the intervals.\footnote{This diagram is adapted from Fig.2 in  \cite{gukov2009surface}.} The metric on $M'_3$ and $M''_3$ can then be written as 
\begin{equation}
    ds^2_{M_3^{',''}} = (dx^B)^2 + f(x^B) (g_C)_{ab} dx^a dx^b,
\end{equation}
where $a,b$ are indices on the Riemann surface $C$, $B$ are indices on $I'$ and $I''$, and $f(x^B)$ is a scalar function along $I'$ and $I''$.

\bigskip\noindent\textit{Topological Invariance of VW Theory and Weyl Rescaling}

Because of the topological invariance of VW theory on $M_4$, we are free to perform a Weyl rescaling of the corresponding Heegaard split metrics on $M_4$ to
\begin{equation}\label{fibrationmetric}
    ds^2_{M_4^{',''}} =\frac{1}{f(x^B)}\bigg[(dx^A)^2+(dx^B)^2\bigg]  +  (g_C)_{ab} dx^a dx^b,
\end{equation}
where $A$ represent indices on $\mathbb{R}^+$. The prefactor is simply a scaling factor on both $\mathbb{R}^+ \times I'$ and $\mathbb{R}^+ \times I''$, whence their topologies are left unchanged. We can thus write $M_4=\big(\mathbb{R}^+ \times I'\times C\big) \cup_{C} \big(\mathbb{R}^+ \times I''\times C\big)$, where $M'_3=I'\times C$ and $M''_3 = I''\times C$. This is illustrated in Fig.~\ref{fig:halfeggandblock} (right), where if $C \to 0$, we indeed have $\mathbb R^+ \times I'$ and $\mathbb R^+ \times I''$.


\subsection{A Vafa-Witten Version of the Atiyah-Floer Correspondence} \label{subsec: VW AF Correspondence}

\bigskip\noindent\textit{An $A$-model on $\mathbb R^+ \times I$}


If $C \to 0$, we end up with an \emph{open} $A$-model in complex structure $I$ (recall from $\S$\ref{sigmareduction}) on $\mathbb{R}^+\times I'$ and $\mathbb{R}^+\times I''$, respectively, with target space $\mathcal{M}^G_{\text{Higgs}}(C)$. It describes open strings with worldsheets $\mathbb R^+ \times I'$ and $\mathbb R^+ \times I''$ that propagate (starting from $t=0)$ in $\mathcal{M}_{\text{Higgs}}^G(C)$ and end on $A$-branes. 
Because we have an $A$-model in complex structure $I$, the admissible branes are those of type $(A, *, *)$, i.e., they are $A$-branes in complex structure $I$, but can be either $A$ or $B$-branes in complex structures $J$ and $K$.  

Specifically, we need an $(A, *, *)$-brane in $\mathcal{M}_{\text{Higgs}}^G(C)$ that corresponds to Higgs pair on $C$ that can be extended to flat complex connections $\cal A$ on $M^{',''}_3$ -- recall from $\S$\ref{SQM interpret of Bd VW} that the partition function of the underlying boundary VW theory gets contributions from the critical points of the complex Chern-Simons functional, and these are flat complex connections $\cal A$ on $M_3=M'_3 \cup_{C}M''_3$.  

Such an $(A, *, *)$-brane has indeed been obtained in~\cite{kapustin2008note}.\footnote{The 4d theory considered in \cite{kapustin2008note} is not the VW but the GL theory of \cite{kapustin2006electric}, albeit with parameter $t=0$. However, both these 4d theories descend to the same 2d $A$-model with target $\mathcal{M}^G_{\text {Higgs}}(C)$ after dimensional reduction on $C$, and since our $A$-branes of interest are $A$-model objects within $\mathcal{M}^G_{\text {Higgs}}(C)$, the arguments used and examples stated in \cite{kapustin2008note} are applicable here.\label{foot:usage of Kap}} It is an $(A,B,A)$-brane $\alpha_{M^{',''}_3}$, that is simultaneously an $A$-brane in $\mathcal{M}^G_{\text {Higgs}}(C)$ and an $A$-brane in ${\cal M}^G_H(C)$ in complex structure $K$, i.e., ${\cal M}^{G_\mathbb{C}}_{\text{flat}}(C)$, the moduli space of flat $G_\mathbb{C}$-connections on $C$, where it corresponds to flat connections that can be extended to $M_3^{',''}$. It is middle-dimensional, and is therefore a Lagrangian brane. Let us henceforth denote this brane as $L$.   





\begin{figure}[hbt!]
    \centering
    \includegraphics[width=0.9\textwidth]{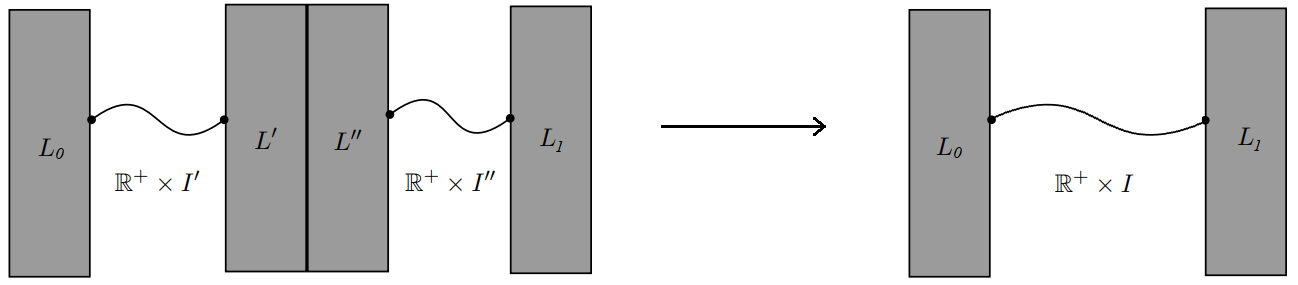} 
    \caption{Identifying $L'$ and $L''$  and gluing them together to form a single open string.}
    \label{fig:gluebranes}
\end{figure}

 Now, with two split pieces $M_4'$ and $M_4''$, when $C \to 0$, we have two strings, each ending on pairs of Lagrangian branes $(L_0, L')$ and $(L'', L_1)$ (see Fig.~\ref{fig:gluebranes}.) 
We then glue the open worldsheets together along their common boundary $L'$ and $L''$, giving us a single $A$-model, with a single string extending from $L_0$ to $L_1$, which is equivalent to gluing $M'_4$ and $M''_4$ along $C \times \mathbb R^+$. (see Fig.~\ref{fig:gluebranes} again.)
%

\bigskip\noindent\textit{The $A$-model on $\mathbb{R}^+ \times I$ as an SQM Model}

Similar to what had been done in $\S$\ref{SQM interpret of Bd VW}, one can recast the $A$-model here as an SQM model, where $\mathbb R^+$ is `time', and the target space is $\mathscr P (L_0, L_1)$, the space of smooth trajectories from $L_0$ to $L_1$ (arising from the interval $I$ that connects them).  

The BPS equations for this $A$-model are \eqref{2dbpsfull}, i.e.,  holomorphic maps from the worldsheet to the target space. The boundary conditions on the worldsheet, however, will impose additional constraints on \eqref{2dbpsfull}, which we will elaborate upon shortly. At any rate, note that \eqref{2dbpsfull} can be written as a gradient flow equation on the worldsheet
\begin{equation}\label{worldsheetgradientflow}
    \frac{\partial Z^l}{\partial t} + i \frac{\partial Z^l}{\partial s} = 0,
\end{equation}
where we have used real coordinates $t$ and $s$ (for $z=t+is$), and here, $Z^l = X^l + Y^l$. 

Comparing \eqref{worldsheetgradientflow} with \eqref{sqminstanton}, one can see that the fixed critical points of the underlying potential of the SQM model that contribute to the partition function are defined by $ \dot{Z^l} ={\partial Z^l / \partial s} = 0$. Since `$s$' is the spatial coordinate of $I$, it would mean that the fixed critical points just correspond to fixed stationary trajectories in $\mathscr P (L_0, L_1)$, i.e., the intersection points of $L_0$ and $L_1$.

Notice that the worldsheet of the (topological) $A$-model can be identified as a disk, $D$, which left and right boundary arcs end on $L_0$ and $L_1$ in $\mathcal{M}^G_{\text {Higgs}}(C)$, respectively. 

Each flow line satisfying \eqref{worldsheetgradientflow} then corresponds to a holomorphic map $Z:D \to \mathcal{M}^G_{\text{Higgs}}(C)$, such that the boundary conditions are 
\begin{equation}\label{wordlsheetflowcondition}
    \begin{aligned}
        Z|_{D_0}\in L_0, &\quad Z|_{D_1}\in L_1,\\
        Z|_S = p, &\quad Z|_N = q,
    \end{aligned}
\end{equation}
where $D_0$, $D_1$ are the left and right boundary arcs of $D$; `$S$' and `$N$' denote the south and north points of $D$, which represent time $t=0$ and $t= \infty$, respectively; and $p$, $q$ are two different points in $\mathcal{M}^G_{\text{Higgs}}(C)$.  

Thus, the partition function of the $A$-model, which, from the SQM model perspective, is given by an algebraic count of the fixed critical points of its underlying potential, will be an algebraic count of the intersection points of $L_0$ and $L_1$, where there are flow lines between the intersection points that obey \eqref{worldsheetgradientflow}. These flow lines  correspond to holomorphic disks with boundary conditions \eqref{wordlsheetflowcondition}, in which $p$ and $q$ are different intersection points of $L_0$ and $L_1$ that the corresponding flow line will start and end at, respectively. In other words, these flow lines correspond to holomorphic Whitney disks. 

\bigskip\noindent\textit{Lagrangian Floer Homology}


Note that from this description of the partition function, we have physically realized the Lagrangian Floer homology first defined in~\cite{Floer1988MorseTF}, where the intersection points of $L_0$
and $L_1$ actually generate the chains of the Lagrangian Floer complex, and the Floer differential, which counts the
number of holomorphic Whitney disks, can be interpreted as the outgoing flow lines at each intersection point of $L_0$ and $L_1$ which number would be the degree of the corresponding chain in the complex. 




Specifically, let $(L_0\cap L_1)_i^{n_i}$ denote the $i^{\text{th}}$ point of the intersection $L_0\cap L_1$ where there are $n_i$ outgoing flow lines, whence we can identify
\begin{equation}
 (L_0\cap L_1)_i^{n_i} \in  \text{HF}^{\text{Lagr}}_{n_i}\big(\mathcal{M}^G_{\text{Higgs}}(C), L_0, L_1\big), 
\end{equation}
where $\text{HF}^{\text{Lagr}}_{n_i} \big(\mathcal{M}^G_{\text{Higgs}}(C), L_0, L_1 \big)$ is the Lagrangian Floer homology of $(L_0, L_1)$ on $\mathcal{M}^G_{\text{Higgs}}(C)$ of degree $n_i$. Then, the partition function of the $A$-model will be given by


\begin{equation} \label{ZA = HF^lag}
    \mathcal{Z}_{A,L}\big(\tau, \mathcal{M}^G_{\text{Higgs}}(C)\big) = \sum_{i} \text{HF}^{\text{Lagr}}_{n_i}\big(\mathcal{M}^G_{\text{Higgs}}(C), L_0, L_1, \tau\big),
\end{equation}
A $\tau$-dependency appears here because of a $\tau$-dependent term in the $A$-model action (see \eqref{2dfinalaction2}).

\bigskip\noindent\textit{A Novel Vafa-Witten Atiyah-Floer Correspondence}

Since the underlying boundary VW theory on $M_4 = M_3 \times \mathbb{R}^+$ is topological, we will have the following equivalence of partition functions:
\begin{equation} \label{Zvw = Z_AL}
      \mathcal{Z}_{\text{VW},M_4}(\tau, G) = \mathcal{Z}_{A,L}\big(\tau, \mathcal{M}^G_{\text{Higgs}}(C)\big), 
\end{equation}
which, from \eqref{4d3dpartitionfinal} and \eqref{ZA = HF^lag}, means that 
\begin{equation}\label{atconj2}
    {\sum_k \text{HF}_{d_k}^{\text{VW}}(M_3, G, \tau)= \sum_{i}\text{HF}_{n_i}^{\text{Lagr}}\big(\mathcal{M}^G_{\text{Higgs}}(C), L_0, L_1, \tau\big).}
\end{equation}

A natural question to ask at this juncture, is whether the gradings in `$d_k$' and `$n_i$' match, whence we would have a degree-by-degree isomorphism of the VW Floer homology and the Lagrangian Floer homology. 




To ascertain this, recall that the VW flow lines between fixed critical points in $\frak A$ are non-fixed solutions to the VW equations \eqref{4dbps} on $M_3 \times \mathbb R^+$. Also, in $\S$\ref{4d bps to 2d bps}, it was shown that the VW equations descend to the worldsheet instanton equations \eqref{2dbpsfull} defining holomorphic maps from the worldsheet to $\mathcal{M}^G_{\text{Higgs}}(C)$, the non-fixed solutions of which are the flow lines between the fixed critical points in $\mathscr P(L_0,L_1)$. Thus, there is a one-to-one correspondence between the flow lines that define $\text{HF}_{*}^{\text{VW}}$ through $d_k$ and underlie the LHS of \eqref{atconj2}, and the flow lines that define $\text{HF}_{*}^{\text{Lagr}}$ through $n_i$ and underlie the RHS of \eqref{atconj2}. 



In other words, the gradings `$d_k$' and `$n_i$' in \eqref{atconj2} do match, and moreover, since `$k$' and `$i$' obviously match, we do have a degree-by-degree isomorphism of the VW Floer homology and the Lagrangian Floer homology, whence we would have a Vafa-Witten Atiyah-Floer correspondence 
\begin{equation}\label{atconj}
   \boxed{ \text{HF}_{*}^{\text{VW}}(M_3, G, \tau) \cong \text{HF}_{*}^{\text{Lagr}}\big(\mathcal{M}^G_{\text{Higgs}}(C), L_0, L_1, \tau\big)} 
\end{equation}

Notice that in the special case that $B=0$ in the underlying VW equations whence they become the instanton equation (see~\eqref{4dbps}) while $\mathcal{M}^G_{\text{Higgs}}(C)$ gets replaced by the moduli space of flat $G$-connections on $C$ (see \eqref{Hitchin eqns}-\eqref{varphi}), \eqref{atconj} just reduces to the celebrated Atiyah-Floer correspondence. Thus, \eqref{atconj} is indeed a consistent generalization thereof.

\subsection{A Physical Proof and Generalization of a Conjecture by Abouzaid-Manolescu about the Hypercohomology of a Perverse Sheaf of Vanishing Cycles}

A hypercohomology $\text{HP}^*(M_3)$ was constructed by Abouzaid-Manolescu in \cite{abouzaid2020sheaf}, where it was conjectured to be isomorphic to instanton Floer homology assigned to $M_3$ for the complex gauge group $SL(2,\mathbb{C})$.

Its construction  was via a Heegaard split of $M_3=M'_3 \cup_{C}M''_3$ along $C$ of genus $g$, and the intersection of the two associated Lagrangians in the moduli space $X_{\text{irr}}(C)$  of irreducible flat $SL(2, \mathbb{C})$-connections on $C$ (that represent solutions extendable to $M'_3$ and $M''_3$, respectively), to which one can associate a perverse sheaf of vanishing cycles. $\text{HP}^*(M_3)$ is then the hypercohomology of this perverse sheaf of vanishing cycles in $X_{\text{irr}}(M_3)$, where it is an invariant of $M_3$ independent of the Heegaard split.


\bigskip\noindent\textit{A Physical Realization of $\text{HP}^*(M_3)$}


Based on the mathematical construction of $\text{HP}^*(M_3)$ described above, it would mean that a physical realization of (the dual of) $\text{HP}^*(M_3)$ ought to be via an open $A$-model with Lagrangian branes $L_0$ and $L_1$ in the target $X_{\text{irr}}(C)$, where the observables contributing to the partition function can be interpreted as classes in the Lagrangian Floer homology $ \text{HF}_{*}^{\text{Lagr}}\big(X_{\text{irr}}(C), L_0, L_1, \tau\big)$. One can argue that this is indeed the case. 

To this end, first, note that there is an isomorphism between $\text{HF}_{*}^{\text{Lagr}}$ and the homology of Lagrangian submanifolds in $X_{\text{irr}}(C)$~\cite[Theorem 11]{Pedroza2018AQV}, i.e., 
\begin{equation}
    \text{HF}_{*}^{\text{Lagr}}\big(X_{\text{irr}}(C), L_0, L_1, \tau\big)\cong\text{H}_*(L, \mathbb{Z}_2)_{\otimes\mathbb{Z}_2}\Lambda,
\end{equation}
where $\Lambda$ is a scalar function over $\mathbb{Z}_2$, called the Novikov field, and $L$ on the RHS can be taken as either $L_0$ or $L_1$. The homology cycles of the  Lagrangian (i.e., middle-dimensional) submanifolds of $X_{\text{irr}}(C)$ have a maximum dimension of $\frac{1}{2}\text{dim}(X_{\text{irr}}(C))$, where $\frac{1}{2}\text{dim}(X_{\text{irr}}(C))=2(3g-3)$.\footnote{It is a fact that $\text{dim}(X_{\text{irr}}(C))$ is given by $4 (N^2-1)(g-1)$ for $G_{\mathbb{C}} = SL(N, \mathbb{C})$, where $g$ is the genus of $C$.} 
Including the zero-cycle, the grading of $\text{H}_*(L, \mathbb{Z}_2)_{\otimes\mathbb{Z}_2}\Lambda$ and therefore $\text{HF}_{*}^{\text{Lagr}}\big(X_{\text{irr}}(C), L_0, L_1, \tau\big)$, goes as  $0, 1, \dots, 2(3g-3)$. 


 Second, note that in~\cite[Theorem 1.8]{abouzaid2020sheaf}, it was computed that $\text{HP}^k$ is nonvanishing only if $-3g+3\leq k \leq 3g-3$. In other words, the grading of $\text{HP}^*$ goes as $-(3g-3), \dots, 0, \dots, (3g-3)$. 
 
 These two observations then mean that there is a one-to-one correspondence between the gradings of $\text{HP}^*(M_3)$  and $\text{HF}_{*}^{\text{Lagr}}$. Moreover, the generators of $\text{HP}^*$  and $\text{HF}_{*}^{\text{Lagr}}$ both originate from the intersection points of $L_0$ and $L_1$ in $X_{\text{irr}}(C)$. Hence, we can identify $\text{HP}^*$  with (the dual of) $\text{HF}_{*}^{\text{Lagr}}$, i.e., 
\begin{equation} \label{HP - HF^lag}
  \boxed{   \text{HP}^*(M_3) \cong \text{HF}_{*}^{\text{Lagr}}\big(X_{\text{irr}}(C), L_0, L_1, \tau\big)}
 \end{equation} 
This agrees with~\cite[Remark 6.15]{brav2012symmetries}.

\bigskip\noindent\textit{A Physical Proof of the Abouzaid-Manolescu Conjecture}

Notice from the Morse functional \eqref{HFvw functional} and the gradient flow equation \eqref{HFvw grad flow} that the definition of $\text{HF}_*^{\text{VW}}$ coincides with the definition of the instanton Floer homology in~\cite{floer1988instanton}, albeit for a \emph{complex} gauge group $G_{\mathbb{C}}$. This means that we can also express the LHS of \eqref{atconj} as $\text{HF}_*^{\text{Inst}}\big(M_3, G_{\mathbb{C}}, \tau \big)$, the instanton Floer homology of $G_{\mathbb{C}}$ assigned to $M_3$. 

Also, recall that the Lagrangian branes $L_0$ and $L_1$ on the RHS of \eqref{atconj} are $(A,B,A)$-branes, i.e., they can also be interpreted as Lagrangian branes in ${\cal M}^G_H(C)$ in complex structure $K$, or equivalently, ${\cal M}^{G_{\mathbb{C}}}_{\text{flat}}(C)$, the moduli space of irreducible flat $G_{\mathbb{C}}$-connections on $C$.

These two points then mean that we can also write \eqref{atconj} as
\begin{equation}\label{HF^inst = HF^lag(Mflat)}
   \boxed{ \text{HF}_{*}^{\text{inst}}(M_3, G_{\mathbb{C}}, \tau) \cong \text{HF}_{*}^{\text{Lagr}}\big({\cal M}^{G_{\mathbb{C}}}_{\text{flat}}(C), L_0, L_1, \tau\big)} 
\end{equation}
In other words, the VW Atiyah-Floer correspondence in \eqref{atconj} can also be interpreted as an Atiyah-Floer correspondence for $G_{\mathbb{C}}$-instantons. 
 
It is now clear from \eqref{HF^inst = HF^lag(Mflat)} and \eqref{HP - HF^lag}, that for $G_\mathbb C = SL(2, \mathbb{C})$, we have
\begin{equation} \label{AB-Mano conj}
 \boxed{   \text{HP}^*(M_3) \cong \text{HF}_{*}^{\text{inst}}(M_3, SL(2, \mathbb{C}), \tau) }
\end{equation}
for complex constant $\tau$. This is exactly the conjecture by Abouzaid-Manolescu about $\text{HP}^*(M_3)$ in~\cite{abouzaid2020sheaf}! 

This agrees with their expectations in~\cite[sect.~9.2]{abouzaid2020sheaf} that $\text{HP}^*(M_3)$ ought to be part of 3+1 dimensional TQFT based on the VW equations.

\bigskip\noindent\textit{A Generalization of the Abouzaid-Manolescu Conjecture}

It was argued in~\cite[sect.~9.1]{abouzaid2020sheaf} that the construction of $\text{HP}^*(M_3)$ can be generalized to $SL(N, \mathbb{C})$. The question therefore, is whether  a corresponding generalization of \eqref{AB-Mano conj} exists. Our answer is `yes', and to complex gauge groups $G_\mathbb C$ that are not limited to $SL(N, \mathbb{C})$.   

Indeed, notice that \eqref{HF^inst = HF^lag(Mflat)} implies that there ought to be a $G_\mathbb{C}$ generalization of the Abouzaid-Manolescu conjecture in \eqref{AB-Mano conj} to 
\begin{equation} \label{AB-Mano conj generalized}
 \boxed{   \text{HP}^*(M_3, G_\mathbb{C}) \cong \text{HF}_{*}^{\text{inst}}(M_3, G_\mathbb{C}, \tau) }
\end{equation}
where the hypercohomology $\text{HP}^*(M_3, G_\mathbb C)$ of the perverse sheaf of vanishing cycles in ${\cal M}^{G_\mathbb{C}}_{\text{flat}}(M_3)$ is such that
\begin{equation} \label{HP - HF^lag general}
  \boxed{   \text{HP}^*(M_3, G_\mathbb{C}) \cong \text{HF}_{*}^{\text{Lagr}}\big({\cal M}^{G_\mathbb{C}}_{\text{flat}}(C), L_0, L_1, \tau\big)}
 \end{equation} 
which again agrees with~\cite[Remark 6.15]{brav2012symmetries}. 

\section{Langlands Duality of Vafa-Witten Invariants, Gromov-Witten invariants, Floer Homologies and the Abouzaid-Manolescu Hypercohomology }\label{sec:langlands duality} 

In this section, we will demonstrate a Langlands duality of the invariants, Floer homologies and Abouzaid-Manolescu hypercohomology that we have physically derived hitherto, from the $S$-duality of VW theory. 



\subsection{Langlands Duality of Vafa-Witten Invariants}


It is known that $\mathcal{N}=4$ supersymmetric Yang-Mills theories has a $SL(2, \mathbb{Z})$ symmetry, with $S$- and $T$-duality, as mentioned in $\S$\ref{vwgeneral}. In particular, the theory with complex coupling $\tau$ and gauge group $G$, is $S$-dual to a theory with complex coupling $-\frac{1}{n_{\frak g}\tau}$ and Langlands dual gauge group $^L G$, 
i.e., we have, up to a possible phase factor of modular weights that is just a constant, a duality of VW partition functions
\begin{equation}\label{vwdual}
  \boxed{  \mathcal{Z}_{\text{VW},M_4}(\tau, G) \longleftrightarrow \mathcal{Z}_{\text{VW},M_4}\Big(-\frac{1}{n_{\mathfrak{g}}\tau},\, ^LG \Big)}
\end{equation}
In other words, we have a Langlands duality of VW invariants of $M_4$, given by \eqref{vwdual}.


\subsection{Langlands Duality of Gromov-Witten Invariants}

Note that if $M_4 = \Sigma \times C$, from \eqref{vwdual} and \eqref{VW=GW}, 4d $S$-duality would mean that we have the 2d duality 
\begin{equation}\label{ZGW=ZGW}
   \boxed{ \mathcal{Z}_{GW,\Sigma}\big(\tau, \mathcal{M}^{G}_{\text{Higgs}}(C)\big) 	\longleftrightarrow \mathcal{Z}_{GW,\Sigma}\Big(-\frac{1}{n_{\mathfrak{g}}\tau}, \,\mathcal{M}^{^LG}_{\text{Higgs}}(C)\Big)} 
\end{equation}
 where $\mathcal{M}^{G}_{\text{Higgs}}$ and $\mathcal{M}^{^LG}_{\text{Higgs}}$ are mirror manifolds. 
 
 In other words, we have a Langlands duality of GW invariants 
 that can be interpreted as a mirror symmetry of Higgs bundles, given by \eqref{ZGW=ZGW}.  


\subsection{Langlands Duality of Vafa-Witten Floer Homology}

If $M_4 = M_3 \times \mathbb{R}^+$, from \eqref{4d3dpartitionfinal} and \eqref{vwdual}, we have the duality 
\begin{equation}\label{ZFL to ZFLG}
  \mathcal{Z}^{\text{Floer}}_{\text{VW},M_3}(\tau, G) \longleftrightarrow \mathcal{Z}^{\text{Floer}}_{\text{VW},M_3}\Big(-\frac{1}{n_{\mathfrak{g}}\tau},\, ^LG \Big).
\end{equation}
In turn, from \eqref{4d3dpartitionfinal}, this means that we have the duality
\begin{equation}\label{HFVWG to HFVWLG}
  \boxed { \text{HF}^{\text{VW}}_*(M_3, G, \tau) \longleftrightarrow \text{HF}^{\text{VW}}_*(M_3, {^LG}, -1/n_{\mathfrak{g}}\tau)}
\end{equation}

In other words, we have a Langlands duality of VW Floer homologies assigned to $M_3$, given by \eqref{HFVWG to HFVWLG}.

\subsection{Langlands Duality of Lagrangian Floer Homology}

From \eqref{ZFL to ZFLG} and \eqref{Zvw = Z_AL}, we have the duality
\begin{equation}\label{ZAL to ZAL}
   \mathcal{Z}_{A,L}\big(\tau, \mathcal{M}^G_{\text{Higgs}}(C)\big)
 \longleftrightarrow \mathcal{Z}_{A,L}\Big(-\frac{1}{n_{\mathfrak{g}}\tau},\, ^LG \Big).
\end{equation}
Then, from the RHS of the VW Atiyah-Floer correspondence in \eqref{atconj}, which defines the state spectrum of $\mathcal{Z}_{A,L}$, we have the duality
\begin{equation} \label{HFL to HFL}
\boxed{\text{HF}_{*}^{\text{Lagr}}\big(\mathcal{M}^G_{\text{Higgs}}(C), L_0, L_1, \tau\big) \longleftrightarrow \text{HF}_{*}^{\text{Lagr}}\big(\mathcal{M}^{^LG}_{\text{Higgs}}(C), L_0, L_1, -1/n_{\mathfrak{g}}\tau\big)}
\end{equation}

In other words, we have a Langlands duality of Lagrangian Floer homologies of Higgs bundles, given by \eqref{HFL to HFL}.

\subsection{Langlands Duality of the Abouzaid-Manolescu Hypercohomology}

From \eqref{AB-Mano conj generalized}, the fact that its RHS can be identified with $\text{HF}^{\text{VW}}_*(M_3, G, \tau)$, and the relation \eqref{HFVWG to HFVWLG}, we have the duality
\begin{equation} \label{Abou-Mano Langlands}
 \boxed{   \text{HP}^*(M_3, G_\mathbb{C}, \tau) \longleftrightarrow \text{HP}^*(M_3, ^LG_\mathbb{C}, - 1/n_\frak g\tau) }
\end{equation}

In other words, we have a Langlands duality of the Abouzaid-Manolescu hypercohomologies of a perverse sheaf of vanishing cycles in the moduli space of irreducible flat  complex connections on $M_3$, given by \eqref{Abou-Mano Langlands}.

\section{A Geometric Langlands Correspondence with Purely Imaginary Parameter}\label{sec:geometric langlands}

In this section, we will first derive a quantum geometric Langlands correspondence with purely imaginary parameter from the $S$-duality of VW theory, and then show that it specializes to the classical correspondence in the zero-coupling limit. 
  

\subsection{An Open $A$-model and a Category of $A$-branes}

Consider VW theory on $M_4 = \Sigma_{\text{open}} \times C = I \times \mathbb{R}^+ \times C$.\footnote{We actually need to ``pull down'' interaction terms from the action on $\Sigma_{\text{open}}$ to absorb fermion zero modes in the path integral. That said, they play no role in our proceeding discussions, just as they played no role in the parallel discussions of~\cite{kapustin2006electric}.} Upon dimensional reduction where $C \to 0$, we get an \emph{open} $A$-model (that starts at $t=0$) with target $\mathcal{M}^G_{\text{Higgs}}(C)$. This furnishes us with a (derived) category of $A$-branes in $\mathcal{M}^G_{\text{Higgs}}(C)$. Since we have an $A$-model in complex structure $I$, we can only have branes that are of type $(A,*,*)$ in $\mathcal{M}^G_{\text{Higgs}}(C)$. Because the $A$-model in complex structure $I$ will map to itself under 4d $S$-duality, it will mean that `$S$-dual' branes are also of type $(A,*,*)$ in $\mathcal{M}^{^LG}_{\text{Higgs}}(C)$. Some examples of these $A$-branes are given in \cite{kapustin2008note}.\footnote{See footnote~\ref{foot:usage of Kap}.}



\subsection{From $A$-branes in $\mathcal{M}^G_{\text{Higgs}}(C)$ to Twisted $D$-modules on $\text{Bun}_{G_{\mathbb{C}}}(C)$}

Looking back to the action of the $A$-model in \eqref{2dfinalaction2}, we see that the topological term is of the form 
\begin{equation}\label{bfieldterm}
    i\tau\int_{\Sigma_{\text{open}}}\,\Phi^{*}(\omega_I) = \int_{\Sigma_{\text{open}}}\,\Phi^{*}(\omega-iB).
\end{equation}
Here, $\omega$ is the K\"{a}hler form, and $B$ is the $B$-field on $\mathcal{M}^G_{\text{Higgs}}(C)$. The expression on the RHS of \eqref{bfieldterm} is the usual expression for the topological term in an $A$-model involving the complexified K\"{a}hler class, $\omega-iB$. In relation to the 4d theory, $B$ is the $\theta$-angle in the topological term of \eqref{bosonaction}.
\begin{figure}
    \centering
    \includegraphics[width=0.35\textwidth]{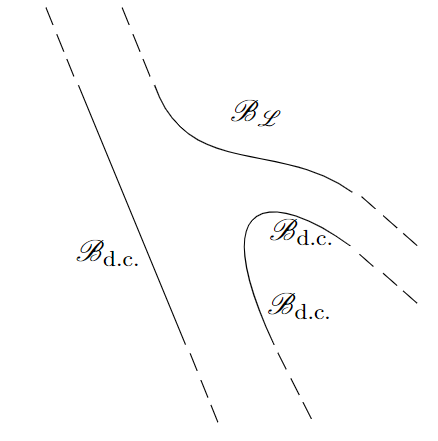} 
    \caption{Merging of string worldsheets along a common boundary $\mathscr{B}_{\text{d.c.}}$, representing the operation  $\mathcal{H}_{\mathscr{B}_{\text{d.c.}}, \mathscr{B}_{\text{d.c.}}}\otimes \mathcal{H}_{\mathscr{B}_{\text{d.c.}}, \mathscr{B}_{\mathscr{L}}}\to \mathcal{H}_{\mathscr{B}_{\text{d.c.}}, \mathscr{B}_{\mathscr{L}}}$.}
    \label{fig:mergesheet}
\end{figure}
It was shown in \cite{kapustin2008note} that if $B=0$, we can have a d.c.-brane (distinguished coisotropic) of type $(A,A,B)$ that is space-filling. Furthermore, it was also argued in~\cite{kapustin2008note} that in this case, the category of $A$-branes in $\mathcal{M}^G_{\text{Higgs}}(C)$ can be identified with a category of twisted $D$-modules on $\text{Bun}_{G_{\mathbb{C}}}(C)$, the moduli space of principal $G_{\mathbb{C}}$ bundles on $C$, where $G_{\mathbb{C}}$ is the complexified version of $G$. This latter claim can be understood as follows.

The $A$-model will have boundary conditions on both sides of the $I \times {\mathbb R}^+$ worldsheet, say boundary conditions 1 and 2, giving us $\mathscr{B}_{1}$ and $\mathscr{B}_{2}$ branes in $\mathcal{M}^G_{\text{Higgs}}(C)$. The strings suspended between these branes define a vector space $\mathcal{H}_{\mathscr{B}_{1}, \mathscr{B}_{2}}$ of $(\mathscr{B}_{1}, \mathscr{B}_{2})$-strings. For arbitrary branes $\mathscr{B}_{1}$, $\mathscr{B}_{2}$ and $\mathscr{B}_{3}$, we can have  $(\mathscr{B}_{1}, \mathscr{B}_{2})$ and $(\mathscr{B}_{2}$, $\mathscr{B}_{3})$-strings, where the operation $\mathcal{H}_{\mathscr{B}_{1}, \mathscr{B}_{2}}\otimes \mathcal{H}_{\mathscr{B}_{2}, \mathscr{B}_{3}}\to \mathcal{H}_{\mathscr{B}_{1}, \mathscr{B}_{3}}$ is physically equivalent to merging $(\mathscr{B}_{1}, \mathscr{B}_{2})$ and $(\mathscr{B}_{2}$, $\mathscr{B}_{3})$-strings along their common  boundary $\mathscr{B}_{2}$ to produce $(\mathscr{B}_{1}$, $\mathscr{B}_{3})$-strings. In particular, if $\mathscr{B}_{1}=\mathscr{B}_{2}=\mathscr{B}_{\text{d.c.}}$, where $\mathscr{B}_{\text{d.c.}}$ is the d.c.-brane, and $\mathscr{B}_{3}=\mathscr{B}_{\mathscr{L}}$, where $\mathscr{B}_{\mathscr{L}}$ is any Lagrangian brane, the operation can be understood physically as in Fig.~\ref{fig:mergesheet}. In this way, one can see that a $(\mathscr{B}_{\text{d.c.}}, \mathscr{B}_{\mathscr{L}})$-string is a module for a $(\mathscr{B}_{\text{d.c.}}$, $\mathscr{B}_{\text{d.c}})$-string. In turn, this means that the category of $A$-branes (spanned by the $\mathscr B_{\mathscr L}$'s) can be identified with the category of modules of $(\mathscr{B}_{\text{d.c.}}$, $\mathscr{B}_{\text{d.c}})$-strings. All that is left to explain is why $(\mathscr{B}_{\text{d.c.}}$, $\mathscr{B}_{\text{d.c}})$-strings can be identified with twisted differential operators on $\text{Bun}_{G_{\mathbb{C}}}(C)$.

To this end, note that at the classical level, the $(\mathscr{B}_{\text{d.c.}}$, $\mathscr{B}_{\text{d.c}})$-strings correspond to holomorphic functions on Hitchin moduli space in complex structure $J$. This space can be identified with the moduli space  of flat $G_{{\mathbb{C}}}$-connections on $C$, $\mathcal{M}^{G_{\mathbb{C}}}_{\text{flat}}(C)$, which is isomorphic to the twisted cotangent bundle ${\cal T}^\star\text{Bun}_{G_{\mathbb{C}}}(C)$~\cite{beilinson1988determinant, faltings1993stable}. In other words, classical $(\mathscr{B}_{\text{d.c.}}$, $\mathscr{B}_{\text{d.c}})$-strings can be interpreted as holomorphic functions on ${\cal T}^\star\text{Bun}_{G_{\mathbb{C}}}(C)$. The quantization of the $(\mathscr{B}_{\text{d.c.}}$, $\mathscr{B}_{\text{d.c}})$-strings then leads to their identification with (the sheaf of) holomorphic differential operators on the line bundle $\mathcal{L}^{-h^{\lor}+q}$ over $\text{Bun}_{G_{\mathbb{C}}}(C)$, where ${\cal L}^{-h^\vee} = K^{1/2}$, and $K$ is the canonical line bundle on $\text{Bun}_{G_{\mathbb{C}}}(C)$. Here, $h^{\lor}$ is the dual Coxeter number of $G$, and the parameter $q=\tau$ is purely imaginary because $B=0$.  

This is how the $\tau$-dependent category $\text{Cat}_{\text{$A$-branes}}\big(\tau, \mathcal{M}^{G}_{\text{Higgs}}(C)\big)$ of $A$-branes in $\mathcal{M}^G_{\text {Higgs}}(C)$, can be identified with a category ${\cal D}^{\textbf{c}}_{-h^\vee}\text{-mod}\big(q, {\text{Bun}_{G_{\mathbb{C}}}}\big) $ of twisted $D$-modules on $\text{Bun}_{G_{\mathbb{C}}}(C)$ with parameter $q$, where `$D$' refers to the differential operator we just described.


\subsection{A Quantum Geometric Langlands Correspondence with Purely Imaginary Parameter}

Note that from \eqref{vwdual} and \eqref{4d2dpartition}, 4d $S$-duality would mean that we have the 2d duality 
\begin{equation}\label{root}
    \mathcal{Z}_{A,\mathscr{B}}\big(\tau, \mathcal{M}^{G}_{\text{Higgs}}(C)\big) 	\longleftrightarrow \mathcal{Z}_{A,\mathscr{B}}\Big(-\frac{1}{n_{\mathfrak{g}}\tau}, \,\mathcal{M}^{^LG}_{\text{Higgs}}(C)\Big), 
\end{equation}
where $\mathcal{Z}_{A,\mathscr{B}}$ is the partition function of the open $A$-model with branes $\mathscr B$. 

In turn, this implies a homological mirror symmetry of the $\tau$-dependent category of $A$-branes:
\begin{equation}\label{dualcata}
    \boxed{\text{Cat}_{\text{$A$-branes}}\big(\tau, \mathcal{M}^{G}_{\text{Higgs}}(C)\big) \longleftrightarrow
    \text{Cat}_{\text{$A$-branes}}\Big(-\frac{1}{n_{\mathfrak{g}}\tau}, \,\mathcal{M}^{^LG}_{\text{Higgs}}(C)\Big) }
\end{equation}
where $\mathcal{M}^{G}_{\text{Higgs}}$ and $\mathcal{M}^{^LG}_{\text{Higgs}}$ are mirror manifolds. 
 
 As explained above, for $\theta = B=0$, the category of $\tau$-dependent $A$-branes can be identified with a category of twisted $D$-modules on $\text{Bun}_{G_{\mathbb{C}}}(C)$ with parameter $q$. Thus, this mirror symmetry would mean that we have 
\begin{equation}\label{dualdmod}
    \boxed{{\cal D}^{\textbf{c}}_{-h^\vee}\text{-mod}\big(q, {\text{Bun}_{G_{\mathbb{C}}}}\big) \longleftrightarrow {\cal D}^{\textbf{c}}_{-^Lh^\vee}\text{-mod}\Big(-\frac{1}{n_{\mathfrak{g}}q}, \,{\text{Bun}_{{^LG}_{\mathbb{C}}}}\Big)}
\end{equation}
This is a quantum geometric Langlands correspondence for $G_{\mathbb{C}}$ with complex curve $C$ and purely imaginary parameter $q$ \cite[eqn.~(6.4)]{frenkel2005lectures}. 

\subsection{A Classical Geometric Langlands Correspondence}

In the zero-coupling, `classical' limit of the 4d theory in $G$ where $\text{Im}(\tau) \to \infty$, we have $q \to \infty$. In this limit, the LHS of \eqref{dualdmod} can be identified with the category $\text{Cat}_{\text{coh}}\big (\mathcal{M}_{\text{flat}}^{G_{\mathbb{C}}}(C) \big )$ of coherent sheaves on $\mathcal{M}_{\text{flat}}^{G_{\mathbb{C}}}(C)$~\cite{frenkel2005lectures}.     

This `classical' limit corresponds to the `ultra-quantum' limit of the $S$-dual 4d theory in $^LG$, where $^Lq = -\frac{1}{n_{\mathfrak{g}}q} \to 0$. In this limit, the RHS of \eqref{dualdmod} can be identified with the category ${\cal D}^{\textbf{c}}_{-{^Lh}^\vee}\text{-mod}\big(0, {\text{Bun}_{{^LG}_{\mathbb{C}}}}\big) $ of critically-twisted $D$-modules on $\text{Bun}_{{^LG}_{\mathbb{C}}}(C)$.

In short, we have 
\begin{equation}\label{classical GL}
    \boxed{\text{Cat}_{\text{coh}}\big (\mathcal{M}_{\text{flat}}^{G_{\mathbb{C}}}(C) \big ) \longleftrightarrow {\cal D}^{\textbf{c}}_{-^Lh^\vee}\text{-mod}\Big(0, \,{\text{Bun}_{{^LG}_{\mathbb{C}}}}\Big)}
    \end{equation}
    This is a classical geometric Langlands correspondence for $G_{\mathbb{C}}$ with complex curve $C$ \cite[eqn.~(6.4)]{frenkel2005lectures}.

\section{A Novel Web of Mathematical Relations, and Categorification}\label{sec:web of dualities}

In this final section, we will show how the dualities, correspondences and identifications between the various mathematical objects we physically derived in $\S$\ref{vwgeneral}--\ref{sec:geometric langlands} starting from VW theory, will lead us to a novel web of mathematical relations. We will then explain how the VW invariant will be systematically categorified in our framework.

\subsection{A Novel Web of Mathematical Relations from Vafa-Witten Theory}

Essentially, from the duality relations \eqref{vwdual}, \eqref{ZGW=ZGW}, \eqref{HFVWG to HFVWLG}, \eqref{HFL to HFL}, the correspondences \eqref{dualcata}, \eqref{dualdmod}, \eqref{classical GL}, and the identifications \eqref{4d2dpartition}, \eqref{4d3dpartition}, \eqref{atconj}, we will get Fig.~\ref{fig:equimath} below.

\begin{figure} 
    \centering
    \begin{tikzcd}[row sep=50, column sep=-32]
\boxed{\text{HF}^{\text{Lagr}}_\ast (\mathcal{M}^G_{\text{Higgs}}(C), L_0, L_1, \tau)} 
\arrow[rrr, leftrightarrow, dashed, "\substack{\text{Lagrangian Floer}\\\text{Langlands}\\\text{ duality}}"]
\arrow[dr, leftrightarrow,  " \substack{\text{VW Atiyah-Floer}\\\text{correspondence}}", " \substack{\textbf{d}:\\\text{ Heegaard split}\\ \text{of $M_3$ along $C$,}\\C\to 0}"']
&&&\boxed{\text{HF}^{\text{Lagr}}_\ast (\mathcal{M}^{^LG}_{\text{Higgs}}(C), L_0, L_1, -1/n_{\mathfrak{g}}\tau)} 
\arrow[dr, leftrightarrow,  " \substack{\text{VW Atiyah-Floer}\\\text{correspondence}}", "\textbf{d}"']
\\
&\boxed{\text{HF}^{\text{VW}}_\ast(M_3, G,\tau)}  
\arrow[rrr, leftrightarrow, dashed,"\substack{\text{VW-Floer}\\\text{Langlands duality}}"]
\arrow[dl, leftrightarrow, "\textbf{b: }M_4=M_3 \times \mathbb{R}^+"]
&&&\boxed{\text{HF}^{\text{VW}}_\ast (M_3,\, ^LG,-1/n_{\mathfrak{g}}\tau)} 
\arrow[dl, leftrightarrow, "\textbf{b }"]\\
\boxed{\bf{\boldsymbol{\mathcal{Z}}_{\textbf{VW},M_4}(\boldsymbol{\tau}, G)} }
\arrow[rrr, leftrightarrow, dashed, "\textbf{Langlands dual}"]
\arrow[dr, leftrightarrow, "\substack{\textbf{a: }\text{$M_4=\Sigma \times C$,}\\ \text{$C\to 0$}}"]
&&&\boxed{\bf{\boldsymbol{\mathcal{Z}}_{\textbf{VW},M_4}\Big(-\frac{1}{n_{\mathfrak{g}}\boldsymbol{\tau}},\, ^LG \Big)} }
\arrow[dr, leftrightarrow, "\textbf{a }"]\\
&\boxed{\mathcal{Z}_{\text{GW},\Sigma}(\tau, \mathcal{M}^G_{\text{Higgs}}(C))}
\arrow[rrr, leftrightarrow, dashed, "\substack{\text{Mirror symmetry}\\\text{of Higgs bundles}}", crossing over]
&&&\boxed{\mathcal{Z}_{\text{GW},\Sigma}(-1/n_{\mathfrak{g}}\tau,\,\mathcal{M}^{^LG}_{\text{Higgs}}(C)) }  \\
\boxed{\text{Cat}_{\text{$A$-branes}}\big(\tau, \mathcal{M}^{G}_{\text{Higgs}}(C)\big)}
\arrow[d,leftrightarrow,  "\text{Re}(\tau) = 0"']
\arrow[uu, leftrightarrow,  " \substack{{\textbf{c: }M_4=I\times\mathbb{R}^+\times C,}\\{ C\to 0}}"] 
\arrow[rrr, leftrightarrow, dashed, "\substack{\text{Homological}\\\text{ mirror symmetry }\\\text{of Higgs}\\\text{ bundles}}", crossing over]
&&&\boxed{\text{Cat}_{\text{$A$-branes}}\Big(-\frac{1}{n_{\mathfrak{g}}\tau}, \,\mathcal{M}^{^LG}_{\text{Higgs}}(C)\Big)}
\arrow[uu, leftrightarrow, "\substack{\\\\\\\\\\\textbf{c }}"]
\arrow[d, leftrightarrow, "\text{Re}(\tau) = 0"]\\
\boxed{{\cal D}^{\textbf{c}}_{-h^\vee}\text{-mod}\big(q, {\text{Bun}_{G_{\mathbb{C}}}}\big)}
\arrow[rrr, leftrightarrow, dashed, "\substack{{\text{Quantum geometric}}\\{\text{ Langlands}}\\{\text{ correspondence}}}"]
\arrow[d, leftrightarrow, "\tau \to \infty"]
&&&\boxed{{\cal D}^{\textbf{c}}_{-^Lh^\vee}\text{-mod}\Big(-\frac{1}{n_{\mathfrak{g}}q}, \,{\text{Bun}_{^L{G_{\mathbb{C}}}}}\Big)}
\arrow[d, leftrightarrow, "\tau \to \infty"]\\
\boxed{\text{Cat}_{\text{coh}}\big (\mathcal{M}_{\text{flat}}^{G_{\mathbb{C}}}(C) \big )}
\arrow[rrr, leftrightarrow, dashed, "\substack{{\text{Classical geometric}}\\{\text{ Langlands}}\\{\text{ correspondence}}}"]
&&&\boxed{{\cal D}^{\textbf{c}}_{-^Lh^\vee}\text{-mod}\Big(0, \,{\text{Bun}_{{^LG}_{\mathbb{C}}}}\Big)}
\end{tikzcd}
\caption{A novel web of mathematical relations stemming from Vafa-Witten theory.}
\label{fig:equimath}
\end{figure}
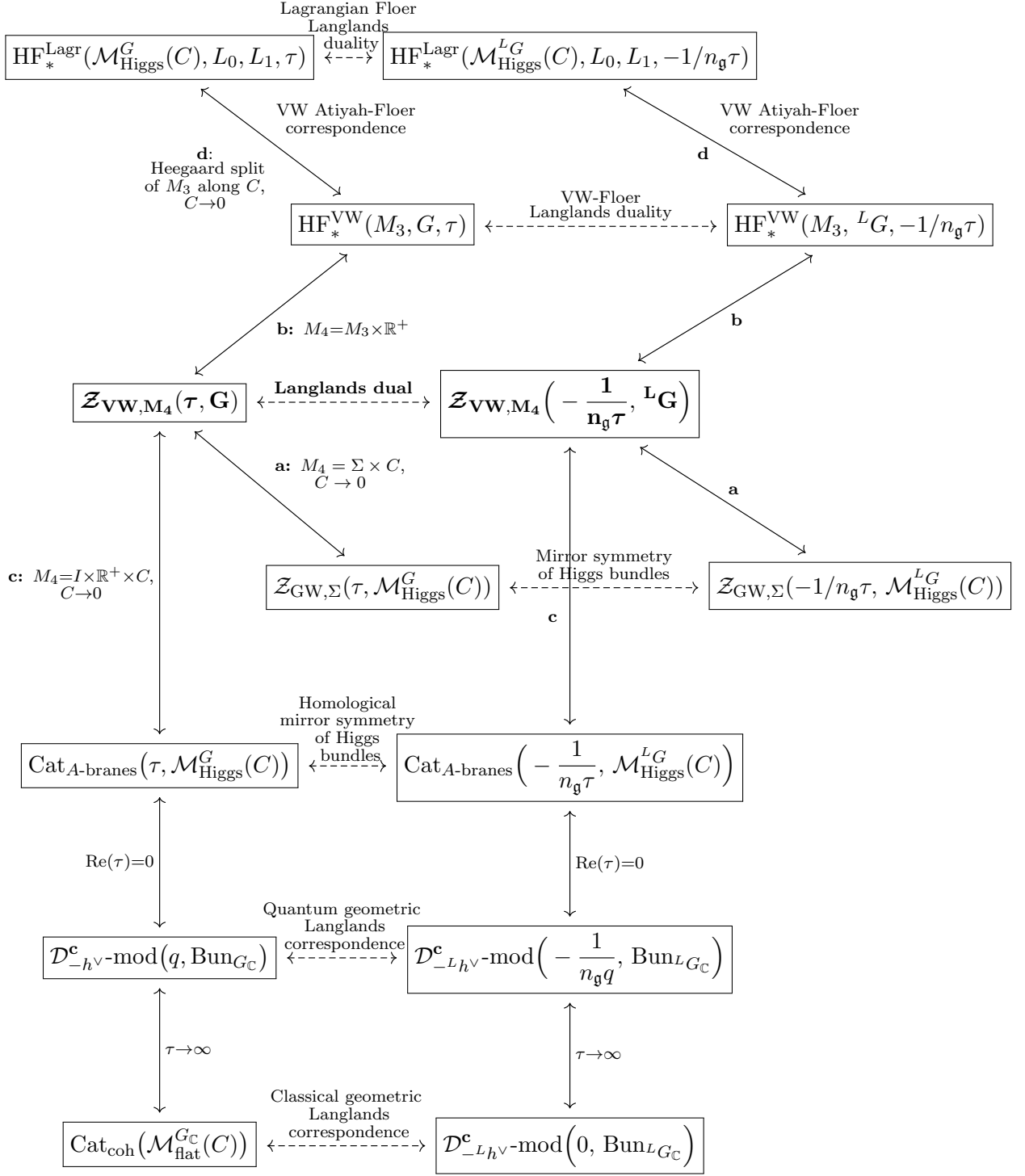

\subsection{Categorifying the Vafa-Witten Invariant}

Categorification is a mathematical procedure that turns a number into a vector space, a vector space into a category, a category into a 2-category, and so on:
\begin{equation}
{\text{number}} \xrightarrow{\text{categorification}} {\text{vector space}} \xrightarrow{\text{categorification}} {\text{category}} \xrightarrow{\text{categorification}} {\text{2-category}}  \xrightarrow{\text{categorification}} \cdots
\end{equation}

From Fig.~\ref{fig:equimath}, one can see that this mathematical procedure is actually realized in our physical framework. Specifically, via the arrows $\bf b$ and $\bf c$, and the fact that the VW invariant is a number, the VW Floer homology is a vector (space), and the $A$-branes span a category of objects, we find that \footnote{This perspective of categorifying topological invariants by successively introducing boundaries to the manifold was first pointed out in \cite{gukov2009surface}.}
\begin{equation}\label{category of objects}
    \begin{aligned}
        &\text{VW theory on } M_4 \quad &\leadsto  \quad& \text{number }  \qquad\mathcal{Z}_{\text{VW}} \\
        &\text{VW theory on } \mathbb{R}^+ \times M_3 \quad &\leadsto  \quad& \text{vector }  \qquad\text{HF}^{\text {VW}}_*\\
        &\text{VW theory on } \mathbb{R}^+ \times I \times C \quad &\leadsto  \quad& \text{1-category } \text{Cat}_{A\text{-branes}}\quad .
    \end{aligned}
\end{equation}

In other words, we have  
\begin{equation} \label{categorifying VW}
\boxed{{{\cal Z}_{\text{VW}}} \xrightarrow{\text{categorification}} {{\text {HF}}^{\text {VW}}_*} \xrightarrow{\text{categorification}} {\text{Cat}_{A\text{-branes}}}}
\end{equation}
a categorification of ${{\cal Z}_{\text{VW}}}$, the VW invariant of $M_4$.

From \eqref{category of objects}, it is clear that categorification can be physically understood as flattening a direction and then ending it on a boundary or boundaries. Explicitly in our case, the first step of categorification involves flattening a direction in $M_4$ and then ending it on an $M_3$ boundary, while the second step involves flattening a direction in $M_3$ and then ending it on two $C$ boundaries. Therefore, one can also understand the procedure of categorifying as computing relative invariants\footnote{A relative invariant is an invariant of an open manifold which was originally defined for a closed manifold.} -- computing the relative invariant of ${\cal Z}_{\text {VW}}$ give us ${\text {HF}}^{\text {VW}}_*$, and further computing the relative invariant of ${\text {HF}}^{\text {VW}}_*$ gives us $\text{Cat}_{A\text{-branes}}$.

All this is also consistent with the fact pointed out in~\cite{kapustin2010topological} that an $n$-dimensional TQFT assigns a $k$-category to a closed $n-k-1$-manifold $M$. Here in our case, we have $n=4$, and when $k = 0$ and $1$, we have the 0-category ${\text {HF}}^{\text {VW}}_*$ assigned to a closed 3-manifold $M_3$ and the 1-category $\text{Cat}_{A\text{-branes}}$ assigned to a closed 2-manifold $C$, respectively.

\subsection{Higher Categories from Vafa-Witten Theory}

\bigskip\noindent\textit{A 2-category from Vafa-Witten Theory}

One could continue to further categorify the VW invariant of $M_4$ by flattening a direction along $C$ and ending it on $S^1$ boundaries, i.e., let $C = I' \times S^1$. This should give us a 2-category, 2-Cat, consisting of objects, morphisms between these objects, and 2-morphisms between these morphisms. Thus, we have an extension of \eqref{category of objects} to 
\begin{equation}\label{category of objects- extend}
    \begin{aligned}
        &\text{VW theory on } M_4 \quad &\leadsto  \quad& \text{number } \quad \qquad\mathcal{Z}_{\text{VW}} \\
        &\text{VW theory on } \mathbb{R}^+ \times M_3 \quad &\leadsto  \quad& \text{vector } \quad \qquad\text {HF}^{\text {VW}}_*\\
        &\text{VW theory on } \mathbb{R}^+ \times I \times C \quad &\leadsto  \quad& \text{1-category } \quad\text{Cat}_{A\text{-branes}}\\
        &\text{VW theory on } \mathbb{R}^+ \times I \times I' \times S^1 \quad &\leadsto  \quad& \text{2-category } \quad \text{2-Cat}\quad .
    \end{aligned}
\end{equation}

Let us now determine 
what this 2-category ought to be. 

First, note that now, we have VW theory on $\mathbb{R}^+ \times I \times I' \times S^1$ -- in other words, we have VW theory compactified on $S^1$ to a 3d TQFT on a semi-infinite block starting at $t=0$ with $\mathbb{R}^+ \times I$ boundaries. The sought-after 2-category is then the 2-category of boundary conditions of this 3d TQFT.\footnote{Just as the 1-category discussed in the previous subsection is the 1-category of boundary conditions of the 2d $A$-model.} 

Second, notice that the aforementioned boundary conditions can be realized by surface defects in VW theory that lie along the $\mathbb{R}^+ \times I \, (I')$ boundaries of the 3d TQFT. In other words, the 2-category we seek is the 2-category of these surface defects in VW theory. From this viewpoint, the surface defects can be interpreted as objects; loop defects on the surface running around $I \times I'$ can be interpreted as morphisms between these objects; while opposing pairs of point defects on the loops can be interpreted as 2-morphisms between these morphisms. 

Third, note that the 3d TQFT in question is a 3d gauged $A$-model described in \cite[sect. 7]{kapustin2010surface},\footnote{In \cite[sect. 7]{kapustin2010surface}, the GL theory at $t=0$ was considered, but it was shown in \cite[sect. 5.2-5.3]{setter2013topological} that this theory compactified on $S^1$ is the same as VW theory compactified on $S^1$. Hence, their results are applicable to us.} and for abelian $G$ and $\text{Re}(\tau)=0$, the 2-category of surface defects have been explicitly determined in $\it{loc.\,cit.}$ to be the 2-category $\text{2-Cat}_{\text{mod-cat}} \big({\text{FF-cat}}(T^2)\big)$ of module categories over the Fukaya-Floer category of $T^2$.\footnote{The 3d gauged $A$-model has a gauge and matter sector, where each sector can either have Dirichlet (D) or Neumann (N) boundary conditions. We have stated the result for the DD case, as this choice of boundary conditions allows us to describe the situation where line defects lie along the surface defects, which is the one relevant to us.} Therefore, we have, for abelian $G$ and $\text{Re}(\tau)=0$, an extension of \eqref{categorifying VW} to   
\begin{equation} \label{categorifying VW - extend}
\boxed{   {{{\cal Z}_{\text{VW}}} \xrightarrow{\text{categorification}} {{\text {HF}}^{\text {VW}}_*} \xrightarrow{\text{categorification}} {\text{Cat}_{A\text{-branes}}}}
\xrightarrow{\text{categorification}} {\text{2-Cat}_{\text{mod-cat}} \big({\text{FF-cat}}(T^2)\big)} }
\end{equation}

Notice that in this case, we have $n=4$ and $k=2$ in our discussion at the end of the previous subsection, whence we ought to have a 2-category assigned to the closed 1-manifold $S^1$. Indeed, as is clear from \eqref{category of objects- extend} we have a 2-category of surface defects that are assigned to a closed 1-manifold $S^1$.  

\bigskip\noindent\textit{Langlands Duality of a 2-category}

Observe from \eqref{category of objects- extend} and Fig.~\ref{fig:equimath} that from 4d $S$-duality, we have a Langlands duality of the 0-category ${\text{HF}}^{\text{VW}}_*$, and a Langlands duality (mirror symmetry) of the 1-category $\text{Cat}_{A \text{-branes}}$. Do we then also have a Langlands duality of the 2-category 2-Cat from 4d $S$-duality? The answer is `yes'. 

According to \cite[sect. 7.4.1]{kapustin2010surface}, 4d $S$-duality, which maps abelian $G$ to its Langlands dual that is itself, will transform the symplectic area $\cal A$ of $T^2$ as
\begin{equation}
    {\cal A} \to {^L{\cal A}} = {4 \pi^2 \over {\cal A}},
\end{equation}
where ${^L{\cal A}}$ is the symplectic area of a torus $^LT^2$ that can be obtained from $T^2$ by inverting the radii of its two circles from $R \to {\alpha'/ R}$ for some constant $\alpha'$. In other words, $^LT^2$ is the $T$-dual torus to $T^2$, and FF-cat($T^2$), which is realized by a 2d open $A$-model with target $T^2$, will be invariant under $T$-duality of the target, i.e., FF-cat($T^2$) $\cong$ FF-cat($^LT^2$). Thus, we have
\begin{equation} \label{2- Cat to 2-Cat}
 \boxed{   \text{2-Cat}_{\text{mod-cat}} \big({\text{FF-cat}}(T^2)\big) \longleftrightarrow \text{2-Cat}_{\text{mod-cat}} \big({\text{FF-cat}}(^LT^2)\big) }
\end{equation}

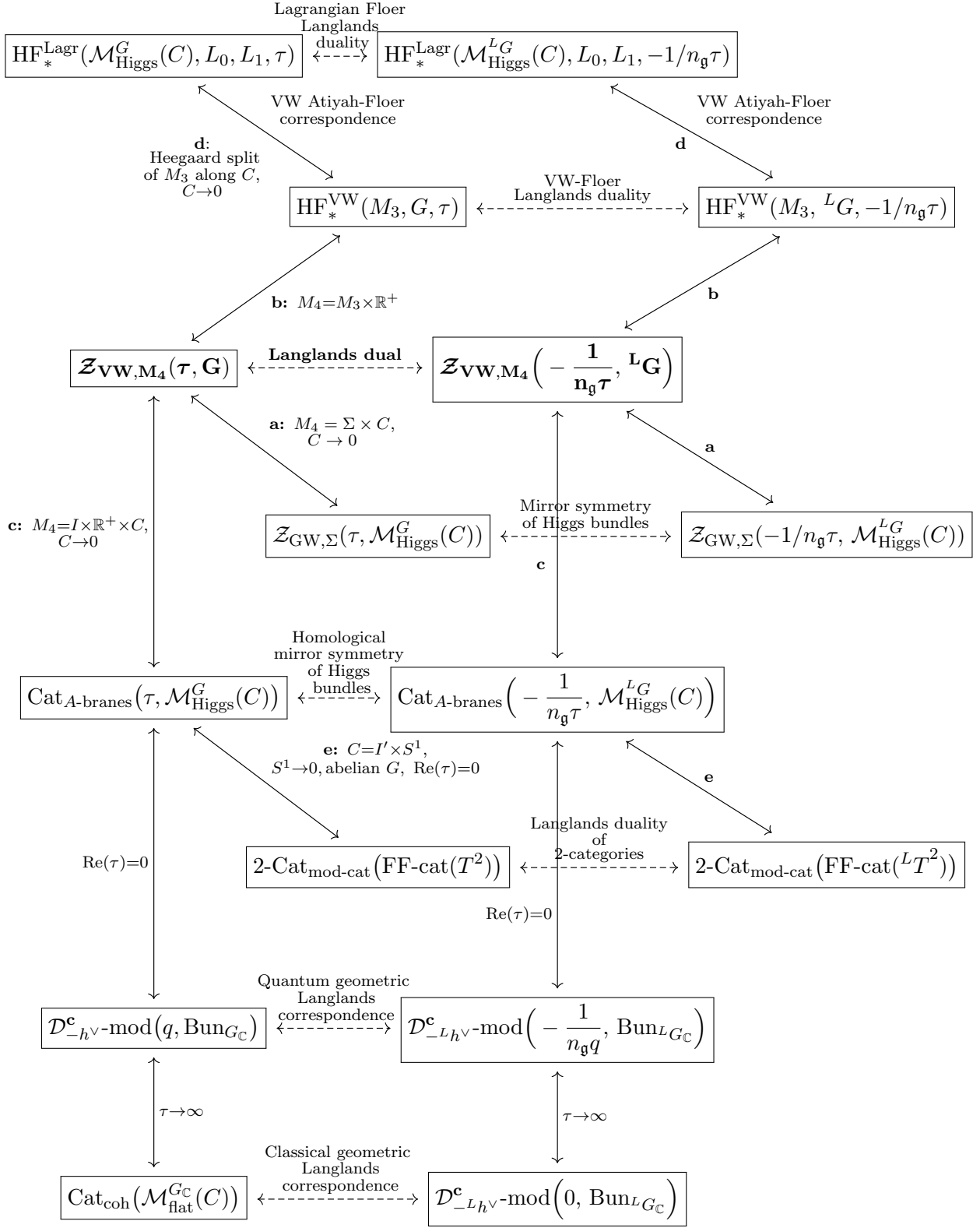
\begin{figure} 
    \centering
    \begin{tikzcd}[row sep=45, column sep=-37]
\boxed{\text{HF}^{\text{Lagr}}_\ast (\mathcal{M}^G_{\text{Higgs}}(C), L_0, L_1, \tau)} 
\arrow[rrr, leftrightarrow, dashed, "\substack{\text{Lagrangian Floer}\\\text{Langlands}\\\text{ duality}}"]
\arrow[dr, leftrightarrow,  " \substack{\text{VW Atiyah-Floer}\\\text{correspondence}}", " \substack{\textbf{d}:\\\text{ Heegaard split}\\ \text{of $M_3$ along $C$,}\\C\to 0}"']
&&&\boxed{\text{HF}^{\text{Lagr}}_\ast (\mathcal{M}^{^LG}_{\text{Higgs}}(C), L_0, L_1, -1/n_{\mathfrak{g}}\tau)} 
\arrow[dr, leftrightarrow,  " \substack{\text{VW Atiyah-Floer}\\\text{correspondence}}", "\textbf{d}"']
\\
&\boxed{\text{HF}^{\text{VW}}_\ast(M_3, G, \tau)}  
\arrow[rrr, leftrightarrow, dashed,"\substack{\text{VW-Floer}\\\text{Langlands duality}}"]
\arrow[dl, leftrightarrow, "\textbf{b: }M_4=M_3 \times \mathbb{R}^+"]
&&&\boxed{\text{HF}^{\text{VW}}_\ast (M_3,\, ^LG, -1/n_{\mathfrak{g}}\tau)} 
\arrow[dl, leftrightarrow, "\textbf{b }"]\\
\boxed{\bf{\boldsymbol{\mathcal{Z}}_{\textbf{VW},M_4}(\boldsymbol{\tau}, G)} }
\arrow[rrr, leftrightarrow, dashed, "\textbf{Langlands dual}"]
\arrow[dr, leftrightarrow, "\substack{\textbf{a: }\text{$M_4=\Sigma \times C$,}\\ \text{$C\to 0$}}"]
&&&\boxed{\bf{\boldsymbol{\mathcal{Z}}_{\textbf{VW},M_4}\Big(-\frac{1}{n_{\mathfrak{g}}\boldsymbol{\tau}},\, ^LG \Big)} }
\arrow[dr, leftrightarrow, "\textbf{a }"]\\
&\boxed{\mathcal{Z}_{\text{GW},\Sigma}(\tau, \mathcal{M}^G_{\text{Higgs}}(C))}
\arrow[rrr, leftrightarrow, dashed, "\substack{\text{Mirror symmetry}\\\text{of Higgs bundles}}", crossing over]
&&&\boxed{\mathcal{Z}_{\text{GW},\Sigma}(-1/n_{\mathfrak{g}}\tau,\,\mathcal{M}^{^LG}_{\text{Higgs}}(C)) }  \\
\boxed{\text{Cat}_{\text{$A$-branes}}\big(\tau, \mathcal{M}^{G}_{\text{Higgs}}(C)\big)}
\arrow[dd,leftrightarrow,  "\text{Re}(\tau) = 0"']
\arrow[dr, leftrightarrow, " \substack{\textbf{e: } C = I' \times S^1,\\ S^1 \to 0,\, \text{abelian $G$, }\,\text{Re}(\tau) = 0}"]
\arrow[uu, leftrightarrow,  " \substack{{\textbf{c: }M_4=I\times\mathbb{R}^+\times C,}\\{ C\to 0}}"] 
\arrow[rrr, leftrightarrow, dashed, "\substack{\text{Homological}\\\text{ mirror symmetry }\\\text{of Higgs}\\\text{ bundles}}", crossing over]
&&&\boxed{\text{Cat}_{\text{$A$-branes}}\Big(-\frac{1}{n_{\mathfrak{g}}\tau}, \,\mathcal{M}^{^LG}_{\text{Higgs}}(C)\Big)}
\arrow[uu, leftrightarrow, "\substack{\\\\\\\\\\\textbf{c }}"]
\arrow[dd, leftrightarrow, "\substack{\\\\\\\\\\\\\\\\\text{Re}(\tau) = 0}"']
\arrow[dr, leftrightarrow, " \textbf{e}"]\\
&\boxed{\text{2-Cat}_{\text{mod-cat}} \big({\text{FF-cat}}(T^2)\big)}
\arrow[rrr, leftrightarrow, dashed, "\substack{{\text{Langlands duality}}\\{\text{ of }}\\{\text{2-categories}}}"]
&&&\boxed{\text{2-Cat}_{\text{mod-cat}} \big({\text{FF-cat}}({^LT}^2)\big)}\\
\boxed{{\cal D}^{\textbf{c}}_{-h^\vee}\text{-mod}\big(q, {\text{Bun}_{G_{\mathbb{C}}}}\big)}
\arrow[rrr, leftrightarrow, dashed, "\substack{{\text{Quantum geometric}}\\{\text{ Langlands}}\\{\text{ correspondence}}}"]
\arrow[d, leftrightarrow, "\tau \to \infty"]
&&&\boxed{{\cal D}^{\textbf{c}}_{-^Lh^\vee}\text{-mod}\Big(-\frac{1}{n_{\mathfrak{g}}q}, \,{\text{Bun}_{^L{G_{\mathbb{C}}}}}\Big)}
\arrow[d, leftrightarrow, "\tau \to \infty"]\\
\boxed{\text{Cat}_{\text{coh}}\big (\mathcal{M}_{\text{flat}}^{G_{\mathbb{C}}}(C) \big )}
\arrow[rrr, leftrightarrow, dashed, "\substack{{\text{Classical geometric}}\\{\text{ Langlands}}\\{\text{ correspondence}}}"]
&&&\boxed{{\cal D}^{\textbf{c}}_{-^Lh^\vee}\text{-mod}\Big(0, \,{\text{Bun}_{{^LG}_{\mathbb{C}}}}\Big)}
\end{tikzcd}
\caption{A novel web of mathematical relations stemming from Vafa-Witten theory that also involves higher categories.}
\label{fig:equimath - cat}
\end{figure}

Hence, Fig.~\ref{fig:equimath} will be enhanced to Fig.~\ref{fig:equimath - cat}. 


\bigskip\noindent\textit{A 3-category from Vafa-Witten Theory?}

We could take one last step to further categorify the VW invariant of $M_4$ by flattening $S^1$ and ending it on point boundaries, i.e., let $S^1 = [0,1]$. This should give us a 3-category, 3-Cat, consisting of objects, morphisms between these objects, 2-morphisms between these morphisms, and 3-morphisms between these 2-morphisms. Thus, we have yet another extension of \eqref{category of objects} to 
\begin{equation}\label{category of objects-extendII}
    \begin{aligned}
        &\text{VW theory on } M_4 \quad &\leadsto  \quad& \text{number }  \quad\qquad\mathcal{Z}_{\text{VW}} \\
        &\text{VW theory on } \mathbb{R}^+ \times M_3 \quad &\leadsto  \quad& \text{vector } \quad\qquad \text {HF}^{\text {VW}}_*\\
        &\text{VW theory on } \mathbb{R}^+ \times I \times C \quad &\leadsto  \quad& \text{1-category } \quad\text{Cat}_{A\text{-branes}}\\
        &\text{VW theory on } \mathbb{R}^+ \times I \times I' \times S^1 \quad &\leadsto  \quad& \text{2-category } \quad\text{2-Cat}\quad\\ 
        &\text{VW theory on } \mathbb{R}^+ \times I \times I' \times [0,1] \quad &\leadsto  \quad& \text{3-category } \quad\text{3-Cat}\quad.
    \end{aligned}
\end{equation}
That is, we have a 3-category of 3d boundary conditions of  VW theory along $\mathbb R^+ \times I \times I'$ which is assigned to a point.


These 3d boundary conditions can be realized by domain walls. So, the sought-after 3-category has domain walls along $\mathbb{R}^+ \times I \times I'$ as objects; surface defects within the domain walls along $I \times I'$ as morphisms between these objects; line defects on the surfaces in the $I$ or $I'$ direction as 2-morphisms of these morphisms; and point defects on the lines as 3-morphisms of these 2-morphisms.

Determining the classification of such domain walls in VW theory is beyond the scope of this paper, and we shall leave it for future work. In short, we can summarize how ${\cal Z}_{\text{VW}}$, the VW invariant of $M_4$, can be completely categorified as
\begin{equation} \label{categorifying VW - extend II}
\hspace{0.0cm} \boxed{   {{{\cal Z}_{\text{VW}}} \xrightarrow{\text{categorify}} {{\text {HF}}^{\text {VW}}_*} \xrightarrow{\text{categorify}} {\text{Cat}_{A\text{-branes}}}}
\xrightarrow{\text{categorify}} {\text{2-Cat}_{\text{mod-cat}} \big({\text{FF-cat}}(T^2)\big)}
\xrightarrow{\text{categorify}} {\text{3-Cat (?)}}
}
\end{equation}
where 2-Cat remains to be determined for non-abelian $G$, while 3-Cat has yet to be determined for any $G$. 
\printbibliography
\end{document}